\documentclass[twocolumn,trackchanges]{aastex631}

\newcommand{\Msun}{\ensuremath{\textrm{\,$M$}_{\odot}}}
\newcommand{\bcep}{\ensuremath{\textrm{\,$\beta$}}~Cep}
\usepackage{amsmath}
\usepackage{booktabs}
\usepackage{hyperref}
\usepackage{natbib}

\shorttitle{\bcep\ Catalog}
\shortauthors{Shitrit et al.}

\graphicspath{{./}}

\begin{document}

\title{A Catalog of \bcep\ Stars and Their Characterization}

\author[0000-0001-9511-6054]{Noi Shitrit}
\affiliation{The School of Physics and Astronomy, Tel Aviv University, Tel Aviv 69978, Israel}
\author[0000-0001-7756-1568]{Gerald Handler}
\affiliation{Nicolaus Copernicus Astronomical Center (NCAC), Polish Academy of Sciences, 00-716 Warsaw, Poland}
\author[0000-0001-7090-4898]{Iair Arcavi}
\affiliation{The School of Physics and Astronomy, Tel Aviv University, Tel Aviv 69978, Israel}
\author[0000-0002-9514-7210]{Filiz Kahraman Ali\c{c}avu\c{s}}
\affiliation{Çanakkale Onsekiz Mart University, Science Faculty, Physics Department, 17100, Canakkale, Türkiye}
\affiliation{Çanakkale Onsekiz Mart University, Astrophysics Research Center and Ulupınar Observatory, TR-17100, Çanakkale, Türkiye}
\author[0000-0002-0420-2473]{Mathijs Vanrespaille}
\affiliation{Institute of Astronomy, KU Leuven, Celestijnenlaan 200D, B-3001 Leuven, Belgium}
\author[0000-0003-1822-7126]{Conny Aerts}
\affiliation{Institute of Astronomy, KU Leuven, Celestijnenlaan 200D, B-3001 Leuven, Belgium}
\affiliation{Department of Astrophysics, IMAPP, Radboud University Nijmegen, PO Box 9010, 6500 GL Nijmegen, The Netherlands}
\affiliation{Max Planck Institut für Astronomie, Königstuhl 17, 69117 Heidelberg, Germany}
\author[0000-0001-7444-5131]{Sowgata Chowdhury}
\affiliation{Nicolaus Copernicus Astronomical Center (NCAC), Polish Academy of Sciences, 00-716 Warsaw, Poland}
\author[0000-0003-2488-6726]{Andrzej Pigulski}
\affiliation{Astronomical Institute Wroc{\l}aw University, ul. Kopernika 11, 51-622 Wroc{\l}aw, Poland}
\author{Micha{\l} Nowak}
\affiliation{Warsaw University Observatory, 00-478 Warsaw, Poland}
\author{John Hoffman}
\affiliation{Department of Astrophysical Sciences, Princeton University, NJ 08544, USA}
\author[0000-0002-2919-6786]{Jonathan Labadie-Bartz}
\affiliation{Lehigh University, Physics Department, Bethlehem, PA, 18015, USA}
\author[0000-0002-6495-0676]{Grzegorz Pojma\'nski}
\affiliation{Warsaw University Observatory, 00-478 Warsaw, Poland}
\author[0000-0001-7204-6727]{Gaspar \'A. Bakos}\
\affiliation{Department of Astrophysical Sciences, Princeton University, NJ 08544, USA}
\author[0000-0002-7777-0842]{Igor Soszy\'nski}
\affiliation{Astronomical Observatory, University of Warsaw, Al. Ujazdowskie 4, 00-478 Warszawa, Poland}
\author[0000-0001-7402-3852]{Dominic M. Bowman}
\affiliation{School of Mathematics, Statistics and Physics, Newcastle University, Herschel Building,
Newcastle upon Tyne, NE1 7RU, United Kingdom}
\author[0000-0002-2926-5171]{Keegan Thomson-Paressant}
\affiliation{School of Mathematics, Statistics and Physics, Newcastle University, Herschel Building,
Newcastle upon Tyne, NE1 7RU, United Kingdom}
\author[0000-0002-3481-9052]{Keivan G. Stassun}
\affiliation{Vanderbilt University, Department of Physics \& Astronomy, 6301 Stevenson Center Lane, Nashville, TN 37235, USA}

\correspondingauthor{Noi Shitrit}
\email{noish26@gmail.com}

\begin{abstract}
$\beta$~Cep stars are pulsating late O- and early B-type stars. Their oscillations can be used, through asteroseismology, to obtain constraints on the internal structure of massive stars. In the past two decades, space and ground-based surveys have detected hundreds of potential $\beta$~Cep stars. In this work, we collect all $\beta$~Cep stars reported in the literature since the last $\beta$~Cep catalog was published more than two decades ago. We also perform an archival search of ground-based surveys, which, together with the literature search, yields 2,543 potential $\beta$~Cep stars (of which 499 are newly reported here). We collect spectra for 67 stars, which we analyze and model to derive stellar parameters and compare with Gaia estimates. We perform visual classification of the 2,421 stars with available TESS data by inspecting their light curves, periodograms, spectral types and temperatures. We confirm 1,130 $\beta$~Cep pulsators, including hybrids and eclipsing binary systems, of which 180 are newly reported here, and an additional 246 $\beta$~Cep candidate stars, of which 18 are new. This constitutes an order of magnitude increase of confirmed $\beta$~Cep stars since the previous catalog published in 2005. We reject 1,045 stars as $\beta$~Cep stars and reclassify most of them as candidate slowly pulsating B-type, $\delta$~Scuti, or $\gamma$~Doradus pulsators. We conduct frequency analysis on all stars classified as $\beta$~Cep and perform statistical analyses of their stellar and pulsation parameters. This catalog presents the largest compilation of $\beta$~Cep pulsators to date, including pulsation parameters and reliable visually-classified pulsators for use in future studies.
\end{abstract}


\section{Introduction} \label{sec:intro}
Massive stars play a fundamental role in astrophysics. They produce strong winds, emit ionizing radiation, and most end their lives in core-collapse supernovae, leaving behind compact remnants. Understanding these phenomena, which is essential for building a complete picture of both stellar and galactic evolution, requires a detailed understanding of the internal structure of massive stars and physical processes occurring in their interiors. 

Asteroseismology provides a way to probe the internal structure and evolution of massive stars by studying their pulsations, which can reveal internal rotation profiles, convective core sizes and the extent of mixing in radiative zones. \bcep\ stars are of late O- and early B-type, generally considered to span a mass range of $\sim$8--25~$\Msun$. Their oscillations are primarily driven by the opacity ($\kappa$-) mechanism of iron-peak elements \citep{Moskalik_Dziembowski_1992,Dziembowski_Pamiatnykh_1993_opacity}. They oscillate in low-radial order pressure ($p$) and low-order gravity ($g$) modes modes with periods between approximately 2 and 8 hours \citep{Stankov_Handler, Aerts_book, Bowman_rev}. Some \bcep\ stars also oscillate in high-order gravity ($g$) modes with longer periods of more than a day \citep{Handler_2006_betaCep, Handler_2009}. Stars exhibiting both \bcep-type pulsations and high-order $g$ modes are referred to as hybrid pulsators.

Pioneering asteroseismic modeling of 16~Lac by \citet{1996AA...306..436D} compared observed pulsation frequencies with theoretical low-degree modes. Assuming no core overshooting, the models suggested an inward increase in rotation rate, although the mode identifications (i.e., identifying the harmonic degree~$l$ and azimuthal order~$m$ of each mode) remained ambiguous, limiting the certainty of the inferred rotation profile. Their subsequent study of 12~Lac explored alternative explanations for its nearly equidistant frequency triplet, highlighting the need for reliable mode identification to constrain seismic models \citep{1999AA...341..480D}.

HD~129929 was later the first \bcep\ star for which a complete mode identification was performed based on long term (21-year) ground-based multi-band observations \cite{Aerts_2003}. These observations were used to constrain the core-overshooting parameter, metallicity, the rotation rate of the star, providing the first robust asteroseismic evidence to rule out rigid rotation in a star (see also \citealt{2004AA...415..251D}). This single-site campaign was soon followed by months-long multi-site campaigns, which led to seismic modeling of $\nu$~Eri \citep{2004MNRAS.350.1022P},  $\theta$~Oph \citep{2007MNRAS.381.1482B}, 12\,Lac \citep{2009MNRAS.396.1460D}, and $\gamma$~Peg \citep{2009MNRAS.398.1339H}.

A more recent example is provided by \cite{Burssens_2023Nat}, who used a long-baseline, semi-continuous light curve of the \bcep\ star HD~192575 obtained with the Transiting Exoplanet Survey Satellite \citep[TESS;][]{TESS} to constrain the mass of the convective core and identify non-rigid rotation in its interior. Later, \cite{2025AA...701A...5V} updated the asteroseismic forward modeling using a longer TESS baseline and performed the first detailed rotation inversion for a \bcep\ pulsator, finding weak radial differential rotation.

The last complete compilation of \bcep\ stars was published over two decades ago by \cite{Stankov_Handler} who reviewed more than 250 papers covering $\sim$100 years of literature and gathered 231 candidates. Of these, 93 were confirmed as \bcep\ stars, 77 were classified as candidate \bcep\ stars due to limited or insufficient data, and the remaining 61 were rejected as \bcep\ candidates. Among the confirmed \bcep\ stars, the pulsation periods ranged between 1.60 to 7.66 hours (corresponding to $3.1348-14.9925$ $d^{-1}$) and the spectral types spanned B0-B2. However, since spectral classifications are not always accurate, \cite{Stankov_Handler} broadened the spectral range defining \bcep\ stars to span from O7 to B3.

Since the \cite{Stankov_Handler} catalog was published, ground-based surveys such as the Optical Gravitational Lensing Experiment \citep[OGLE;][]{OGLE}, the All Sky Automated Survey \citep[ASAS;][]{ASAS_1997,ASAS_2001}, the Hungarian Automated Telescope \citep[HAT;][]{HAT_2004}, the Kilodegree Extremely Little Telescope \citep[KELT;][]{KELT_2007,KELT_2012}, the All-Sky Automated Survey for Supernovae \citep[ASAS-SN;][]{ASAS-SN}, and the Zwicky Transient Facility \citep[ZTF;][]{ZTF} collected and published new data. Additionally, space-based missions such as the Convection, Rotation and planetary Transits mission \citep[CoRoT;][]{CoRoT}, Kepler \citep{Kepler}, TESS and Gaia \citep{Gaia} were launched. These surveys and missions produced large amounts of new photometry, and as a result, new identifications of \bcep\ candidates were published. These surveys also enabled new systematic population studies of pulsating B stars and deepened our physical understanding of \bcep\ pulsations. For example, \citet{2022MNRAS.515..828S} analyzed 119 B-type stars in the Scorpius-Centaurus association and detected pulsations in 81 of them, including \bcep\ stars. In addition, space-based missions, with their long baselines for many stars, allow for more accurate frequency analyses, which can help reclassify potential \bcep\ stars and perform more accurate asteroseismological studies of these stars. New observations also enabled studies of the effects of rapid rotation on \bcep\ pulsations \citep{Aerts_2006_rapid,Handler_2012_rapid}. In some rapidly rotating stars, pulsation periods shorter than about 6 hours can result from high-order $g$ modes, a possibility explored in earlier theoretical work on other B-type pulsators  \citep{Ushomirsky_Bildsten_1998_rapid,Townsend_2000_rapid}.

TESS also expanded the known population of \bcep\ stars in binary systems. \cite{Eze_Handler_2024_EBs} identified 59 new \bcep\ stars in eclipsing binary (EB) systems using TESS data. This sample provides a foundation for follow-up observations and modeling of \bcep\ stars in binaries and enables meaningful comparisons with their single-star counterparts. This is especially important given that EBs offer model independent dynamical masses and radii, and when combined with asteroseismic modeling, provide tight constraints on interior physics (see review by \citealt{2026ARAA..64..223S}).

While new data are enabling new discoveries, there are only a handful of \bcep\ stars with complete asteroseismic modeling, including mode identification for all the dominant pulsation modes of the star (e.g., \citealt{Aerts_2000_6_beta_cep,Aerts_2003,DeRidder_2004_NuEri,2005MNRAS.362..619B,Handler_multi_band_theta_Oph, Handler_2006_betaCep_12Lac, Handler_2012_V986_Oph_fast_rot,2012MNRAS.427..483B,2013MNRAS.431.3396D,2013AJ....146..102S,2015MNRAS.454..724J,2019AA...632A..95M}). These type of studies usually require multi-band data (not currently available from space, and hence obtained from the ground) to enable mode identification of the dominant modes, without relying on rotational splitting, which is not always available. 

\cite{Fritzewski_Gaia_modeID} used multi-band data from Gaia together with a probabilistic framework based on the known dominant modes of six previously studied \bcep\ stars to perform mode identification for 119 Gaia-discovered and TESS-observed \bcep\ stars, using observed amplitude ratios to determine the spherical degree, $l$, of the dominant pulsation mode in each case. They also searched for stars showing rotationally split multiplets and in other cases identified additional modes as radial when their light curves showed a ``standstill'' phenomena (i.e., a brief pause in the increase in brightness near the time of maximum radial velocity, followed by further brightening to peak brightness and then a rapid decline). This work represents a significant step toward large-scale asteroseismology of \bcep\ stars using modern TESS and Gaia data combined with earlier ground-based studies. \cite{2026arXiv260710834V} extended this work by performing forward modeling and asteroseismic analysis for stars in which at least three modes could be identified (not necessarily all of the dominant modes), in some cases aided by rotational splitting. They modeled 36 stars: 31 from the \cite{Fritzewski_Gaia_modeID} sample and five additional stars from the literature.

The Global Asteroseismology Project (GAP) aims to triple the number of studied massive stars with a complete asteroseismic analysis, combining space- and ground-based continuous photometry \citep{Shitrit_Arcavi_GAP}. The project was awarded 9000 observing hours over three years on the 0.4\;m telescopes of the Las Cumbres Observatory network \citep{LasCumbresObservatory}. GAP is collecting multi-band, high-cadence data of $\sim~20$~\bcep\ stars to complement the single-band TESS data. This will allow complete seismological studies, including mode identification, of a sample of massive stars.

New and upcoming surveys and modeling techniques such as those described above motivate the need for an updated \bcep\ catalog. Here, we provide such a catalog by incorporating new observations and analyses of known \bcep\ stars, together with candidates newly identified in this work and in the literature published since the compilation by \cite{Stankov_Handler}.

In Section~\ref{sec:comp_the_catalog} we explain how our catalog was constructed. In Section~\ref{sec:photo_data} we describe the sources of the photometric data used, together with the light-curve and frequency analysis. In Section~\ref{sec:spectra} we present our spectroscopic observations, along with the spectral analysis and modeling. In Section~\ref{sec:visual_inspection} we outline the visual inspection and classification process. In Section~\ref{sec:class_res} we present the classification results. In Section~\ref{sec:res_stat} we present a statistical analysis of the stars in the catalog. We summarize our work in Section~\ref{sec:conclusions}.

\section{Compiling the Catalog} \label{sec:comp_the_catalog}
We compile objects for our catalog through two channels: a literature review of reported \bcep\ stars, and our own search for new \bcep\ stars through light-curve analysis of archival data.

\subsection{Literature Review} \label{sec:literature_rev}
We review the literature published since the \citet{Stankov_Handler} catalog (including a few \bcep\ publications from before the \citealt{Stankov_Handler} catalog) until January 2026\footnote{Several papers published since identify potential \bcep\ stars that may not be included in this work \citep[e.g.,][]{2026MNRAS.551g1493T}.}, focusing on papers that discuss, in their titles and abstracts, \bcep\ stars, asteroseismology of massive stars and classification of pulsators. We find 218~papers (including \citealt{Stankov_Handler}), among which 73~papers report newly identified \bcep\ stars. Additionally, we include all the \bcep\ classification designations from the SIMBAD Astronomical Database \citep{SIMBAD} that do not include a reference to their classification, as well as stars classified as \bcep\ stars in the International Variable Star Index \citep[VSX;][]{2015yCat....102027W}. In total, after removing duplicates, the literature search yields 2,044~potential \bcep\ stars. We present the full list of references and the number of objects in each reference in Table~\ref{tab:sources} in Appendix~\ref{sec:literature_rev_app}.
We retrieve data for these stars, including their spectral type, $V$-band magnitude, and pulsation classification from SIMBAD \citep{SIMBAD} when available.

\subsection{Archival Data Search} \label{sec:new_ident}
In addition to our literature survey, we inspect light curves from ASAS \citep{1997AcA....47..467P,2001ASPC..246...53P}, the HAT system \citep{2004PASP..116..266B} and the fourth phase of OGLE \citep[OGLE-IV;][]{2015AcA....65....1U} sky surveys for new possible \bcep\ stars. 

We complement our search with the results from \cite{2020AJ....160...32L}, who performed an archival search of the KELT data. We follow the same protocol as \cite{2020AJ....160...32L} to select the candidates and briefly summarize it here. We cross-match all stars with spectral classifications between O4--B7 with the survey light curves and examine the matched stars for pulsations that fall within the \bcep\ frequency range \citep[from 1.60 to 7.66 hours;][]{Stankov_Handler} by performing frequency analysis of the light curves.

We check the literature for possible misclassifications, for example cases with uncertain spectral types or published spectroscopic effective temperatures ($T_{\mathrm{eff}}$) inconsistent with \bcep\ stars. In addition, we use Gaia Data Release 3 (DR3) parallaxes \citep{Gaia,gaia_dr3} to estimate distances and then stellar luminosities to reject objects whose luminosities are too low for \bcep\ stars ($log(L/L_{\odot})<2.7$) such as $\delta$~Scuti stars or gravity-mode sub-dwarf B (sdB) pulsators .

This search yields a total of 949~potential \bcep\ stars, of which 499 are newly reported. Of these, 4 do not have a TESS light curve and we do not classify them in this work.
In total, together with the literature and the photometric search, the complete list of potential \bcep\ stars with which we begin our classification process contains 2,543~stars (see Table~\ref{tab:cand_summary} for a summary of the number of objects).

\begin{deluxetable}{lc}
\tablecaption{Summary of the number of potential \bcep\ stars identified in the literature and archival searches.\label{tab:cand_summary}}
\startdata
\\
Literature review (no duplicates) & 2,044\\
Archival review (total) & 949\\
Archival review (newly reported) & 499\\
\hline
Final potential \bcep\ list (no duplicates) & 2,543\\
\enddata
\end{deluxetable}

\section{Photometric Data} \label{sec:photo_data}
\subsection{Cross-Matching with TESS and Gaia}
We first cross match all potential \bcep\ stars with the Gaia DR3 catalog, using the coordinates from SIMBAD, the relevant publication (for the literature search) or the relevant survey data (for the archival search) as available. We use an initial match radius of 1$\arcsec$, which we expand to 2$\arcsec$ if no match is found. For 23 cases, more than one source is found within the applied search radius. We mark them with the \textit{Gaia\_MULT\_MATCH} flag, and select the source with the brightest Gaia $G$-band magnitude as the match.
In total, we find a Gaia DR3 source ID for 2,526 out of our 2,543 potential \bcep\ stars, and adopt their Gaia coordinates for the catalog. For the remaining 17 stars not found in the Gaia archive, we use their SIMBAD coordinates. 

Of the 2,526 Gaia sources, 1,695 have global stellar parameters such as $T_{\mathrm{eff}}$, $\log g$ and metallicity estimated by the Generalized Stellar Parametrizer Photometry (GSP-Phot) pipeline \citep{Gaia_DR3_GSP-Phot}. GSP-Phot derives these parameters from Gaia observables, such as the flux ratio between the blue and red photometric bands, and parallaxes.
In addition, 1,882 stars have outputs from the Extended Stellar Parametrizer for Hot Stars \citep[ESP-HS;][]{Gaia_hot_stars} pipeline. Following \cite{Shi_2023_OB_pulsators_Gaia}, \cite{Fritzewski_Gaia_modeID} and \cite{2026MNRAS.551g1493T}, we adopt the ESP-HS parameters when available, as they are more reliable for hot pulsating main-sequence stars.

Next, we cross match all potential \bcep\ stars with the TESS Input Catalog (TIC). As with Gaia, we use an initial match radius of 1$\arcsec$, which we expand to 2$\arcsec$ if no match is found. For 13 objects, which we mark with the \textit{TIC\_MULT\_MATCH} flag, multiple TIC sources are found within the search radius. We inspect the candidates individually and select the appropriate source (by individually checking TIC identifiers listed in SIMBAD, any large discrepancies between TESS and SIMBAD V band magnitudes, and positional agreement with Gaia coordinates) as the match. We successfully match 2,542 of the 2,543 stars in the catalog to a TIC entry. Of these, 2,421 stars have a TESS light curve in the Mikulski Archive for Space Telescopes (MAST; see Section~\ref{sec:lc_freq_analysis}) which we use to visually classify their pulsations (see Section~\ref{sec:visual_inspection}). We do not visually classify the 122 stars without TESS light curves. These stars are too bright or too faint (fainter than the QLP data faint magnitude limit), or are located in overly crowded fields, suffering from contamination by surrounding stars, prohibiting the extraction of accurate photometry. They remain in the sample as visually unclassified stars and are considered potential \bcep\ stars.

\subsection{Light Curves and Frequency Analysis} \label{sec:lc_freq_analysis}
We use the \textit{Lightkurve} package \citep{Lightkurve_2018} to extract TESS light curves from MAST for each of the 2,421 stars with TESS photometry. Of these, 2,413 have photometry from the Quick-Look Pipeline \citep[QLP;][]{QLP} and 1,292 have photometry from the Science Processing Operations Center \citep[SPOC;][]{SPOC}. The SPOC pipeline generally provides higher-precision light curves, but not all stars have SPOC light curves, and not all sectors in which a star is observed are processed by SPOC. In this work, we use all available light curves of each type. For the statistical analysis (Section \ref{sec:res_stat}), we prioritize SPOC data for stars with both SPOC and QLP light curves. However, we adopt the QLP data when the QLP light curve is of good quality and covers more TESS sectors than the SPOC light curve.

For each light curve, we apply the following processing steps. For each sector, we subtract the sector median from each flux value and divide the result (and the flux uncertainties) by the median. This normalization removes sector-to-sector zero-point offsets introduced by instrumental effects while preserving the relative variability amplitudes within each segment. The result is a light curve normalized around zero, with values given in fractional (relative) flux and uncertainties expressed in relative units.

Since not all sectors are observed with the same cadence, following \cite{Fritzewski_Gaia_modeID}, we bin the shorter cadence sectors to the longest exposure time available for that star to prevent inhomogeneous data sampling, which could bias the frequency analysis. Binning to the longest exposure time, up to 30~min, is suitable for \bcep\ stars because the corresponding Nyquist frequency of 24~d$^{-1}$ lies above the \bcep\ frequency regime considered here (although finite integration time can attenuate the pulsation amplitudes, particularly for the shortest-period modes). For our sample, we find that cadence equalization has a negligible effect on the recovered dominant frequencies, while the amplitudes agree for 62.1\% of the stars. When differences occur, the amplitudes derived from the unbinned data tend to be slightly larger. We adopt the unbinned data for the statistical analysis to avoid amplitude attenuation. The full comparison between binned and unbinned data can be found in Appendix~\ref{sec:binning_vs_nobinning}.

During the visual inspection process, we further perform sigma-clipping of the data, globally or per sector (depending on the star), using the median and standard deviation (as the central-value and scatter estimators, respectively). The values of the sigma-clipping for each star are set and recorded during the visual classification process (Section~\ref{sec:visual_inspection}). 
 
For the frequency analysis, we extract the pulsation frequencies, amplitudes and phases using a prewhitening process. Before applying this process, we subtract the mean time value from the complete time stamp array to reduce correlations between phases and frequencies, which improves subsequent frequency extraction. 

For the prewhitening process, we use STAR SHADOW \citep{2024AA...685A..62I}, which is an automated time-domain framework designed originally for modeling of EB variability together with intrinsic stellar oscillations in space-based photometry such as from TESS. The code constructs a global regression model in the time domain consisting of harmonic components describing any orbital signal (if present) together with additional sinusoidal terms representing intrinsic variability. Candidate periodicities are identified using both a Lomb-Scargle algorithm and phase-dispersion minimization (PDM), after which the full multi-component model is optimized directly in the time domain using non-linear regression. We use the ``star\_shadow\_asteroseismology'' version\footnote{\url{https://github.com/Mathijs-Vanrespaille/star_shadow_asteroseismology/tree/mathijs_changes}} by M. Vanrespaille, which is updated to multiperiodic oscillations rather than binarity. For additional information about STAR SHADOW and how we use it see Appendix~\ref{sec:prewhitening_app}.

\subsection{Bolometric Corrections and Luminosities} \label{sec:bol_par}
To estimate the bolometric luminosity of each star and compute the bolometric corrections for the Gaia G, BP, and RP bands, we follow ``Model 3'' of \cite{Pedersen2020_BC} which expresses the bolometric correction as a function of $T_{\mathrm{eff}}$ and $\log g$ (taken from the ESP-HS pipeline) and metallicity (taken from the GSP-Phot pipeline). For all the calculations described below, we derive one set of values using the available GSP-Phot metallicities and another assuming solar metallicity. This allows us to assess the effect of the metallicity assumption and obtain luminosity estimates for stars without GSP-Phot metallicities. We compare the resulting luminosity and extinction estimates, together with their available sample sizes, in Appendix~\ref{sec:luminosity_method_comparisons}.

To correct the Gaia magnitudes for interstellar extinction, we obtain three-dimensional reddening estimates from the Bayestar19 dust map \citep{2019ApJ...887...93G} using the Gaia coordinates and inverse-parallax distances. We adopt the median of the posterior distribution of the color excess $E(B-V)$ as the reddening estimate, and half the difference between the 84th and 16th percentiles as its uncertainty. We convert the reddening into the extinctions $A_G$, $A_{BP}$, and $A_{RP}$ using the coefficients from \cite{Pedersen2020_BC}.

Following \cite{Pedersen2020_BC}, we apply the corresponding bolometric correction to each Gaia extinction-corrected apparent magnitude to calculate G-, BP-, and RP-band bolometric magnitudes. We adopt their inverse-variance-weighted mean as our three-band bolometric magnitude. 

Because reliable Bayestar19 reddening estimates are not available for every star, we also calculate a G-band bolometric magnitude using $A_G$ provided directly by ESP-HS. This method requires only the ESP-HS parameters and Gaia G-band photometry, thereby providing an independent extinction-based luminosity estimate for a larger sample of stars.

We convert the resulting bolometric magnitudes into $\log(L/L_\odot)$ using inverse-parallax distances. We restrict the luminosity calculations to stars with reliable Gaia astrometry and positive parallaxes, defined as Renormalized Unit Weight ($\mathrm{RUWE})~<~1.4$ and \textit{parallax\_over\_error}~$>~5$. We propagate the available uncertainties in the Gaia photometry, ESP-HS $T_{\mathrm{eff}}$ and $\log g$, GSP-Phot metallicity, extinction, and parallax.

Our catalog includes the Bayestar19 extinction, extinction-corrected absolute magnitudes, and final luminosity estimates computed and used in this work.

\section{Spectroscopic Observations and Analysis} \label{sec:spectra}
We collect spectra for 67 southern-hemisphere potential \bcep\ stars using the High-Resolution Spectrograph \citep[HRS;][]{SALT_HRS_2010,SALT_HRS_2012,SALT_HRS_2014} mounted on the South African Large Telescope \citep[SALT;][]{SALT_2006}. We select these stars by considering all southern-hemisphere stars in our catalog and assigning each star a ranking based on the number of pulsation modes detected from the ground, its observability with TESS, and its apparent magnitude (and hence expected variability detection level). We select the 67 highest-ranked stars that are observable with SALT and lack high-resolution spectra in public databases. Here, we discuss the spectroscopic observations and the subsequent data reduction, analysis, and modeling.

The HRS operates in three modes: low (resolving power $R~\approx~14,000$), medium ($R~\approx~40,000$) and high ($R~\approx~65,000$) resolution. Among the 67 stars in our sample, we observe 45 in high resolution, 18 in medium resolution, 3 in low resolution, and one star in both medium and low resolution modes. In total, we obtain 101 spectra for the 67 stars. For 40 stars, we obtain at least two spectral epochs to perform a preliminary check for possible radial-velocity variations. We present the spectral log in Table~\ref{tab:spectra} in Appendix~\ref{sec:spectra_app}.

We reduce the spectra using the PySALT User Package\footnote{\url{http://pysalt.salt.ac.za/}} \citep{SALT_pipeline_2010}.
We then normalize the spectra using the \textit{continuum} task of the National Optical Astronomy Observatories (NOAO) optical astronomy package within the Image Reduction and Analysis Facility (IRAF)\footnote{\url{https://iraf.readthedocs.io/en/doc-autoupdate/tasks/noao/}}. 

We check the normalization level of each spectrum by comparing it with theoretical spectra, which we calculate using the non-local thermodynamic equilibrium (NLTE) TLUSTY~BSTAR2006 \citep{TLUSTY_BSTAR2006} grid assuming solar abundances. We generate the initial synthetic spectra using atmospheric parameters
estimated from the spectral types listed in Table~\ref{tab:spectra}
in Appendix~\ref{sec:spectra_app}. For this initial comparison, we
broaden the synthetic spectra using trial projected rotational velocity ($v \sin i$) values consistent with measurements for B-type stars of similar spectral types \citep{2002ApJ...573..359A}. We consider the continuum placement satisfactory when the differences between the observed and theoretical continua remain within the noise level of the spectrum. In cases where noticeable deviations are present, we re-normalize the spectra until we achieve a consistent continuum level.

We further compare the normalized spectra with the same initial theoretical spectra to investigate the possible presence of flux contributions from secondary components. In addition, we examine radial-velocity shifts from the theoretical spectra using the cross-correlation technique. Through the detection of two distinct sets of spectral lines belonging to the two components, we identify ALS~1135 (identified previously by \citealt{2013MNRAS.429.1354M}), ALS~4680, HD~81370, HD~116538, HD~171141, and HD~172367 as candidate double-lined spectroscopic binary (SB2) systems. We subsequently analyze the remaining stars under the assumption of single-star spectra because they do not show clear evidence of multiplicity within the available spectra.
For stars with more than one spectrum, we combine the spectra (after correcting any measured radial-velocity shifts) to improve the signal to noise ratio (SNR) and to reduce possible pulsational effects on the line profiles, recognizing that this can affect the $v \sin i$ derivation \citep{2014AA...569A.118A}. 

To determine the value of $v \sin i$, we apply the profile-fitting method \citep{Gray_book_2005} to the profiles of the \ion{He}{2} ($\lambda = 4026$ Å) and \ion{He}{1} ($\lambda = 4387$ Å) lines, using limb-darkened and instrumentally broadened models calculated in steps of $1~\mathrm{km\,s^{-1}}$. We calculate the intrinsic synthetic He-line profiles using SYNSPEC \citep{2011ascl.soft09022H,2017arXiv170601859H}, including its approximate treatment of Stark broadening, and subsequently apply rotational and instrumental broadening using ROTIN3 \citep{2017arXiv170601859H}. We select the microturbulent velocities individually based on approximate $T_{\mathrm{eff}}$ inferred from the spectral types and keep them fixed during the determination of $v\sin i$. We adopt values ranging from 2 to $14~\mathrm{km\,s^{-1}}$, consistent with the range reported for comparable B-type stars by \cite{2025AA...701A.297M}. We do not explicitly include macroturbulent broadening, which may lead to overestimated $v\sin i$, particularly for slowly rotating stars. Its relative contribution is smaller when rotational broadening dominates \citep{2017AA...597A..22S}. The quoted
fitting uncertainties do not include this potential systematic effect. We derive $v\sin i$ by minimizing the differences between the observed and theoretical spectra, following the procedure described in \cite{Balmer_lines_2004}. We estimate the uncertainties in $v \sin i$ using the $\Delta\chi^2$ criterion by considering the range of models around the minimum $\chi^2$ solution corresponding to the 1$\sigma$ confidence interval.

To determine the stellar atmospheric parameters, namely $T_{\mathrm{eff}}$ and $\log g$, we compare the observed $H{\beta}$ and $H{\gamma}$ Balmer line profiles with the NLTE TLUSTY BSTAR2006 \citep{TLUSTY_BSTAR2006} model grid assuming solar abundances. We adopt the $v \sin i$ values derived in the previous step as input parameters for the synthetic spectra. The model grid covers a $T_{\mathrm{eff}}$ range of $15,000$--$30,000$~K in steps of $1,000$~K and a $\log g$ range of $1.75$--$4.75$~dex in steps of 0.25~dex.

\begin{figure*}
    \centering
    \includegraphics[width=1\linewidth]{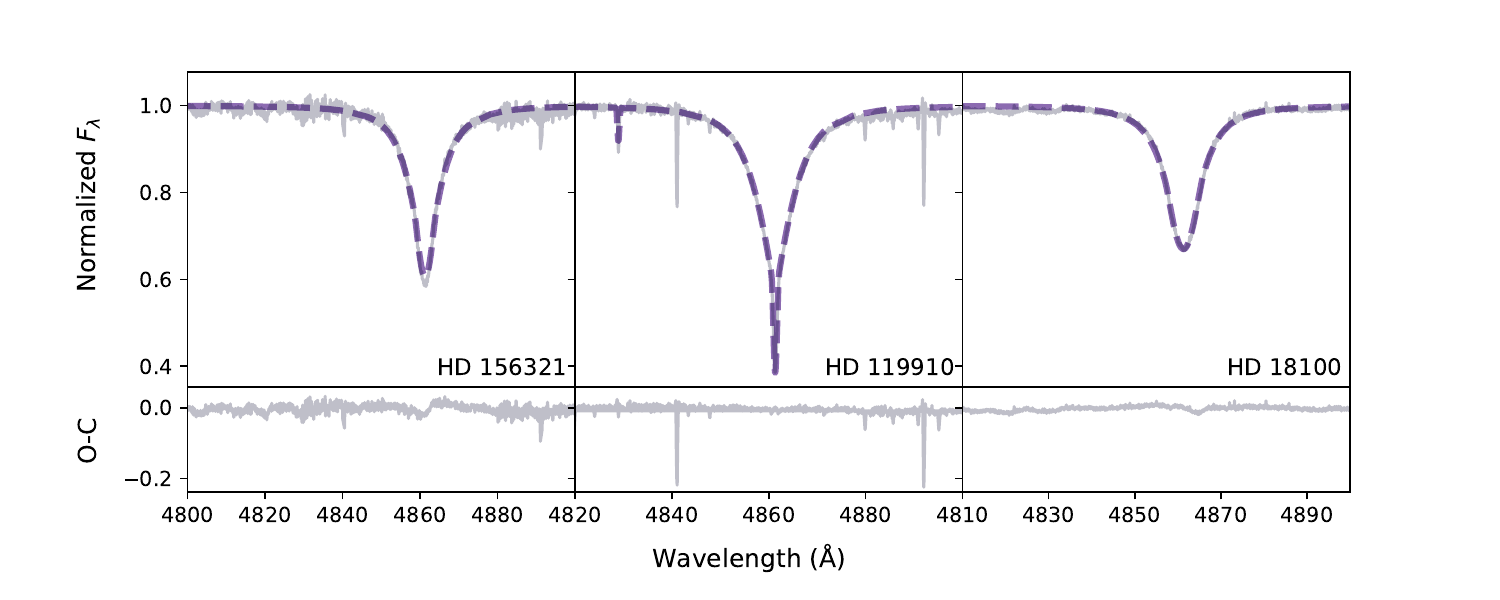}
    \caption{Top: Theoretical fits (dashed purple lines) to the observed $H{\beta}$ line profiles (solid gray lines) for three stars with different atmospheric parameters and $v \sin i$ values. Bottom: Residuals between the observed ($O$) and calculated ($C$) spectra.}
    \label{fig:Balmer_fit}
\end{figure*}

We derive the atmospheric parameters using the same profile-fitting approach that we apply for the $v \sin i$ determination, by minimizing the differences between the observed and theoretical Balmer line profiles. The best-fit model within the TLUSTY~BSTAR2006 grid provides the initial estimates of $T_{\mathrm{eff}}$ and $\log g$, after which we perform linear interpolation between neighboring models in both $T_{\mathrm{eff}}$ and $\log g$, using an interpolated search grid with steps of 500~K and 0.1~dex, respectively. We estimate the uncertainties in $T_{\mathrm{eff}}$ and $\log g$ from the fitting-statistic profiles around the best-fitting solution to determine the parameter bounds corresponding to the 1$\sigma$ confidence interval, similarly to computing the $v\sin i$ uncertainties. We list the derived atmospheric parameters and $v\sin i$, together with their uncertainties, in Table~\ref{tab:spectra} in Appendix~\ref{sec:spectra_app}. Figure~\ref{fig:Balmer_fit} shows three examples of the resulting theoretical fits to the observed $H{\beta}$ line profiles. 

We find two stars with $T_{\mathrm{eff}}$ below 15,000~K (HD~73918, and HD~151654). For these two stars, the agreement between the observed and theoretical Balmer-line profiles continues to improve toward the lower-temperature boundary of the BSTAR2006 grid. We therefore estimate their $T_{\mathrm{eff}}$ by linearly extrapolating the lowest-temperature synthetic line profiles. Only $T_{\mathrm{eff}}$ is obtained through this extrapolation and we determine $\log g$ separately using the same Balmer-line fitting procedure described for the other stars. We adopt an uncertainty of $1,000$~K for their $T_{\mathrm{eff}}$ and estimate their $\log g$ uncertainties using the same $\Delta\chi^2$ criterion as for the remainder of the sample. These $T_{\mathrm{eff}}$ values are significantly lower than expected for $\beta$ Cep stars. Additionally, HD~151654 is classified as F0V in the literature \citep{1999MSS...C05....0H}, whereas our analysis yields $T_{\mathrm{eff}}$ of $13,000$~K, significantly higher than expected for an F0V star, indicating that its published spectral classification is incorrect. Additionally, there is one star, HD~156172, with $T_{\mathrm{eff}}=30,000$~K, which lies at the upper boundary of the BSTAR2006 grid. We adopt a temperature uncertainty of $3,000$~K, twice the initial uncertainty estimate of $1,500$~K, to account approximately for this boundary limitation.

\section{Visual Inspection and Classification} \label{sec:visual_inspection}
Identifying \bcep\ stars based on automated frequency-range selection from the prewhitening analysis can lead to misclassifications of impostors (see Figure~\ref{fig:impostors} in Appendix~\ref{sec:vis_class_app}). To avoid this, we perform systematic visual inspection of all potential \bcep\ stars using a custom developed graphical user interface (GUI)\footnote{https://github.com/NoiShitrit/visual-stellar-classification-GUI}. The GUI displays the TESS light curve and the Lomb–Scargle periodogram for each star, together with its prior classification, spectral type and Gaia $T_{\mathrm{eff}}$ when available. The GUI allows us to sigma-clip each light curve, recalculate the periodogram following sigma clipping, and save user notes for each object. We show an example screenshot of the GUI in Figure~\ref{fig:visual_inspection_HD67600} in Appendix~\ref{sec:vis_class_app}.

\subsection{Possible Impostors}
One type of \bcep\ impostor we wish to exclude are Slowly Pulsating B-type (SPB) stars. SPB stars have a mass of approximately 3--8~\Msun\ and pulsation periods of $\sim$0.3--5~days \citep{Aerts_book}, but some also display \bcep-type pulsations and are classified as hybrid \bcep\ stars. However, in some cases, the pulsations in the \bcep\ range are actually harmonics and combination frequencies of the SPB pulsations, making these stars \bcep\ impostors \citep{Kurtz_2015_freq_spect,2017AA...598A..74P}. In some cases, these harmonic frequencies reach higher amplitudes than the intrinsic low-frequency modes, which increases the risk of misclassifying the star as a \bcep\ pulsator. An example of such an impostor we encountered in our classification process is HD~67600, which was previously classified as a \bcep\ candidate \citep{Pigulski_HD67600_example}, but TESS data reveal that the pulsations within the \bcep\ frequency range could be in fact harmonics of the SPB pulsations (Figure~\ref{fig:visual_inspection_HD67600} in Appendix~\ref{sec:vis_class_app}). Another example is shown in the top panel of Figure~\ref{fig:impostors} in Appendix~\ref{sec:vis_class_app}. Cases where frequencies in the \bcep\ range appear to be harmonics and combination frequencies (i.e., frequencies corresponding to sums or differences of other detected pulsation frequencies) of SPB-range frequencies are noted with the flag \textit{LOW\_FREQ\_HARMONICS}. We identify such impostors in this paper as part of the visual classification process; we do not perform detailed calculations or fitting procedures for each suspected impostor.

A similar type of impostor is the $\gamma$~Doradus ($\gamma$~Dor; GDOR) type pulsator. $\gamma$~Dor stars are late A- to early F-type stars, with pulsation periods of $\sim$0.3--3~days. Without a clear spectral type, they can be confused with \bcep\ stars due to harmonics and combination frequencies, similar to the case of SPB impostors (see the second panel from the top of Figure~\ref{fig:impostors} in Appendix~\ref{sec:vis_class_app}). Cases where the target itself is identified as a possible $\gamma$~Dor pulsator are noted with the flag \textit{GDOR}.

Another type of \bcep\ impostor is the class of $\delta$~Scuti ($\delta$~Sct; DSCT) stars, which are A- or F-type pulsators with pulsation periods of $\sim$0.5--7 hours that overlap with the \bcep\ period range \citep{Aerts_book}. Reliable absolute magnitudes or spectral classifications generally distinguish $\delta$~Sct from \bcep\ stars because of their substantially different luminosities or spectral types. However, incorrect cross-matches or blending of nearby sources in the TESS CCD images may associate the detected variability with the wrong source, and a nearby $\delta$~Sct star may contaminate the TESS light curve of a genuinely hot star. Therefore, $\delta$~Sct variability can still cause an object to be mistaken for a potential \bcep\ star. The frequency spectrum can provide additional evidence, as \bcep\ stars usually display a sparser frequency spectrum with only a few strong peaks and a narrower range of excited frequencies, whereas $\delta$~Sct stars exhibit a denser spectrum with many independent modes and harmonics (see the third and fourth panels from the top of Figure~\ref{fig:impostors} in Appendix~\ref{sec:vis_class_app}). Such evidence is, however, not conclusive, but suggests a need for further scrutiny. Cases where the target itself is identified as a possible $\delta$~Sct pulsator are noted with the flag \textit{DSCT}; when we suspect that there is $\delta$~Sct variability in the signal that is attributed to a nearby, blended star, we record this only in the notes and do not assign the \textit{DSCT} flag to the target.

In the visual classification we also check the $T_{\mathrm{eff}}$ and spectral type when available. This makes it easier to identify impostors. Cases where Gaia ESP-HS $T_{\mathrm{eff}}$ values are inconsistent with the spectral types reported in the SIMBAD database are noted with the flag \textit{SPEC\_WRONG}.

\subsection{Classification Flags}
While performing the visual classification we are able to identify other phenomena visible in the light curve or the periodogram, on top of the pulsation type, mentioned below.

Stars with light curves and periodograms that do not show signs of significant variability are noted with the flag \textit{CONST}.

We identify \bcep, hybrid and SPB stars that also display stochastic low-frequency (SLF) variability. SLF is characterized by a broad power excess at low frequencies with no stable, coherent peaks. Stars displaying SLF variability accompanying another variability classification are noted with the flag \textit{SLF}. We also identify a subgroup of O–B stars previously classified as \bcep\ pulsators that, in the TESS data we analyze here, display only SLF variability. We note these stars with the flag \textit{ONLY\_SLF} (the \textit{SLF} and \textit{ONLY\_SLF} categories are mutually exclusive). The physical origin of the SLF variability is uncertain without a more detailed study of these specific stars and may involve different mechanisms across the Hertzsprung-Russell (HR) diagram \citep{2026MNRAS.551g1493T}. One proposed explanation is damped internal gravity waves stochastically excited by turbulent convection, which have been associated observationally with SLF variability in blue supergiants \citep{2019NatAs...3..760B}. 

We also identify stars displaying possible combination frequencies among their detected variability signals. We note these cases with the flag \textit{COMB}.

Rotational modulation may arise from surface inhomogeneities such as chemical spots or magnetic features that co-rotate with the star, producing periodic variability at the rotation frequency and its harmonics and combination frequencies (see \citealt{1950MNRAS.110..395S}). In the frequency spectrum, this typically appears as a low-frequency peak distinct from the pulsation modes. We note such cases with the flag \textit{ROT}.

We also identify three types of binary-related variability. EB systems are noted with the flag \textit{EB}. Ellipsoidal variable (ELL) systems, in which tidal distortion produces brightness variations typically at twice the orbital frequency, are noted with the flag \textit{ELL}. We note eccentric ellipsoidal binary systems exhibiting the characteristic ``heartbeat'' variability around periastron with the flag \textit{HEARTBEAT}.

Rotational and ellipsoidal variables can be separated by means of harmonic structure in the periodogram. Ellipsoidal variables often show a subharmonic peak at half of the dominant frequency, whereas rotational variables can show higher-order harmonics. In case only a single low frequency peak is observed, no distinction can be made.

\subsection{Photometric Quality Flags}
Because of the relatively low spatial resolution of TESS, contamination from nearby stars may affect the photometry. Suspected cases of blending based on the visual classification are noted with the flag \textit{BLEND}. When the possible contaminating source is also included in the catalog and the origin of the detected variability is unclear, we include it in the Notes column and provide the TIC numbers of the relevant stars. We note such cases with the flag \textit{NEARBY\_STAR}.

Noisy QLP photometry cases are noted with the flag \textit{QLP\_BAD}. We note cases where the photometry is affected by saturation-induced systematics for stars near the TESS saturation limit ($\sim$ 6 mag), with the flag \textit{SAT}. As noted above, we use the SPOC data when available, unless the QLP light curve is of good quality and contains more TESS sectors than the SPOC light curve. In the latter case, we mark the star with the flag \textit{QLP\_OVER\_SPOC}.

For all flags, we assign a value of 1 when we identify the corresponding phenomenon or data-quality issue with high confidence, a value of 0.5 when the identification is tentative, and a value of 0 when it is not identified.

\section{Classification Results} \label{sec:class_res}
We find a total of 1,130 \bcep\ stars, classified as ``BCEP'' (displaying only or mostly \bcep-type pulsations), ``hybrid'' (displaying both \bcep- and SPB-type pulsations), ``BCEP+EB'', and ``hybrid+EB''. We classify an additional 246 stars as candidates of these types. These stars, additionally labeled as ``candidate'',  have TESS data of insufficient quality to confirm or reject a \bcep\ classification. They exhibit light-curve morphologies consistent with the assigned variability type, but lack ESP-HS $T_{\mathrm{eff}}$ and spectral-type information. The 122 stars without TESS light curves remain unclassified. Of stars reported as displaying or suspected to display \bcep\ pulsations, we reject 1,045 stars (some we classify as candidate SPB stars or flag as DSCT/GDOR). We classify stars that are neither \bcep\ nor SPB candidates as ``Non-BCEP'' (these may include candidate DSCT or GDOR stars). 

We summarize our classification results in Table~\ref{tab:classification_summary}. Of the 499 possible \bcep\ stars newly reported from our archival search, we classify 180 as confirmed \bcep\ stars, 18 as candidates, and 297 as rejected; the remaining four do not have TESS light curves and remain unclassified. The complete catalog, including the classifications, flags, and derived parameters, is provided electronically. An excerpt of the catalog can be seen in Table~\ref{tab:catalog}.

\begin{deluxetable}{lcc}
\tablecaption{Results of the visual classification process, ordered by group size and separated into the confirmed BCEP, candidates BCEP and non-BCEP groups. \label{tab:classification_summary}}
\tablehead{
\colhead{Classification} &
\colhead{Number of Stars} & \colhead{New}} 
\startdata
Hybrid & 580 & 101 \\
BCEP & 457 & 64\\
BCEP+EB & 80 & 12\\
Hybrid+EB & 13 & 3 \\
\hline
EB & 60 & 7 \\
Hybrid Candidate & 60 & 5 \\
Candidate\tablenotemark{a} & 50 & 1\\
BCEP Candidate & 47 & 5 \\
BCEP+EB Candidate & 26 & 0 \\
Hybrid+EB Candidate & 2 & 0 \\
Candidate + EB & 1 & 0\\
\hline
SPB Candidate & 600 & 183 \\
Non-BCEP & 410 & 112 \\
SPB+EB  Candidate & 28 & 2\\
EB+Non-BCEP & 7 & 0\\
\hline
No TESS light curve & 122 & 4 \\
\enddata
\tablenotetext{a}{The classification ``candidate'' denotes stars with TESS data that are of insufficient quality to confirm or reject a \bcep\ classification.}
\tablecomments{BCEP+EB candidate, Hybrid+EB candidate and SPB+EB candidate stars should be considered both EB candidates and candidates for their respective pulsation classes (BCEP, Hybrid, or SPB).}
\end{deluxetable}

We identify 28 BCEP/hybrid/+candidate stars that display unusually large numbers of frequencies, sometimes 3 or 4 times more than the typical number of $\sim$~6 dominant \bcep\ frequencies observed in \bcep\ stars. To note these objects we intoduce a new flag, \textit{MANY\_FREQS}. We also identify 25 BCEP/hybrid/+candidate stars that display high frequencies ($ > 15$ $d^{-1}$) that do not appear to be harmonics and combination frequencies of lower-frequency signals. For these objects, we add the flag \textit{HIGH\_FREQS}. For some of these stars, we identify a possible contamination source nearby, while for others we identify no such source. A similarly complex frequency spectrum was detected in the \bcep\ star EPIC~202929357, for which \citet{2019MNRAS.489.1304B} identified 23 frequencies. Additional data might help determine the nature of the oscillation frequencies seen in these stars. Both flags are based on the visual inspection and are therefore treated as qualitative indicators rather than quantitative classifications.

Table~\ref{tab:flag_counts} presents the numbers of stars assigned to each visual-classification flag (see Section~\ref{sec:visual_inspection}) both across the entire catalog and among stars with confirmed or candidate \bcep\ classifications and EB systems. 

A description of the catalog and its columns is presented in Appendix~\ref{sec:catalog_description}.

\begin{deluxetable*}{lcccc}
\tablecaption{Numbers of secure and tentative visual-classification flag assignments.\label{tab:flag_counts}}
\tablehead{
\colhead{Flag} &
\multicolumn{2}{c}{Entire catalog} &
\multicolumn{2}{c}{Confirmed + candidate \bcep} \\
\colhead{} &
\colhead{Secure ($1$)} &
\colhead{Tentative ($0.5$)} &
\colhead{Secure ($1$)} &
\colhead{Tentative ($0.5$)}
}
\startdata
\textit{ROT} & 121 & 130 & 36 & 57  \\
\textit{ELL} & 29 & 67 & 17 & 33  \\
\textit{EB} & 15 & 18 & 10 & 15   \\
\textit{HEARTBEAT} & 4 & 35 & 2 & 19 \\
\textit{ONLY\_SLF} & 0 & 45 & 0 & 5   \\
\textit{SLF} & 47 & 141 & 14 & 104 \\
\textit{DSCT} & 25 & 39 & 3 & 18 \\
\textit{GDOR} & 6 & 6 & 1 & 0 \\
\textit{CONST} & 21 & 17 & 2 & 3   \\
\textit{QLP\_BAD} & 179 & 12 & 164 & 7  \\
\textit{SAT} & 8 & 7 & 7 & 6   \\
\textit{SPEC\_WRONG} & 36 & 20 & 33 & 15  \\
\textit{BLEND} & 164 & 128 & 121 & 88  \\
\textit{NEARBY\_STAR} & 23 & 5 & 12 & 3   \\
\textit{LOW\_FREQ\_HARMONICS} & 26 & 4 & 24 & 2   \\
\textit{MANY\_FREQS}                  & 32  & 17  & 28  & 0   \\
\textit{HIGH\_FREQS}                  & 34  & 16  & 25  & 0   \\
\textit{COMB} & 21 & 3 & 1 & 1   \\
\textit{QLP\_OVER\_SPOC} & 47 & 0 & 46 & 0 \\
\enddata
\tablecomments{Secure and tentative assignments correspond to flag values of 1 and 0.5, respectively. The final two columns include confirmed and candidate \bcep\ classifications, including their combinations with EB classifications.}
\end{deluxetable*}

The \textit{Notes\_classification} column in the catalog provides additional information for individual stars. For objects classified as candidate BCEP or candidate, we specify the reason that we do not confirm the star as a \bcep\ pulsator and indicate possible alternative classifications, such as DSCT or GDOR. We also include notes on EB, ELL, heartbeat, and SLF variability.

\section{Statistical Analysis} \label{sec:res_stat}
We analyze the pulsation parameters of all stars classified as BCEP, hybrid, and their combinations with EB systems, excluding candidate classifications. We use SPOC light curves when available; otherwise, we retain QLP entries only if they satisfy the quality criterion ($\mathrm{QLP\_BAD}=0$). When both products are available, we adopt the SPOC solution by default, but use the QLP solution for a predefined subset of TICs when it contains more TESS sectors and satisfies $\mathrm{QLP\_BAD}=0$. This selection ensures that the frequency analysis uses high-quality light curves, as data sufficient for visual classification may not always be adequate for quantitative frequency extraction. Here we use the frequency analysis results from the unbinned data (see Appendix~\ref{sec:binning_vs_nobinning}).

\subsection{Subsamples} \label{sec:subsamples}
To determine the lower-frequency boundary for dominant \bcep\ pulsations, we manually inspect all BCEP and hybrid stars with low dominant frequencies (all stars with dominant frequency below 4.5~$d^{-1}$). We exclude cases in which the dominant low-frequency signal is attributable to SPB-like variability or to EB/ELL orbital modulation rather than to a \bcep\ p-mode. The lowest dominant frequency we retain as a \bcep\ pulsation is 3.28~$d^{-1}$ for the stars in our sample. From our inspection, frequencies below this threshold are generally associated with hybrid stars dominated by SPB-like variability, or systems where \bcep\ stars are part of EB or ELL configurations and the dominant variability is associated with orbital modulation rather than \bcep\ pulsation. Based on this inspection and the discussion presented in Section~\ref{sec:stellar_pulsation_stat}, we define Subsample~\textit{I} as stars with the dominant frequency $\geq$~3.28~$d^{-1}$.

We define Subsample~\textit{II} as stars with reliable Gaia astrometry using the Renormalized Unit Weight Error (RUWE), requiring $RUWE<1.4$ \citep{2018A&A...616A...9L}, and parallaxes with high SNR, requiring \textit{parallax\_over\_error}~$> 5$ \citep{2015PASP..127..994B,2021A&A...649A...4L}, for which the inferred distances are expected to be reliable. This places an upper limit on the distances of the targets in the Subsample.

For the parameters extracted directly from Gaia, namely $T_{\mathrm{eff}}$ and $\log g$, we examine stars from Subsample~\textit{II}. Additionally, for the parameters derived in this work from Gaia quantities, including the solar-metallicity three-band luminosity (see Section~\ref{sec:bol_par}), distance and $M_G$, $M_{BP}$ and $M_{RP}$ (the absolute Gaia magnitudes for the G, BP and RP bands, respectively) corrected for extinction, we also examine stars from Subsample~\textit{II}. For luminosity and the corrected absolute Gaia magnitudes, we additionally require Bayestar19 reddening estimates with both the \textit{bayestar\_converged} (a converged dust-map query) and \textit{bayestar\_reliable\_dist} (a reliable queried distance) indicators set to True. For the pulsation constant $Q$, which depends on both Gaia-derived quantities and the pulsation frequencies, we examine stars that fall within the intersection of Subsamples \textit{I} and \textit{II}. $Q$ is the pulsation period normalized to the stellar mean density and represents the period that the same oscillation would have in a star with solar mean density:
\begin{equation*}
Q=P\sqrt{\frac{\bar{\rho}}{\rho_\odot}}=P\sqrt{\frac{g/g_\odot}{R/R_\odot}},
\end{equation*}
where we calculate $P$ as the inverse of the adopted dominant frequency, $f_1$. We derive $R/R_\odot$ from the solar-metallicity three-band $\log(L/L_\odot)$ derived in Section~\ref{sec:bol_par} and the ESP-HS $T_{\mathrm{eff}}$ using the Stefan--Boltzmann relation. For the dominant frequency and its corresponding amplitude, we examine stars from Subsample~\textit{I}. For the $V$-band apparent magnitude, we examine the entire sample, since this information comes from SIMBAD, therefore, it does not depend on the reliability of Gaia data or on frequency-based calculations. We list the subsample chosen for analysis of each parameter and the number of stars within it in Table~\ref{tab:param_conditions}.

\begin{deluxetable*}{lccccccc}
\tablecaption{Number of stars analyzed for each parameter, with the applied subsample and quality requirements.\label{tab:param_conditions}}
\tablehead{
\colhead{Parameter} & \colhead{Subsample~\textit{I}} & \colhead{Subsample~\textit{II}} & \colhead{Bayestar Quality} & \colhead{$N_{\mathrm{BCEP}}$} & \colhead{$N_{\mathrm{Hybrid}}$} & \colhead{$N_{\mathrm{BCEP+EB}}$} & \colhead{$N_{\mathrm{Hybrid+EB}}$}
}
\startdata
Distance &  & V & & 329 & 454 & 51 & 9\\
$m_V$ &  &  & & 354 & 527 & 78 & 13\\
$T_{\mathrm{eff}}$ &  & V & & 278 & 380 & 40 & 7 \\
$\log g$ &  & V & & 278 & 380 & 40 & 7  \\
$\log(L/L_\odot)$ &  & V & V & 120 & 159 & 11 & 4 \\
$Q$ & V & V & & 125 & 135 & 3 & 3\\
$M_G$, $M_{BP}$, $M_{RP}$ & & V & V & 143 & 185 & 16 & 5 \\
$f_1$ ($d^{-1}$) & V & & & 342 & 421 & 22 & 6\\
$A_1$ & V & & & 342 & 421 & 22 & 6\\
\enddata
\tablecomments{V in a column indicates which subsample is used for the statistical analysis of each parameter. Where V appears in both columns, the intersection of the two subsamples is used. 
The Bayestar Quality selection requires both Bayestar indicators to be true.}
\end{deluxetable*}

\subsection{Distribution of Stellar and Pulsation Parameters} \label{sec:stellar_pulsation_stat}
\begin{figure*}
    \centering
    \includegraphics[width=1\linewidth]{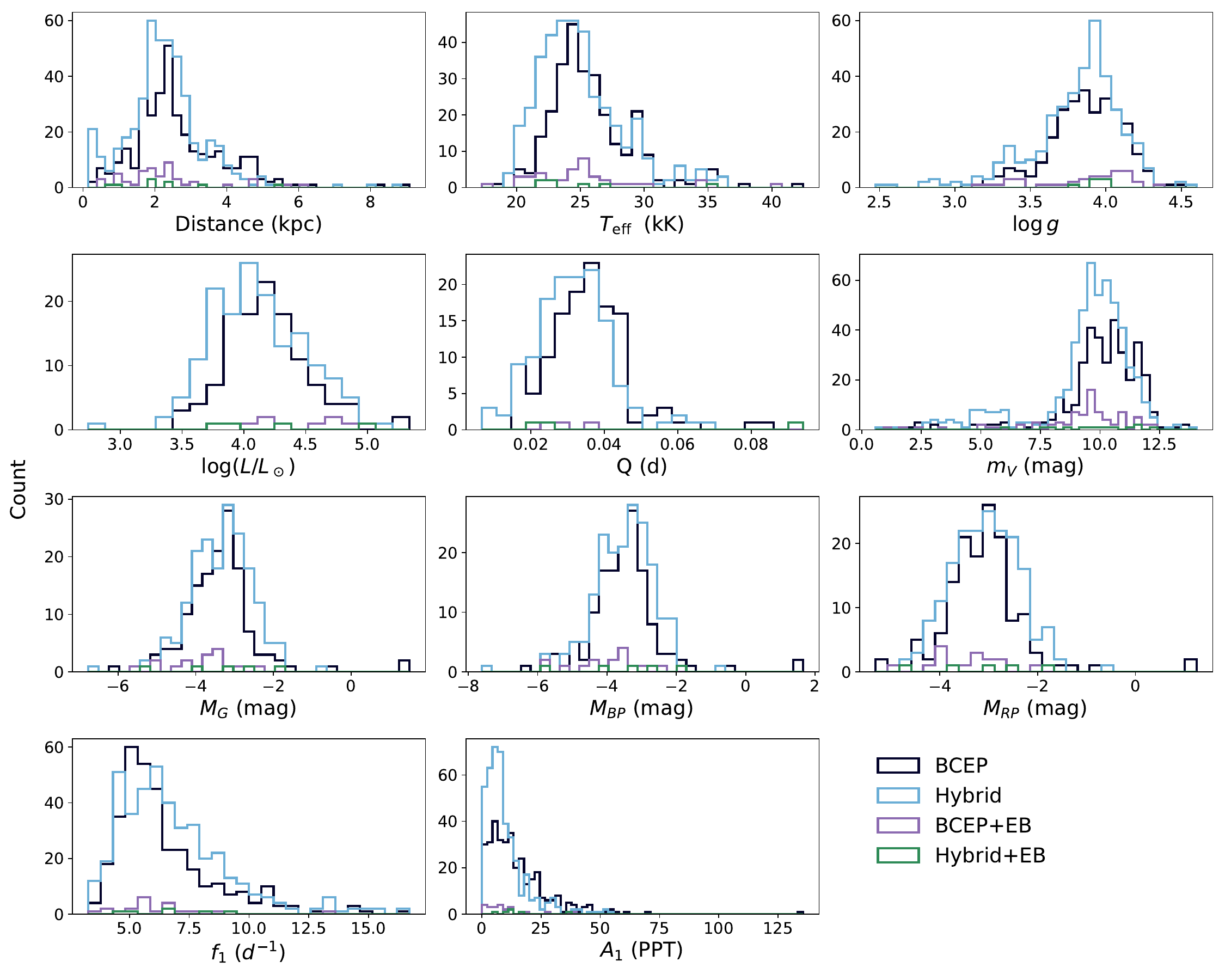}
    \caption{Distributions of key observational, stellar and variability parameters for the different variability classes. $M_G$, $M_{BP}$ and $M_{RP}$ are the absolute Gaia magnitudes in the G, BP and RP bands (respectively) corrected for reddening and $m_V$ is the apparent magnitude from SIMBAD. $Q$ is the pulsation constant. $f_1$ denotes the dominant frequency and $A_1$ its corresponding amplitude.}
    \label{fig:parameters_hist}
\end{figure*}

We present the distributions of the main observational, stellar and variability parameters for the different variability classes in Figure~\ref{fig:parameters_hist}. For each parameter, we determine the histogram bins from the combined sample using the Freedman-Diaconis rule \citep{Freedman_Diaconis1981}, which selects the bin width based on the sample size and interquartile range and later we apply the same bin edges to all classifications. 

The distributions presented here may reflect selection effects of the heterogeneous data sources used to construct the catalog. These sources rely on different telescopes, observing strategies, detection sensitivities, analysis techniques, and classification methods. We therefore cannot quantify or correct for these selection effects consistently, and we interpret the observed distributions as properties of the compiled sample rather than of the intrinsic Galactic \bcep\ population.

We perform several statistical tests on the distributions. Here, we report only the significant results; the full results, including the ranges, medians, and highest-bin centers for each parameter and classification are presented in Appendix~\ref{sec:stellar_star_app}.

The dominant-frequency distributions of the BCEP and hybrid classes are broadly similar, although the hybrid distribution has a higher median (6.32 versus 5.84~$d^{-1}$) and highest-bin center (6.12 versus 5.08~$d^{-1}$). Both frequency distributions show evidence of internal structure and possible bimodality. The BCEP stars have higher dominant amplitudes than the hybrid stars, with medians of 11.51 and 7.11~ppt, respectively, and a more extended high-amplitude tail.

The BCEP stars in EB systems also tend to have lower dominant amplitudes than the apparently single BCEP stars. No corresponding tendency is apparent for the hybrid stars in EB systems (but the EB subsamples are small). Several effects may contribute to the lower observed amplitudes of BCEP stars in EB systems. First, flux dilution by the companion can reduce the observed pulsation amplitudes. Moreover, in close binaries, tidal effects and/or past binary interaction may modify the intrinsic pulsation properties and the observed mode amplitudes (see \citealt{2026ARAA..64..223S}); depending on the system configuration, such effects could either enhance specific modes or contribute to lower amplitudes. 

The $T_{\mathrm{eff}}$ distributions overlap substantially and share a main peak near 24~kK, although BCEP stars have a slightly higher median temperature than hybrid stars (25.29 versus 24.44~kK) and extend to higher temperatures (42.45 versus 36.26~kK).

The $\log g$ distributions of the BCEP and hybrid stars overlap substantially where the hybrid distribution has a lower-gravity tail. Twenty stars have ESP-HS surface gravities below $\log g=3.2$, with 15 being hybrid stars, four are BCEP stars and one is a BCEP in EB system. \citet{2011AA...530A.115B} identify an empirical transition near $\log g=3.2$, which they interpret as the end of the main sequence for stars with masses of approximately 10--20\Msun. Therefore, these stars may have evolved beyond the main sequence. We list them in Table~\ref{tab:low_logg_stars}. However, several of the lowest-gravity values have comparatively large reported uncertainties, so we interpret their evolutionary status cautiously. Spectroscopic follow-up is required to verify their surface gravities and evolutionary states.

\begin{deluxetable}{lcccc}
\tablecaption{Confirmed \bcep\ stars with Gaia ESP-HS surface gravities below $\log g=3.2$, listed in order of increasing $\log g$. \label{tab:low_logg_stars}}
\tablehead{
\colhead{TIC} &
\colhead{Classification} &
\colhead{$\log g$} &
\colhead{$\sigma_{\log g}$} & \colhead{Spectral Type}
}
\startdata
139385056 & Hybrid & 2.474 & 0.622 & B0.5IVe \\
29277573 & Hybrid & 2.588 & 0.511 & B1V C\\
209036736 & Hybrid & 2.763 & 0.030 & B3II \\
232846315 & Hybrid & 2.815 & 0.009 & B0.5Vn\\
378726058 & Hybrid & 2.869 & 0.655 & B1II\\
400902778 & Hybrid & 2.886 & 0.014 & B0.5/1III\\
34803944 & Hybrid & 2.892 & 0.015 & B2\\
372674359 & Hybrid & 2.952 & 0.018 & B0.5III\\
390098366 & Hybrid & 2.998 & 0.009 & B1III/Vnn\\
295435513 & Hybrid & 3.018 & 0.021 & B1(III)n\\
330241794 & Hybrid & 3.109 & 0.024 & B0IVn\\
251637508 & BCEP & 3.109 & 0.023 & B2Ib/II\\
321947833 & BCEP+EB & 3.114 & 0.013 & B0.5/1III\\
458831281 & Hybrid & 3.131 & 0.029 & B1/2II/III\\
348023621 & Hybrid & 3.153 & 0.007 & O9.5I\\
442170794 & Hybrid & 3.167 & 0.027 & B0III:nn\\
155530282 & BCEP & 3.175 & 0.030 & B0.5III\\
120859681 & Hybrid & 3.181 & 0.026 & B1/2Ib/II\\
348506748 & BCEP & 3.193 & 0.011 & B2.0I\\
465555235 & BCEP & 3.195 & 0.031 & B2\\
\enddata
\tablecomments{The spectral types are taken from Simbad.}
\end{deluxetable}

\subsection{Comparison Between Spectroscopic Parameters and Gaia Parameters}
\begin{figure}
    \centering
    \includegraphics[width=1\linewidth]{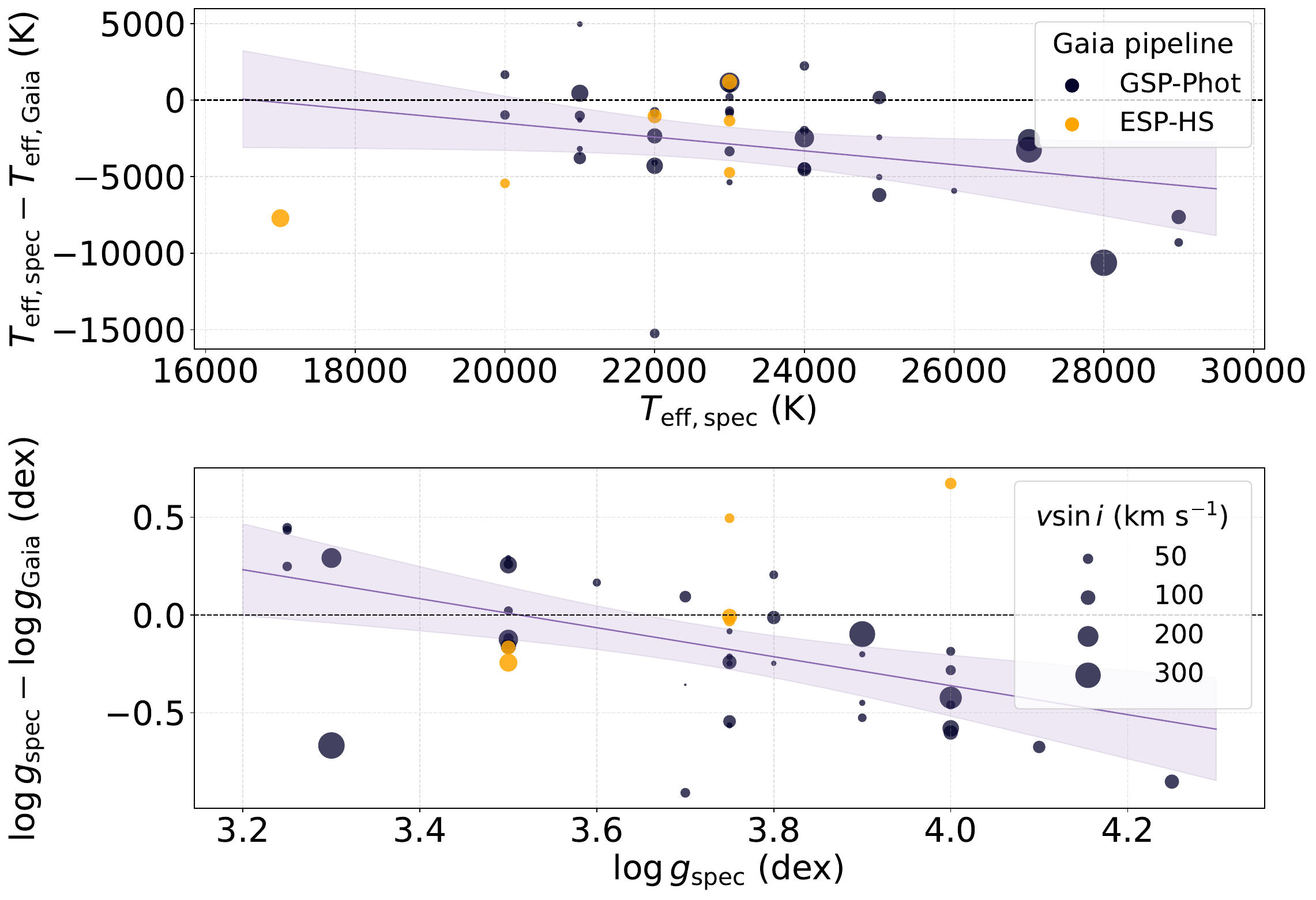}
    \caption{Comparison between the spectroscopically derived atmospheric parameters and the corresponding Gaia values for the studied stars. The upper and lower panels show the differences in $T_{\mathrm{eff}}$ and $\log g$, respectively, as functions of the corresponding spectroscopic values. The solid purple lines represent linear fits to the distributions in the upper and lower panels, while the shaded regions correspond to the 95\% confidence intervals of those fits. The size of each symbol is scaled according to the $v \sin i$ with larger symbols indicating more rapidly rotating stars. Orange points indicate differences calculated using Gaia ESP-HS parameters, whereas dark gray points indicate differences calculated using Gaia GSP-Phot parameters.}
    \label{fig:gaia_comp}
\end{figure}

We compare the spectroscopically derived $T_{\mathrm{eff}}$ and $\log g$ values with the corresponding Gaia parameters \citep{Gaia_DeRidder_pulsations}, and present our results in Figure~\ref{fig:gaia_comp}. The differences between the spectroscopic and Gaia-derived parameters become increasingly negative toward higher $T_{\mathrm{eff}}$ and higher $\log g$ values. We note, however, that a few outlying points, particularly in the $T_{\mathrm{eff}}$ comparison, may influence the apparent trends, and we therefore interpret them with caution. Given that the ESP-HS pipeline is specifically designed for hot stars, we expect its results to show better agreement with our spectroscopic measurements than those obtained from the Gaia GSP-Phot pipeline. The GSP-Phot pipeline is not specifically optimized for hot stars, and the limited temperature sensitivity of its low-resolution BP/RP spectra results in a strong degeneracy between $T_{\mathrm{eff}}$ and extinction. This degeneracy may introduce systematic errors in both parameters and thus contribute to the observed discrepancies \citep{Gaia_hot_stars}. While we do not observe significantly better agreement with our spectroscopic measurements for ESP-HS than for GSP-Phot, only six stars from our spectroscopic sample have ESP-HS measurements available, which limits our ability to make a meaningful comparison between the two Gaia pipelines.

\begin{figure*}
    \centering
    \includegraphics[width=1\linewidth]{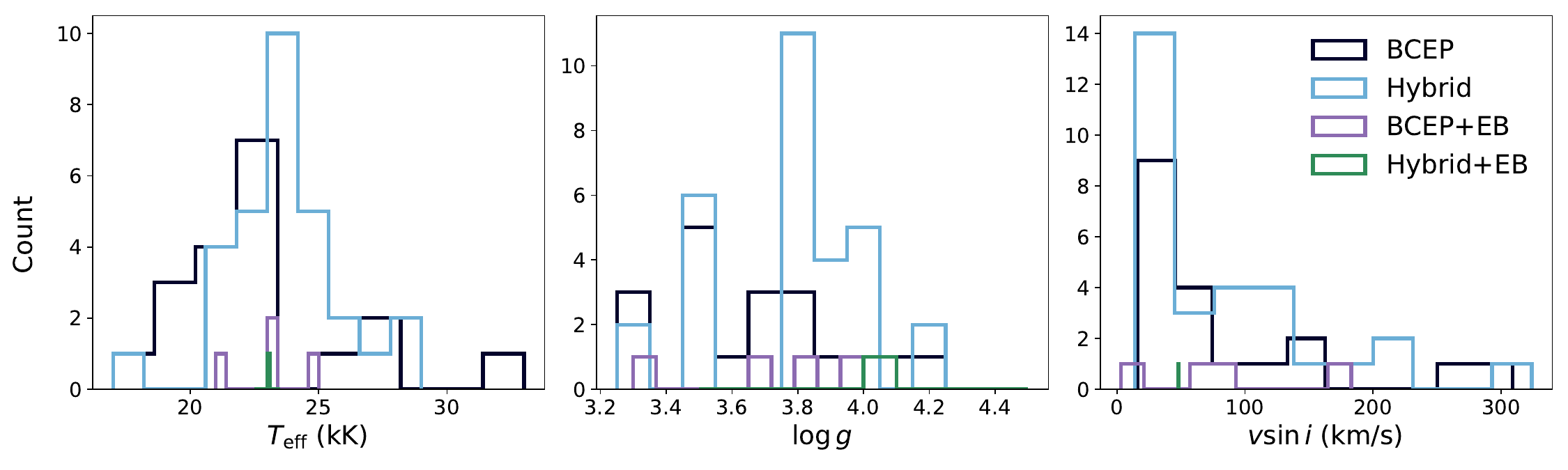}
    \caption{Same as Figure~\ref{fig:parameters_hist} but for parameters derived from the SALT spectra collected in this work for stars classified as \bcep\ stars.}
    \label{fig:histograms_spectra}
\end{figure*}

In Figure~\ref{fig:histograms_spectra} we present the distributions of the stellar parameters derived from the SALT spectra collected in this work. We plot 55 stars, excluding stars that we reclassify in this work as non-\bcep\ pulsators, stars that display spectral features potentially originating from another object (see Section~\ref{sec:spectra}), and stars that we do not reclassify because they do not have TESS light curves.

The $T_{\mathrm{eff}}$ distribution spans 17,000--33,000~K, compared with $\sim$17,290--42,450~K for the Gaia sample. The identification of late O-type \bcep\ pulsators in the previous section, together with Gaia temperatures reaching up to 42,450~K, suggests that the upper limit derived from the SALT sample does not reflect the full temperature range of the \bcep\ population, but rather the limited size and coverage of the spectroscopic sample. Indeed, the small size of the spectroscopic sample limits the conclusions that can be drawn from comparing these distributions with those of the Gaia sample. Additionally, there are selection effects for the spectroscopic sample, as these stars were not randomly chosen, but were prioritized based on certain criteria that could bias these ranges (see Section \ref{sec:spectra}).

\subsection{Positions on the HR Diagram} \label{sec:HRD}
\begin{figure*}
    \centering
    \includegraphics[width=\linewidth]{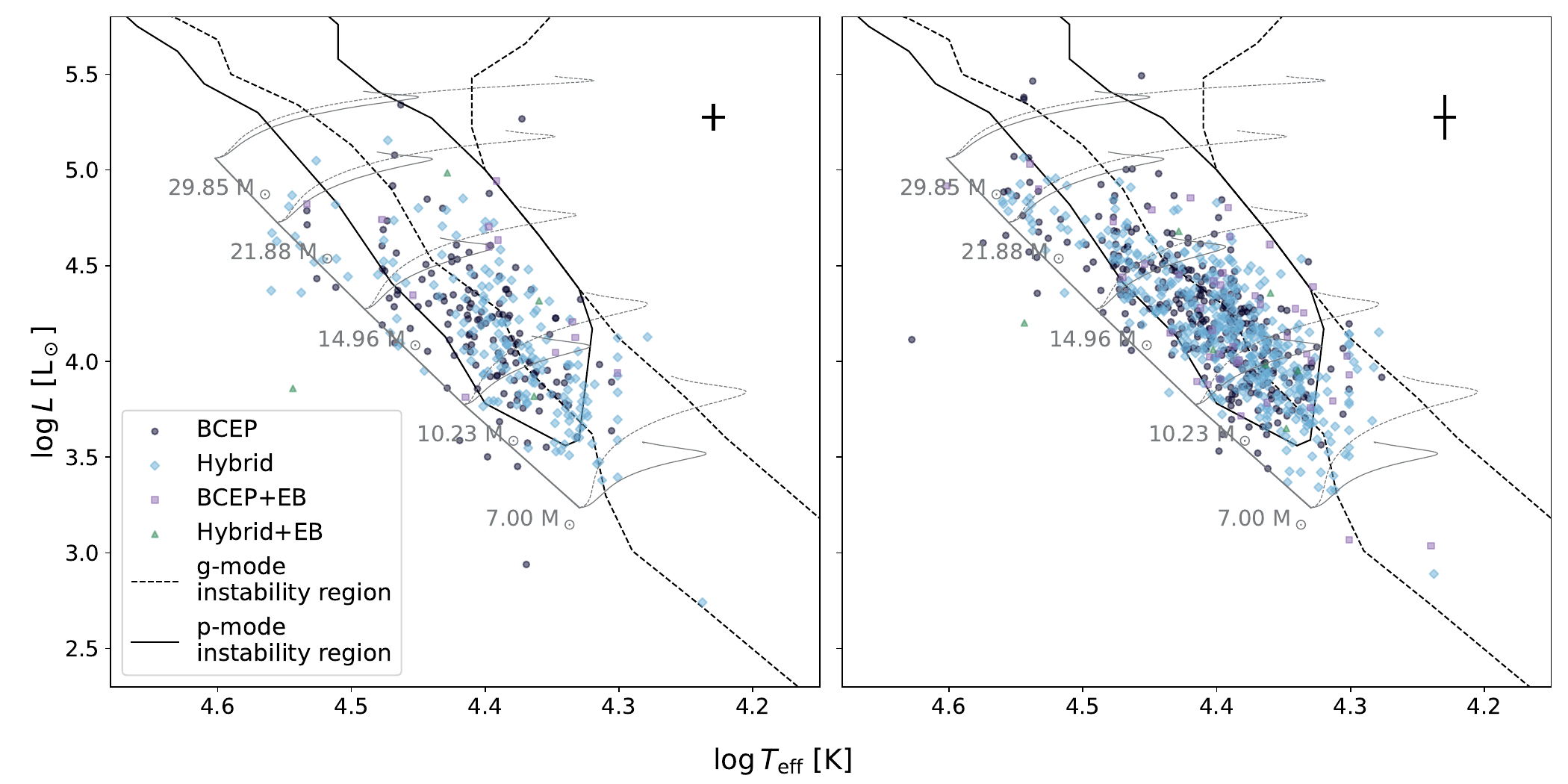}
    \caption{HR diagram for the confirmed \bcep\ stars. The left panel shows 323 stars with Gaia ESP-HS $T_{\mathrm{eff}}$ and weighted three-band Bayestar luminosities. The right panel shows 783 stars with Gaia ESP-HS $T_{\mathrm{eff}}$ and $G$-only luminosities calculated using the Gaia ESP-HS $A_G$ values. Values in both panels assume solar metallicity. Grey lines show stellar tracks from \cite{2026arXiv260710834V} for five different masses as noted (solid lines represent tracks with minimum mixing and dashed lines represent tracks with maximum mixing as included in the grid). The p- and g-mode instability regions, adopted from \cite{2020AA...639A..81B}, are shown within the black solid and dashed lines, respectively. The black crosses show the median propagated uncertainties in each panel.}
    \label{fig:HRD_classic}
\end{figure*}

Figure~\ref{fig:HRD_classic} presents the HR diagram of the confirmed \bcep\ pulsators in our catalog. We compare the weighted three-band Bayestar luminosities with the $G$-only luminosities based on Gaia ESP-HS $A_G$ values to increase the number of stars for which luminosities can be derived. A $G$-only Bayestar luminosity would retain the same 323 stars as the weighted three-band method. In contrast, the $G$-only ESP-HS-$A_G$ method includes all 323 stars in the Bayestar sample and 460 additional stars; the comparison and resulting luminosity choices are described in Appendix~\ref{sec:luminosity_method_comparisons}. For the stars common to both samples, we use the dispersion between the $G$-only and three-band luminosities to estimate the uncertainty of the single-band method, and add this empirical scatter in quadrature to the $G$-only luminosity uncertainties. The larger typical error cross in the right panel reflects this additional uncertainty. 

We also plot solar-metallicity stellar tracks from the model grid of \cite{2026arXiv260710834V} for five different masses (7.00, 10.23, 14.96, 21.88, and 29.85~\Msun), spanning the full mass range of the grid. For each mass, we show models computed with the minimum and maximum mixing values included in the grid: $\log(D) = 1, \; f_{ov} = 0.005$ and $\log(D) = 5, \; f_{ov} = 0.035$, respectively. Here, $D$ denotes the diffusion coefficient (in $\mathrm{cm}^2\mathrm{s}^{-1}$), and $f_{ov}$ is the dimensionless overshooting parameter expressed in units of the pressure scale height. Additionally, instability regions for p-modes and g-modes assuming solar metallicity from \cite{2020AA...639A..81B} are shown.

Figure~\ref{fig:HRD_classic} shows that most \bcep\ stars are concentrated in the mass range of 7–15 \Msun, with only a few objects extending towards lower or higher masses. Most of the stars located below the lower boundary of the region of overlapping p- and g-mode instability, as well as those in the region where only g-modes are expected to be unstable, are hybrid stars or hybrids in EB systems. However, these regions are not exclusively occupied by hybrid stars and a few ``pure'' \bcep\ pulsators also lie outside the displayed instability regions. This is not surprising, given that the instability strips are valid for only one particular choice of input physics; they notably ignore internal rotation. The extended sample shows the same overall temperature--luminosity distribution, while revealing additional lower luminosity stars and a slightly broader luminosity distribution (since it includes more outliers). This broader spread is somewhat expected because its luminosities are derived from one Gaia band rather than from the weighted three-band Bayestar calculation.

Our concentration of \bcep\ stars at 7--15~\Msun\ lies within the broader 5--20~\Msun\ interval containing at least 66\% of the \bcep\ population studied by \cite{2026MNRAS.551g1493T}. Their sample includes lower-mass objects at 5--7~\Msun\ and higher mass outliers reaching approximately 60~\Msun, and it finds \bcep\ pulsators outside the adopted instability regions as well. The overlap between their empirical \bcep\ and SPB distributions is qualitatively consistent with the presence of hybrids across the overlapping instability regions in our diagram. 

\begin{figure}
    \centering
    \includegraphics[width=\linewidth]{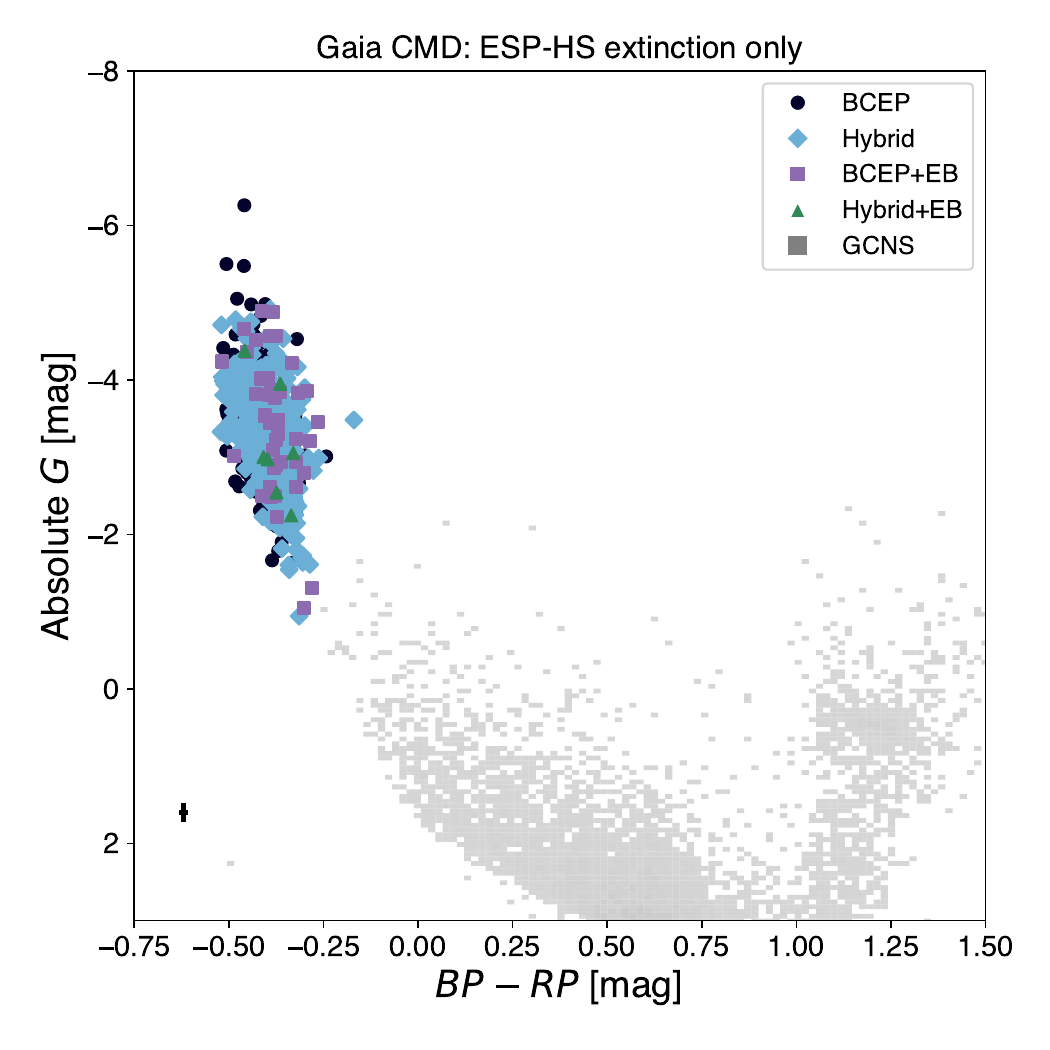}
    \caption{Gaia color-magnitude diagram for the extinction- and reddening-corrected stars in the catalog. The black cross in the bottom-left region shows the mean propagated uncertainty, including the Gaia photometric, parallax, and ESP-HS extinction or reddening uncertainties. The upper-left boundary of the GCNS sample is shown in grey.}
    \label{fig:Gaia_HR_diagram}
\end{figure}

In Figure~\ref{fig:Gaia_HR_diagram}, we display the Gaia color-magnitude diagram for 761 confirmed \bcep\ stars with available Gaia DR3 ESP-HS extinction and reddening values, overlaid on the upper-left edge of the Gaia Catalog of Nearby Stars \citep[GCNS;][]{Gaia2021_GCNS}. We retain only stars with reliable Gaia astrometry, as defined for Subsample~\textit{II}. We directly adopt the extinction and reddening values provided by the ESP-HS pipeline. We further exclude stars with extinction-corrected colors $(BP-RP)_0<-0.521$, because these values are bluer than the approximate physical limit for stellar colors and likely result from over-corrected reddening. This removes 22 stars from the ESP-HS sample. The catalog stars occupy the blue, luminous upper-main-sequence region expected for \bcep\ pulsators, with substantial overlap between the BCEP and hybrid populations.

\subsection{Spectral Types}
Most of the spectral types in our catalog come from the SIMBAD database, which compiles classifications from different literature sources. Some of these spectral classifications are inconsistent with the corresponding ESP-HS $T_{\mathrm{eff}}$. For example, V1155 Cas, is classified as B5 (with $T_{\mathrm{eff}}$ approximately from 14,500 to 16,700 K; \citealt{2013ApJS..208....9P}) but has an ESP-HS $T_{\mathrm{eff}}$ of 29,456 K.

To determine which published spectral classifications are sufficiently reliable for constructing the spectral-type distribution of the \bcep\ stars, we test their consistency with the Gaia ESP-HS $T_{\mathrm{eff}}$. We perform this test using reference spectral-type--effective-temperature relations constructed from NLTE analyses of O- and B-type stars. We do not use the ESP-HS temperatures to infer or assign new spectral types.

For O-type stars, we adopt the mean $T_{\mathrm{eff}}$ for each spectral subtype and luminosity class listed in Table D2 of \cite{2025AA...703A.175H}. For early- and mid-B dwarfs and giants, we associate the spectral classifications listed in Table 1 of \cite{2012AA...539A.143N} with the $T_{\mathrm{eff}}$ derived for the same stars in their Table 5. We then group these stars by spectral subtype and luminosity class and calculate the mean $T_{\mathrm{eff}}$ of each group. We apply the same grouping and averaging procedure to the B-type supergiants with atmospheric parameters listed in Table 3 of \cite{2008AA...478..823M}. Together, these values form the reference spectral-type--effective-temperature relations used in our consistency test.

Before performing the consistency test, we standardize the published spectral classifications.  We ignore suffixes (e.g., n for rotational broadening, e for emission lines, and p for peculiar spectral features). We exclude binary systems with multiple spectral types. We also do not include ambiguous classifications such as ``O'' or ``OB''. We then decompose each valid spectral type into three parts: the spectral class (O or B), the numerical subtype and the luminosity class.

For each star, we test whether its ESP-HS $T_{\mathrm{eff}}$ is consistent with its published spectral subtype. Because the published subtype and luminosity class may be uncertain, we consider all calibration entries for reference subtypes within $\pm$1.5 subtype numbers and include all available luminosity classes. For example, for a star classified as B2, we examine the calibrated temperatures for subtypes B0.5--B3.5 across all available luminosity classes. We define the acceptance interval from the minimum and maximum calibrated mean $T_{\mathrm{eff}}$ values among all these entries and extend it by 1,500~K at both ends.
This extension accounts for discrepancies between $T_{\mathrm{eff}}$ and those expected from spectral types. Although such discrepancies may exceed 1,500~K \citep[as is the case, for example, in Be stars;][]{2026MNRAS.551g1493T}, we consider the spectral classification to be consistent with the ESP-HS temperature if the latter falls within this interval.

The consistency test includes 545 stars, of which 433 have ESP-HS temperatures consistent with their published spectral classifications under these criteria.

\begin{figure*}
    \centering
    \includegraphics[width=1\linewidth]{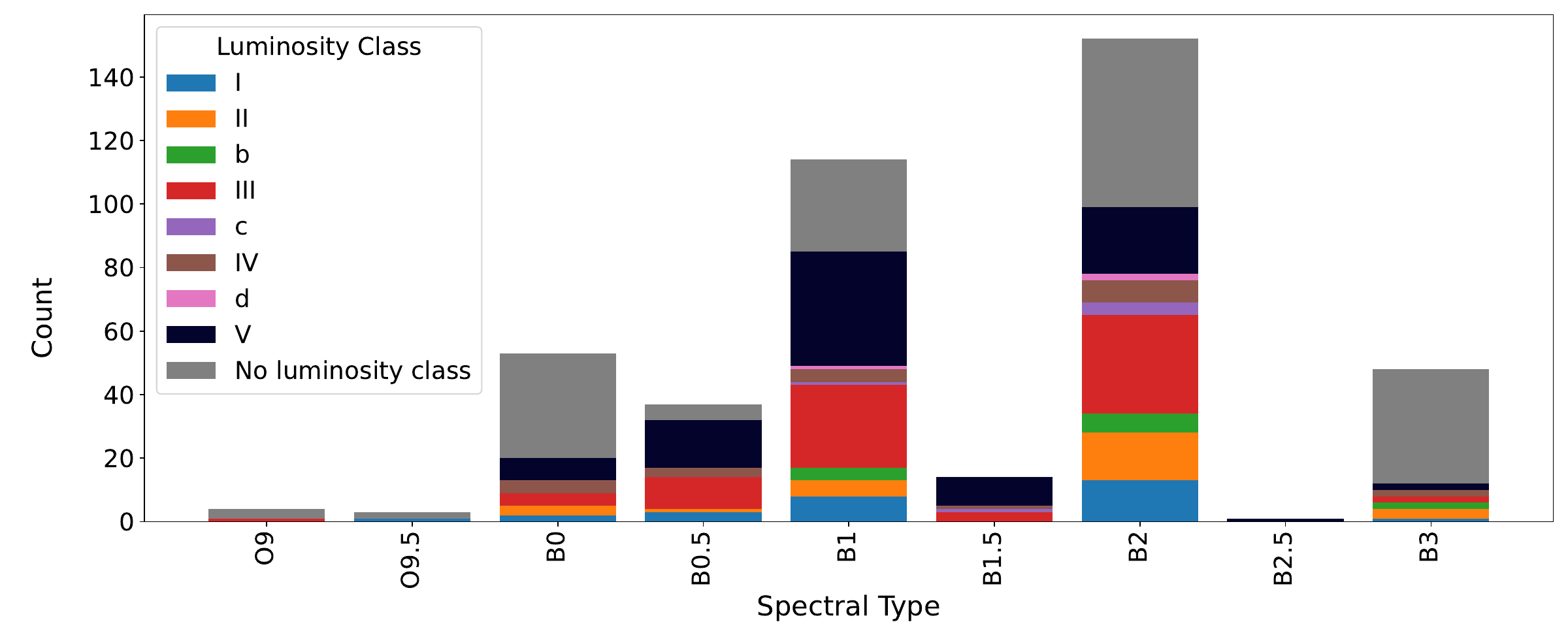}
    \caption{Distribution of the matched spectral types for the confirmed \bcep\ stars in the catalog. The distribution is stacked by reported luminosity class (I--V and intermediate classes), with stars lacking a reported luminosity class shown in grey. Spectral subtypes are grouped into bins centered at half-subtype intervals.}\label{fig:spectype}
\end{figure*}

We present the stacked distribution of the matched spectral types in Figure~\ref{fig:spectype}. The spectral types that pass the consistency test span O9--B3. Most \bcep\ stars have spectral types B1 or B2, consistent with \cite{Stankov_Handler}. The most common luminosity class is V, as expected for \bcep\ stars, which are primarily main-sequence (MS) stars. However, the sample also includes subgiants (IV), giants (III), bright giants (II) and supergiants (I). We do not consider this finding concerning because luminosity classes should be seen as statistical relationships with absolute magnitudes that exhibit large scatter \citep{1998A&A...330..619J}.

Among the stars for which we collect SALT spectra (Table~\ref{tab:spectra} in Appendix~\ref{sec:spectra_app}), the star with the earliest spectral type and a corresponding $T_{\mathrm{eff}}$ that is approximately consistent with its spectral classification is HD~156172, an O9V star. The stars with the latest spectral types and corresponding $T_{\mathrm{eff}}$ values consistent with their spectral classifications have spectral type B2. The resulting O9--B2 range lies within the O9--B3 range obtained from the consistency test.

The temperature range inferred from the temperature distribution shown in Section~\ref{sec:stellar_pulsation_stat} is $\sim$17,290--42,450~K, which corresponds to spectral types O5--B3. This range encompasses both the O9--B2 spectral-type range identified from the SALT spectra and the O9--B3 range obtained from the spectral-type matching analysis. However, the temperature distribution suggests that the population may extend to earlier spectral types, potentially reaching O5. This was also found by \cite{2020AA...639A..81B} and \cite{2026MNRAS.551g1493T}, indicating that there may not be a strict upper limit in $T_\mathrm{eff}$ for \bcep\ stars, although they are rarer at spectral types earlier than O9. The observed upper limit may reflect the rarity of very massive stars and the difficulty of detecting their pulsations, for example, in the presence of strong winds (e.g. \citealt{2010AA...519A..38D}).

\subsection{Coordinates and Velocities} \label{sec:coords_and_vel}
\begin{figure}
    \centering
    \includegraphics[width=\linewidth]{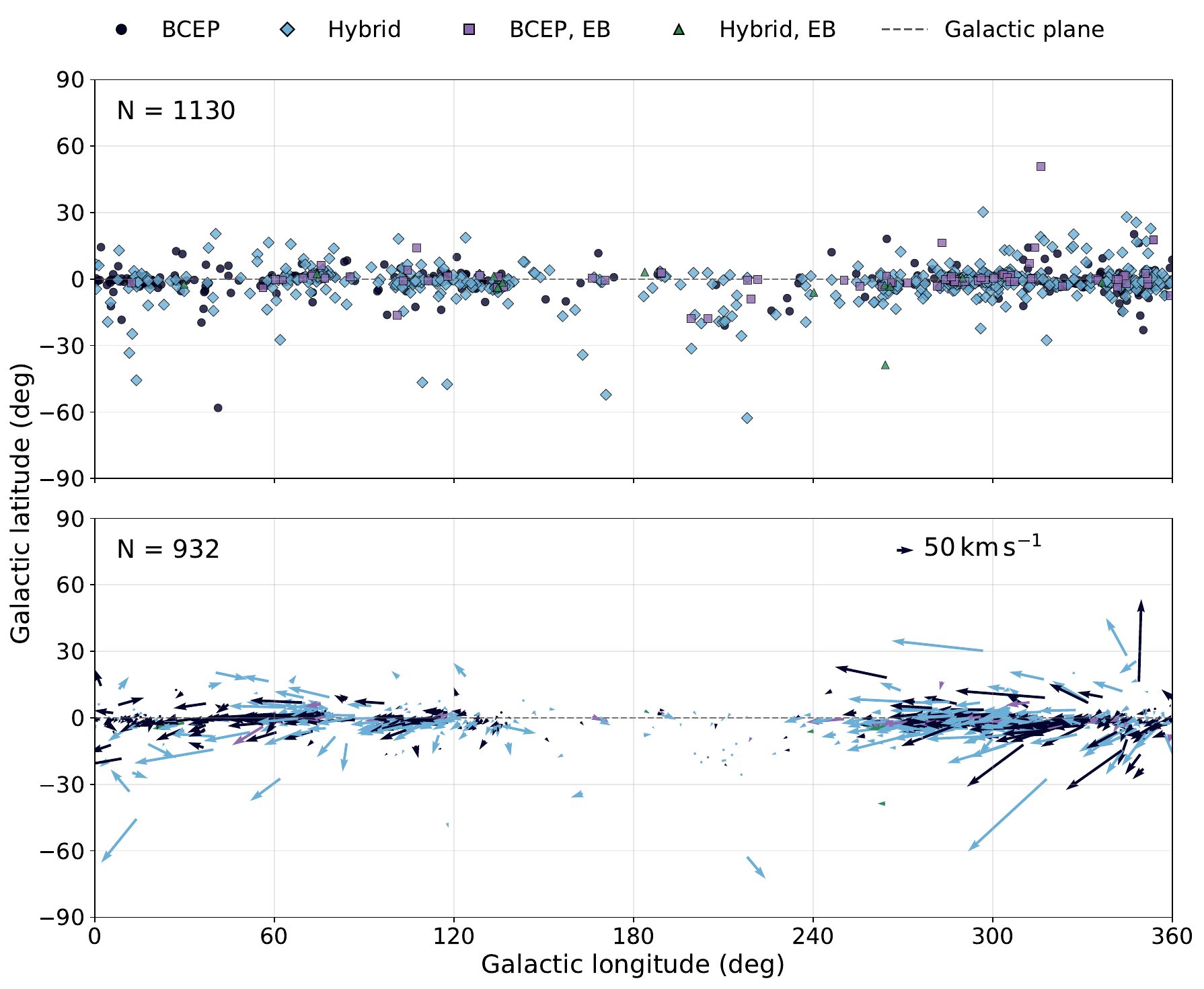}
    \caption{Galactic positions (top; 1,130 stars) and tangential velocity vectors (bottom; 932 stars in Subsample~\textit{II}) of confirmed \bcep\ stars. Dashed horizontal lines mark the Galactic plane. The reference arrow in the bottom panel indicates a tangential velocity of $50~\mathrm{km\,s^{-1}}$.}
    \label{fig:positions_vel}
\end{figure}

Figure~\ref{fig:positions_vel} presents the Galactic positions and tangential velocity vectors of the 1,130 confirmed \bcep\ stars in the upper panel and tangential velocity vectors for the 932 stars in Subsample~\textit{II} in the lower panel. We use Gaia coordinates for 1,121 stars and SIMBAD coordinates for the remaining nine stars, which lack Gaia coordinates. Most of the stars are distributed along the Galactic plane, as expected, with some scatter around it. We compute the tangential velocities from the Gaia DR3 proper motion components and parallax based distances, transforming the proper motion components into Galactic coordinates. 

The velocities exhibit a coherent large-scale pattern, while some stars show clear deviations. We identify 16 stars with exceptional tangential velocities, defined as values exceeding $183.51$ $km \; s^{-1}$ (the sample median plus $3 \sigma$). We additionally identify 39 stars with absolute directional deviations exceeding $46.30^{\circ}$ (the sample median plus $3 \sigma$) and tangential velocities greater than 20 $km \; s^{-1}$. We adopt the latter velocity limit after visually examining the stars with exceptional directions, in order to exclude objects whose large directional angles result from very small velocity components parallel to the Galactic plane despite their low total tangential velocities. We list these stars in Table~\ref{tab:tangential_velocity_candidates} in Appendix~\ref{sec:app_vel}. These selections identify empirical outliers within our catalog; because we do not correct the velocities for Galactic rotation or Solar motion, we do not interpret them as confirmed departures from a Galactic-kinematic model or as confirmed runaway stars. However, these stars may be interesting targets for future investigation.

\begin{figure*}
    \centering
    \includegraphics[width=1\textwidth]{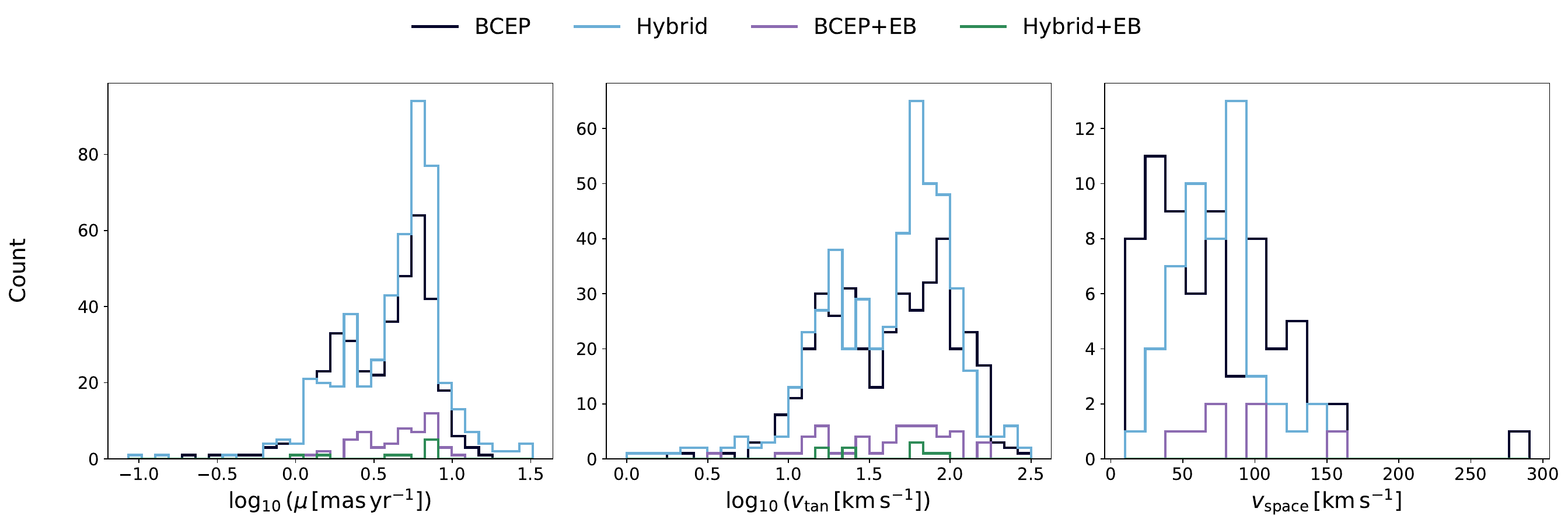}
    \caption{Distributions of proper motion (left), tangential velocity (middle), and total space velocity (right) for the confirmed \bcep\ stars with reliable Gaia astrometry. The colors identify the different classifications. The space-velocity panel includes only stars with available radial-velocity measurements.}
    \label{fig:vel_distributions}
\end{figure*}

In Figure~\ref{fig:vel_distributions} we present the distributions of proper motion, tangential velocity, and space velocity for the stars in our catalog. We convert the Gaia DR3 proper motion measurements into physical plane-of-sky velocities using the parallax based distances. We compute the space velocities as the norm of the tangential and radial-velocity components when radial velocities are available. Consequently, the space-velocity panel contains only 127 stars, compared with 932 stars in the proper-motion and tangential-velocity panels.

For the proper-motion and tangential-velocity distributions in Figure~\ref{fig:vel_distributions}, both the BCEP and hybrid classes show internal structure consistent with bimodality. The kernel density estimates (KDEs) in Figure~\ref{fig:velocity_kdes} in Appendix~\ref{sec:app_vel} similarly show two broad components. The Bayesian Information Criterion (BIC) strongly favors two components for the BCEP proper-motion and tangential-velocity distributions ($\Delta\mathrm{BIC}=51.10$ and $25.22$, respectively) and for the corresponding hybrid distributions ($\Delta\mathrm{BIC}=28.18$ and $38.03$, respectively). Most of the stars with radial-velocity measurements have heliocentric space velocities below approximately $150\,\mathrm{km,s^{-1}}$, while a small number reach higher velocities.

\section{Summary} \label{sec:conclusions}
We collect 2,543 potential \bcep\ stars from the literature and from analysis of archival data (Section~\ref{sec:comp_the_catalog}). Of these, 2,421 have TESS photometry and 2,526 have Gaia data. Based on the TESS light curves, we perform frequency analysis (Section~\ref{sec:lc_freq_analysis}) and visually classify each star (Section~\ref{sec:visual_inspection}). We confirm 1,130 of these potential \bcep\ stars (Section~\ref{sec:class_res}), of which 180 are newly reported here. The confirmed \bcep\ sample includes 593 hybrid stars, displaying both \bcep\ and SPB pulsations, and 93 stars with clear EB signatures in their light curves and/or periodograms. In addition to these 1,130 confirmed \bcep\ stars, we classify 246 stars as candidate \bcep\ / hybrid or EB systems because of noisy light curves that prevent a definitive classification or a lack of reliable spectral types or $T_{\mathrm{eff}}$; 18 of these stars are newly reported here.

We find \bcep\ and hybrid stars within ELL and heartbeat systems, and detect signs of rotational modulation (Table~\ref{tab:flag_counts}). We also find stars with SLF variability on top of their pulsation signals, as well as stars that display only SLF variability with no detected pulsation signal. We further identify 28 \bcep\ stars that display three to four times more frequencies than typically observed in this class, along with 25 stars that, in addition to their \bcep\ pulsations, display high frequencies (greater than 15~$d^{-1}$) that do not appear to be harmonics and combination frequencies of lower frequencies. We discuss the possibility that these signals arise from contamination by nearby stars given the TESS spatial resolution. We indeed identify a candidate contamination source for some of these stars, with the rest requiring further investigation.

We also collect spectra for 67 selected southern-hemisphere potential \bcep\ stars based on their ranking according to the number of pulsation modes detected from the ground, their observability with TESS and their apparent magnitude. We analyze and model these spectra to derive $T_{\mathrm{eff}}$, $\log g$, and projected rotational velocity. We compare the $T_{\mathrm{eff}}$ and surface gravities with the values estimated by Gaia. The differences between the spectroscopically and Gaia-derived parameters become more substantial toward higher $T_{\mathrm{eff}}$ and higher $\log g$, although a few outliers may influence these apparent trends. The limited number of stars with ESP-HS measurements prevents a robust comparison between the Gaia pipelines and the parameters derived from the spectra.

We perform a statistical study (Section~\ref{sec:res_stat}) of the 1,130 confirmed \bcep\ stars. The dominant frequencies of stars classified as \bcep\ and hybrid stars span 3.28--16.63 and 3.31--16.71~$d^{-1}$, respectively, with medians of 5.84 and 6.32~$d^{-1}$. We interpret these distributions with caution because of possible selection effects discussed in Section~\ref{sec:stellar_pulsation_stat}. We classify 246 additional stars as candidate \bcep/hybrid or EB systems and 1,045 stars as candidate SPB, DSCT, GDOR, or unknown-variability-type objects, or as stars without a detectable pulsation signal in the available TESS data.

We find that \bcep\ stars span the spectral types O9--B3, with the majority of the population concentrated in the B1--B2 types. Across all four confirmed populations, the $T_{\mathrm{eff}}$ span 17.29--42.45~kK, with a median of 24.76~kK. While the spectral classifications directly support an O9--B3 range, the temperature distribution suggests that the population may extend to earlier spectral types, potentially reaching O5. Across all four populations, the $\log g$ values span 2.47--4.60, with a median of 3.88. Twenty stars have ESP-HS surface gravities below $\log g=3.2$ (Table~\ref{tab:low_logg_stars}), hinting that they may have evolved beyond the main sequence. However, spectroscopic follow-up is required to verify their surface gravities and evolutionary states.

On the HR diagrams (Figure~\ref{fig:HRD_classic}), most \bcep\ stars are located within the mass range of 7--15~$\Msun$, with relatively few objects at lower or higher masses. Most stars located outside the lower boundary of the region of overlapping p- and g-mode instability, as well as within the region where only g modes are expected to be unstable, are hybrid pulsators or hybrids in EB systems. However, these regions are not exclusively occupied by hybrid pulsators, and a few \bcep\ pulsators also lie outside the displayed instability regions.

This catalog represents the largest compilation of \bcep\ stars to date, allowing us to study them statistically as a group. Additionally, this collection of reliably classified \bcep\ pulsators can be used for selecting targets for follow-up observations for complete asteroseismic studies, such as the GAP \citep{Shitrit_Arcavi_GAP}, and for ensemble studies such as those of \cite{Fritzewski_Gaia_modeID} and \cite{2026arXiv260710834V}. we therefore hope that this catalog provides an essential step in advancing our understanding of the interiors and evolution of massive stars.

\begin{acknowledgments}
We are grateful to Joshua Pepper for his work on the KELT data in the archival data search part of this work.
N.S. acknowledges support from the Israel Science Foundation (grant number 2312/24) and from the Pazy foundation (grant number 216312).
I.A. acknowledges support from the European Research Council (ERC) under the European Union’s Horizon 2020 research and innovation program (grant agreement number 852097) and from the United States - Israel Binational Science Foundation (BSF; grant number 2024812).
D.M.M. gratefully acknowledges UK Research and Innovation (UKRI) in the form of a Frontier Research grant under the UK government's ERC Horizon Europe funding guarantee (grant number EP/Y031059/1) and a Royal Society University Research Fellowship (grant number URF{\textbackslash}R1{\textbackslash}231631). 
This research was supported in part by the Polish National Center for Science (NCN; grant numbers 2015/18/A/ST9/00578 and 2021/43/B/ST9/02972) and by the National Science Centre, Poland (NCN; grant number 2022/45/B/ST9/03862). Polish participation in SALT is funded by MNiSW (grant number 2026/WK/02). 
Some of the observations reported in this paper were obtained with the Southern African Large Telescope (SALT) under programs 2017-2-SCI-029, 2019-1-SCI-036 and 2022-1-SCI-018 (PI: Handler).

\end{acknowledgments}

\software{\textsc{astropy} \citep{astropy:2013,astropy:2018,astropy:2022},
          \textsc{Lightkurve} \citep{Lightkurve_2018}, \textsc{dustmap} \citep{dustmaps_python}, \textsc{SciPy} \citep{2020SciPy-NMeth}, \textsc{scikit-learn} \citep{scikit-learn}, \textsc{STAR SHADOW} \citep{2024AA...685A..62I}
          \textsc{PySALT} \citep{2010SPIE.7737E..25C}}

\bibliography{bib}{}
\bibliographystyle{aasjournal}

\appendix

\section{Literature Review Sources} \label{sec:literature_rev_app}
In Table~\ref{tab:sources} we present the sources from which the literature-review \bcep\ stars were collected (Section~\ref{sec:literature_rev}).

\startlongtable
\begin{deluxetable*}{lcc}
\tablecaption{Summary of the literature-review sources. \label{tab:sources}}
\tablehead{
\colhead{Source} &
\colhead{Total Number of potential \bcep\ stars} &
\colhead{Missing TESS Light Curve}
}
\startdata
SIMBAD; \cite{SIMBAD} & 7 & 2 \\
VSX; \cite{2015yCat....102027W} & 1401 & 27 \\
\cite{1967ApJS...14..301H} & 1 & 0 \\
\cite{1997AcA....47..365P} & 1 & 0 \\
\cite{2003AstL...29..468S} & 1 & 0 \\
\cite{2004ASPC..318..288B} & 2 & 0 \\
\cite{2005IBVS.5667....1H} & 5 & 0 \\
\cite{Stankov_Handler} & 191 & 11 \\
\cite{2006CoAst.147...48D} & 1 & 0 \\
\cite{2006MmSAI..77..342N} & 1 & 1 \\
\cite{2006AcA....56..267P} & 1 & 0 \\
\cite{2007AcA....57...61P} & 2 & 0 \\
\cite{2007IAUS..240..456P} & 4 & 0 \\
\cite{2007ASPC..361...64T} & 100 & 4 \\
\cite{2008JPhCS.118a2068M} & 6 & 0 \\
\cite{Pigulski_HD67600_example} & 120 & 0 \\
\cite{2008AA...477..907P} & 1 & 0 \\
\cite{2008JPhCS.118a2013W} & 1 & 0 \\
\cite{2009AIPC.1170..373A} & 1 & 0 \\
\cite{2009AA...506..471D} & 1 & 0 \\
\cite{2009AN....330..317H} & 2 & 0 \\
\cite{2009IBVS.5881....1J} & 1 & 0 \\
\cite{2009AcA....59..349M} & 1 & 0 \\
\cite{2011MNRAS.413.2403B} & 5 & 0 \\
\cite{2011AA...533A..70H} & 6 & 0 \\
\cite{2011PASP..123..671H} & 1 & 0 \\
\cite{2011AA...526A.124L} & 2 & 0 \\
\cite{2011sf2a.conf..231N} & 1 & 0 \\
\cite{2011AA...531A.135P} & 1 & 0 \\
\cite{2011AA...534A.125U} & 1 & 0 \\
\cite{2012AJ....143..101M} & 2 & 2 \\
\cite{2012AJ....144...86Z} & 1 & 0 \\
\cite{2013AcA....63..339D} & 2 & 0 \\
\cite{2013AA...559A.108M} & 3 & 0 \\
\cite{2013AA...553A.127P} & 2 & 0 \\
\cite{2013AcA....63..379P} & 48 & 47 \\
\cite{2013AA...550A.120S} & 15 & 5 \\
\cite{2013NewA...23...55U} & 1 & 0 \\
\cite{2014AJ....147..149B} & 1 & 0 \\
\cite{2014AA...568A..32D} & 3 & 0 \\
\cite{2014MNRAS.442..273L} & 3 & 0 \\
\cite{2014IAUS..301..479S} & 6 & 0 \\
\cite{2015MNRAS.451.1445B} & 6 & 1 \\
\cite{2016MNRAS.459.3721B} & 1 & 0 \\
\cite{2016MNRAS.456.2505L} & 1 & 0 \\
\cite{2017AA...598A.108L} & 11 & 1 \\
\cite{2017ARep...61...80S} & 2 & 0 \\
\cite{2017OEJV..185....1S} & 1 & 1 \\
\cite{2018MNRAS.478.2676B} & 1 & 0 \\
\cite{2019MNRAS.489.1304B} & 2 & 0 \\
\cite{2019AJ....158...68L} & 1 & 0 \\
\cite{2019AA...632A..95M} & 4 & 0 \\
\cite{2019MNRAS.482..658X} & 5 & 0 \\
\cite{2020MNRAS.493.2528B} & 4 & 0 \\
\cite{2020AA...639A..81B} & 7 & 0\\
\cite{2020AJ....160...32L} & 52 & 0 \\
\cite{2020MNRAS.497L..19S} & 1 & 0 \\
\cite{2021AA...647A..12K} & 2 & 0 \\
\cite{2021AJ....161...32L} & 1 & 0 \\
\cite{2021ApJ...910..133S} & 1 & 0 \\
\cite{2021RAA....21..227Z} & 2 & 0 \\
\cite{2022AA...658A..96B} & 5 & 0 \\
\cite{2022MNRAS.513.3191S} & 11 & 0 \\
\cite{2022ApJS..263...34C} & 5 & 0 \\
\cite{Burssens_2023Nat} & 1 & 0 \\
\cite{2023AA...674A..14R} & 1450 & 29 \\
\cite{Shi_2023_OB_pulsators_Gaia} & 14 & 0 \\
\cite{Eze_Handler_2024_EBs} & 90 & 0 \\
\cite{2024AA...688A..93H} & 165 & 2 \\
\cite{2024ApJS..271...28S} & 99 & 0 \\
\cite{2024AA...683A..49Z} & 17 & 4 \\
\cite{2025MNRAS.543.2958C} & 2 & 0 \\
\cite{2025MNRAS.541.1866C} & 16 & 13 \\
\cite{2025AA...698A.253F} & 27 & 1 \\
\cite{2025AA...704A.280K} & 3 & 0 \\
\cite{2025AA...703A.239N} & 4 & 1 \\
\cite{2025ApJS..281...17S} & 77 & 0 \\
\cite{2026AA...705A..60Z} & 1 & 0 \\
\enddata
\tablecomments{The numbers include duplicates between different literature sources.}
\end{deluxetable*}

\section{Effect of Cadence Equalization (Binning) on Pulsation-Parameter Extraction} \label{sec:binning_vs_nobinning}
We investigate the impact of binning on the recovery of pulsation parameters in TESS light curves, focusing on the frequencies and amplitudes derived from iterative prewhitening using STAR SHADOW (see Section~\ref{sec:lc_freq_analysis}).

The impact of binning depends on the ratio between the oscillation period $P$ and the bin width $\Delta t$. For modes with $P \gg \Delta t$ (equivalently $f \ll 1/\Delta t$), as is the case for the \bcep\ pulsations analyzed here, the signal varies little within each bin, and the measured pulsation properties are not expected to be significantly affected. In contrast, for modes with $P \sim \Delta t$ or $P < \Delta t$ (i.e. $f \gtrsim 1/\Delta t$), the signal changes significantly across the bin, and a larger effect is expected, particularly the attenuation of measured amplitudes caused by temporal averaging \citep{2012MNRAS.422..665M}. 

\cite{2012MNRAS.422..665M} examines cadence effects in Kepler short- and long-cadence data for individual targets, whereas we quantify the effects of cadence equalization across 1,014 TESS targets from our sample, using the same frequency-extraction procedure for the binned and unbinned data. For each star, we compare the dominant frequency and its corresponding amplitude derived from the binned light curve with those derived from the original, unbinned data.

\begin{figure}
    \centering
    \includegraphics[width=0.6\linewidth]{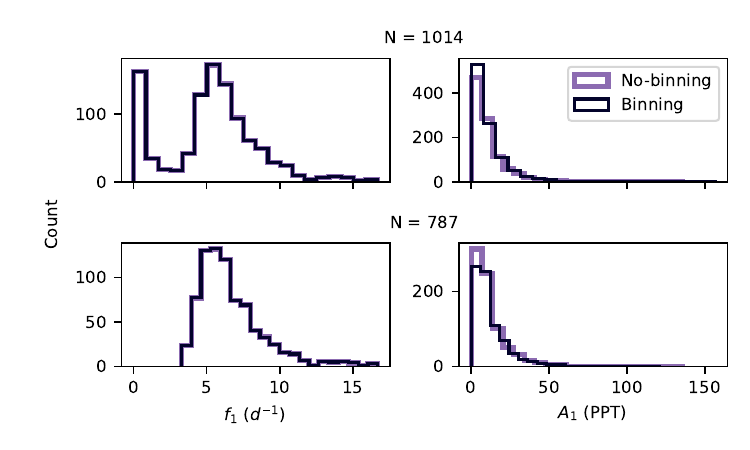}
    \caption{Histograms of the dominant frequency ($f_1$) and its corresponding amplitude ($A_1$) for binned and unbinned light curves and for the complete sample (top) and Subsample \textit{I} (bottom). The number above each row indicates the number of stars included in each panel. While the frequencies remain consistent between the binned and unbinned data, some differences in derived amplitudes are more apparent.}
    \label{fig:binning_VS_nobinning}
\end{figure}

Figure~\ref{fig:binning_VS_nobinning} presents the results for the complete sample of 1,014 stars and for the 787 stars in Subsample~\textit{I}. We find that the frequencies are very consistent. Only three stars in the complete sample ($\sim0.3 \%$) and one star in Subsample~\textit{I} ($\sim0.1 \%$) have an absolute frequency difference greater than $0.01$~$d^{-1}$ between the binned and unbinned data. For a threshold of $0.001$~$d^{-1}$, 81 stars in the complete sample ($\sim8.0 \%$) and 50 stars in Subsample~\textit{I} ($\sim6.4 \%$) exceed the threshold. For a threshold of $0.0001$~$d^{-1}$, the corresponding numbers are 107 ($\sim10.6 \%$) and 63 ($\sim8.0 \%$), respectively. These results show that binning does not introduce significant shifts in the recovered frequencies in this dataset, as expected for bin widths substantially smaller than the periods.

\begin{figure}
    \centering
    \includegraphics[width=0.4\linewidth]{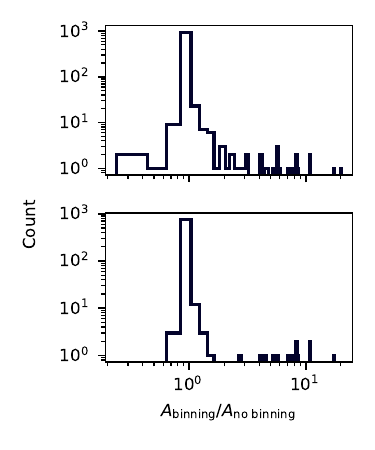}
    \caption{Histograms of the amplitude ratios between binned and unbinned light curves for each star in the complete sample (top) and Subsample~\textit{I} (bottom).}
    \label{fig:binning_amplitude_ratios}
\end{figure}

The binned and unbinned analyses yield identical amplitudes for 630 of the 1,014 stars in the complete sample ($62.1\%$) and 536 of the 787 stars in Subsample~\textit{I} ($68.1\%$). Among the stars with different amplitudes, the unbinned amplitude is larger for 237 of 384 objects ($61.7\%$) and 175 of 251 objects ($69.7\%$), respectively. The ratios between the amplitudes derived from the binned and unbinned data span 0.240--20.195 and 0.678--17.568, respectively, indicating amplitude differences can be substantial for individual cases. However, among stars with different amplitudes, the median ratios are 0.997 and 0.988, indicating that the binned amplitudes are typically only $0.3\%$ and $1.2\%$ lower than the corresponding unbinned values. The amplitude ratio distributions in Figure~\ref{fig:binning_amplitude_ratios} show a strong concentration near unity, indicating that the amplitudes agree closely for most stars. Among stars with different amplitudes, ratios above unity are more common, indicating that the amplitudes calculated from unbinned light curves are more often larger than the ones calculated from binned light curves, as expected. In addition, only a small fraction of stars extend to large ratios.

Overall, the frequencies closely agree for all stars, and the amplitudes closely agree for most stars, with amplitudes derived from the unbinned data on average slightly larger than those derived from binned data. In the unbinned case, sectors with shorter cadence contribute more data points and therefore carry greater statistical weight in the fitting procedure. This can bias the recovered amplitudes toward values characteristic of the high-cadence data. In contrast, the binning procedure equalizes the cadence across sectors, reducing this imbalance, but at the cost of introducing some amplitude attenuation due to temporal averaging. We therefore do not find a global preference for using one method over the other. Here we use the unbinned data for the statistical analysis presented in Section~\ref{sec:stellar_pulsation_stat} in order to avoid the amplitude attenuation introduced by binning.

\section{Prewhitening Process} \label{sec:prewhitening_app}
For the frequency analysis, we compute the pulsation frequencies, amplitudes and phases using the STAR SHADOW code, which performs iterative prewhitening (see Section \ref{sec:lc_freq_analysis}). STAR SHADOW supports two complementary stopping criteria. The first is the BIC: $\mathrm{BIC} = k \ln N - 2 \ln \hat{L},$
where $k$ is the number of free parameters in the time-domain regression model, $N$ is the total number of data points in the light curve, and $\hat{L}$ is the maximum likelihood of the fitted model. In STAR SHADOW, a $\Delta \mathrm{BIC} > 2$ threshold is used by default and is always enabled.
The second criterion is a false-alarm calibrated signal-to-noise (SNR) threshold following \cite{2021AcA....71..113B}.  This threshold considers the Fourier domain, corresponds to a false-alarm probability of $0.1\%$ and depends explicitly on the number of data points. The user can either enable this criterion together with the BIC threshold or disable it.

In the version of STAR SHADOW used here, the BIC threshold can be changed. We adopt $\Delta \mathrm{BIC} > x$, where we set $x=2$ for sources with data from one TESS sector, $x=6$ for two TESS sectors, and $x=10$ for more than two TESS sectors. This increasing BIC threshold requires progressively stronger evidence before adding an additional sinusoid, to limit overfitting and the detection of numerous low-amplitude frequencies in datasets with a larger number of TESS sectors. We use only the BIC threshold as the stopping criterion and disable the SNR stopping criterion, as the latter depends subjectively on the user's choice of the frequency range to compute the SNR.

Since STAR SHADOW is designed to model EB variability on top of the intrinsic stellar oscillations, it also computed 19~EB system parameters for light curves with detected eclipses. 
The intrinsic stellar oscillations are computed in the early stages of the analysis, and the parameters of the EB system are derived at later stages. While early stages are always executed (for a valid light curve), later stages are only reached if the data satisfy specific criteria, such as the presence of significant harmonic structure or eclipse signatures. At some of the later stages, the pulsation parameters (frequencies, amplitudes, and phases) are recomputed based on the EB parameters determined at those stages.

We define an \textit{EB\_SCORE} flag that counts how many of the 19 EB system parameters are computed for each star. We consider stars with \textit{EB\_SCORE}~$\geq~17$ to be EBs (according to STAR SHADOW, regardless of the visual classification), and for these stars, we use the pulsation parameters computed at the final stage of the STAR SHADOW analysis. For stars with \textit{EB\_SCORE}~$<~17$, we adopt the pulsation parameters obtained from the initial frequency extraction stage.

A point to note is that tidally excited oscillations in EBs are classified as orbital harmonics. Tidally excited pulsations naturally occur at integer multiples of the orbital frequency due to resonant tidal forcing (see discussion by \citealt{2026ARAA..64..223S}), and the analysis procedure explicitly enforces this by matching frequencies within a tolerance to exact orbital harmonics. As a result, genuine pulsations located at $n \cdot f_{\mathrm{orb}}$ are absorbed into the harmonic model and excluded from the non-harmonic frequency list, which is a known limitation of the code \citep{2024AA...685A..62I}.

\section{Luminosity and Extinction Method Comparisons}
\label{sec:luminosity_method_comparisons}
We compare the luminosity and extinction estimates calculated in Section~\ref{sec:bol_par} and used in the HR diagrams presented in Section~\ref{sec:HRD} to assess how the choice of luminosity calculation method affects both the derived values and the number of stars that can be included in the HR diagrams. We retain confirmed \bcep\ and hybrid pulsators, including EB systems, with reliable Bayestar19 estimates (\texttt{bayestar\_converged}=True and \texttt{bayestar\_reliable\_dist}=True).

For each comparison, we use only the mutual sample, namely stars with finite values for both quantities being compared:
\begin{enumerate}
\item[(a)] Weighted three-band versus $G$-only Bayestar luminosities, both calculated assuming solar metallicity. This comparison contains 323 stars. The weighted three-band calculation requires reliable Bayestar estimates and Gaia $G$, $BP$, and $RP$ photometry.
\item[(b)] Weighted three-band Bayestar luminosities versus $G$-only luminosities calculated using the ESP-HS $A_G$ values, assuming solar metallicity. This comparison contains 323 stars.
\item[(c)] Bayestar versus ESP-HS $A_G$ estimates. This comparison contains 323 stars.
\item[(d)] Weighted three-band Bayestar luminosities calculated using solar and GSP-Phot metallicities. This comparison contains 245 stars because it additionally requires an available GSP-Phot metallicity.
\end{enumerate}

\begin{figure}
    \centering
    \includegraphics[width=0.7\linewidth]{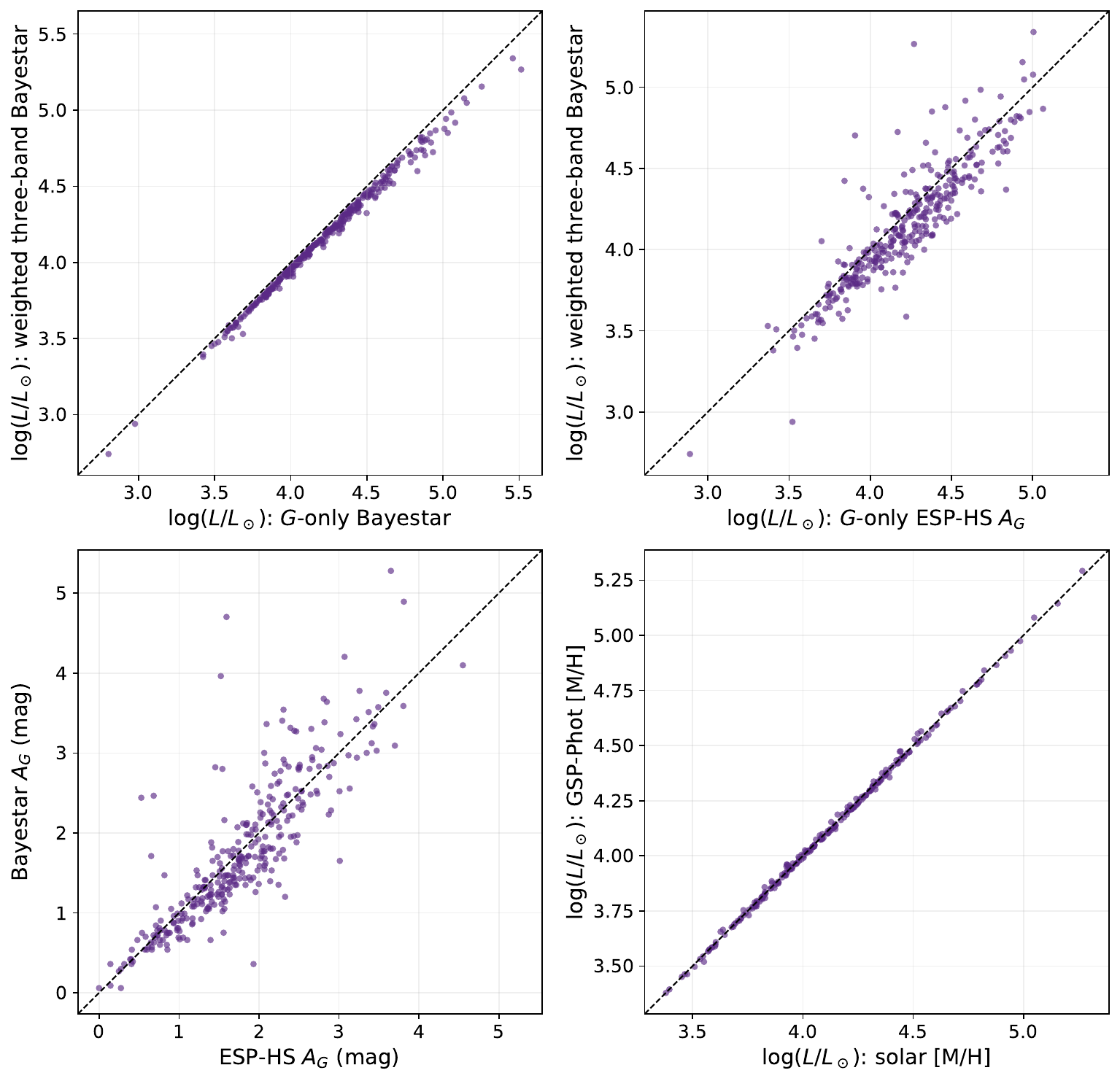}
    \caption{Comparison of the luminosity and extinction estimates used for the HR diagrams. The panels, read from upper left to lower right, show comparisons (a)--(d), respectively: weighted three-band versus $G$-only Bayestar luminosities assuming solar metallicity; weighted three-band Bayestar versus $G$-only ESP-HS-$A_G$ luminosities assuming solar metallicity; Bayestar versus ESP-HS $A_G$ estimates; and weighted three-band Bayestar luminosities calculated using solar and GSP-Phot metallicities. The dashed lines indicate equality.}
    \label{fig:luminosity_method_comparisons}
\end{figure}

Figure~\ref{fig:luminosity_method_comparisons} shows these four comparisons. In (a), the weighted three-band luminosities are lower than the $G$-only Bayestar luminosities by a median of 0.048~dex. In (b), the median difference is $-0.081$~dex and the scatter is larger. In (c), Bayestar gives lower extinction values than ESP-HS by a median of 0.077~mag. Finally, (d) shows a median difference of only $-0.004$~dex, indicating that the solar-metallicity assumption has a negligible effect on the luminosities.

Together with the corresponding sample sizes, these comparisons motivate the two HR diagram luminosity choices. We retain the weighted three-band Bayestar luminosity as the primary estimate for the left panel of Figure~\ref{fig:HRD_classic} because it uses all three Gaia bands and reliable Bayestar19 extinction estimates. Although the corresponding $G$-only Bayestar luminosities agree closely with the weighted three-band Bayestar luminosities, they show a small systematic offset and provide the same 323 stars under the Bayestar-quality selection. Thus, the $G$-only Bayestar method does not extend the HR diagram sample. We retain the $G$-only ESP-HS-$A_G$ luminosity for the extended HR diagram sample in the right panel of Figure~\ref{fig:HRD_classic} because it is available for substantially more stars; it provides 783 stars, including all 323 stars in the Bayestar sample and 460 additional stars. Its broader scatter relative to the weighted three-band values supports treating it as a separate estimate and including the empirical dispersion as an additional uncertainty. The difference between the Bayestar and ESP-HS extinction estimates further supports retaining the two methods separately rather than combining them. Finally, because the solar- and GSP-Phot-metallicity luminosities agree almost exactly for the stars with both estimates, we adopt solar metallicity for the HR diagram luminosities, which also permits the inclusion of more stars without GSP-Phot metallicities.

\section{Stellar Parameters from Spectra} \label{sec:spectra_app}
In Table~\ref{tab:spectra} we present the spectral observation log and the stellar parameters obtained from the spectra (see Section~\ref{sec:spectra}), namely $T_{\mathrm{eff}}$, $\log g$ and $v \sin(i)$. We also present the classification of each star as determined in this work. Stars with no classification in the table do not have TESS light curves and therefore were not reclassified in this work. We note that some of the stars for which we collect spectra were reclassified in this work as non-\bcep\ pulsators, but we still publish their spectroscopic data and analysis here for completeness.

\begingroup
\begin{deluxetable*}{lcccccccccc}
\renewcommand{\arraystretch}{0.95}
\tabletypesize{\scriptsize}
\tablecaption{Spectral observation log and stellar parameters computed from the spectra.} \label{tab:spectra}
\tablehead{
  \multicolumn{1}{l}{Name} &
  \multicolumn{1}{c}{Spec. Type} &
  \multicolumn{1}{c}{Obs. Date} &
  \multicolumn{1}{c}{Num.} &
  \multicolumn{1}{c}{SNR} &
  \multicolumn{1}{c}{Mode\tablenotemark{a}} &
  \multicolumn{1}{c}{$T_{\mathrm{eff}}$} &
  \multicolumn{1}{c}{$\log g$} &
  \multicolumn{1}{c}{$v \sin i$} &
  \multicolumn{1}{c}{Ref.\tablenotemark{b}} &
  \multicolumn{1}{c}{Classification\tablenotemark{c}} \\
  \multicolumn{1}{c}{} &
  \multicolumn{1}{c}{} &
  \multicolumn{1}{c}{} &
  \multicolumn{1}{c}{} &
  \multicolumn{1}{c}{} &
  \multicolumn{1}{c}{} &
  \multicolumn{1}{c}{(K)} &
  \multicolumn{1}{c}{} &
  \multicolumn{1}{c}{($km \; s^{-1}$)} &
  \multicolumn{1}{c}{} &
  \multicolumn{1}{c}{} 
}
\startdata
 BW\,Cru            & B2\,III            & Jan., Apr. 2018           & 2   & 208, 230        & H     & 22,000\,$\pm$\,1000  & 3.50\,$\pm$\,0.15  & 94\,$\pm$\,5 &1 & Hybrid  \\
 HN\,Aqr            &    B2           & June 2019, Jul, Aug 2022  & 4   & 55, 130, 85, 90 & M, L  & 22,000\,$\pm$\,1000  & 3.80\,$\pm$\,0.10  & 36\,$\pm$\,3 &2 &BCEP  \\
 KZ\,Mus            & B2\,III       & Feb. 2018                 & 1   & 130             & H     & 20,000\,$\pm$\,1000  & 3.75\,$\pm$\,0.10  & 44\,$\pm$\,3  &3 & BCEP \\
 V4386\,Sgr         & B1\,II       & Apr. 2018                 & 2   & 170,215         & H     & 27,000\,$\pm$\,1000  & 3.90\,$\pm$\,0.15  & 309\,$\pm$\,17 &4 & BCEP+EB \\
 V916\,Cen          & B0.5\,IVne      & Feb., May 2019            & 2   & 275, 150        & M     & 25,000\,$\pm$\,1000  & 3.80\,$\pm$\,0.10  & 84\,$\pm$\,4  &5 & BCEP+EB \\
 BD\,-02\,4752      & B0.5\,V       & Apr., May 2019            & 2   & 35, 50          & M     & 25,000\,$\pm$\,1000  & 4.25\,$\pm$\,0.15  & 95\,$\pm$\,5  &6 & BCEP \\
 CD\,-31\,17605     &               & Aug. 2022                 & 2   & 100, 65         & L     & 23,000\,$\pm$\,1000  & 4.00\,$\pm$\,0.10  & 38\,$\pm$\,2  &7 & Hybrid \\
 CD\,-38\,4128      & B1\,V         & May 2019                  & 1   & 140             & M     & 21,000\,$\pm$\,1000  & 3.70\,$\pm$\,0.15  & 31\,$\pm$\,2  & 8 & BCEP+EB \\
 CD\,-44\,4596      & B1\,III       & Feb. 2019                 & 1   & 150             & M     & 23,000\,$\pm$\,1000  & 4.00\,$\pm$\,0.10  & 48\,$\pm$\,2  &9 & Hybrid+EB \\
 CD\,-49\,3738      &               & Dec. 2018                 & 1   & 100             & M     & 23,000\,$\pm$\,1000  & 3.75\,$\pm$\,0.10  & 107\,$\pm$\,2 & & Hybrid \\
 HD\,18100          & B5\,II/III    & Nov. 2018                 & 1   & 140             & H     & 27,000\,$\pm$\,1000  & 4.00\,$\pm$\,0.15  & 228\,$\pm$\,6 & 10 & Hybrid \\
 HD\,46149          & O8.5\,V       & Jan. 2018                 & 2   & 75, 110         & H     & 29,000\,$\pm$\,1500  & 4.00\,$\pm$\,0.15  & 97\,$\pm$\,3  &11 & No TESS light curve \\
 HD\,46994          & B2/3\,V         & Sep., Oct. 2019           & 2   & 180, 220        & H     & 21,000\,$\pm$\,1000  & 4.00\,$\pm$\,0.10  & 285\,$\pm$\,8 &12 & SPB \\
 HD\,48553          & B2\,III       & Dec. 2017                 & 1   & 178             & H     & 21,000\,$\pm$\,1000  & 3.80\,$\pm$\,0.20  & 14\,$\pm$\,2 &13 & Hybrid  \\
 HD\,59864          & B1/2\,Ib/II     & Oct. 2018                 & 2   & 165, 220        & H     & 26,000\,$\pm$\,1000  & 3.75\,$\pm$\,0.10  & 18\,$\pm$\,1 &10 & Hybrid  \\
 HD\,61193          & B2\,V         & Apr. 2018                 & 1   & 250             & H     & 23,000\,$\pm$\,1000  & 4.25\,$\pm$\,0.10  & 221\,$\pm$\,6 &14 & Hybrid \\
 HD\,71913          & B1/2\,II      & Dec. 2018, Feb. 2019      & 2   & 230, 170        & H     & 20,000\,$\pm$\,500   & 3.25\,$\pm$\,0.20  & 42\,$\pm$\,2 &10 & BCEP  \\
 HD\,73568          & B2\,III/IV    & Jan. 2018                 & 1   & 100             & H     & 22,000\,$\pm$\,1000  & 3.50\,$\pm$\,0.10  & 86\,$\pm$\,4 &14  & Hybrid \\
 HD\,73918          & B5\,III       & May 2019                  & 1   & 220             & M     & 14,000\,$\pm$\,1000   & 3.70\,$\pm$\,0.10  & 121\,$\pm$\,3 &10 & Non-BCEP \\
 HD\,74339          & B2/3\,II/III    & Feb. 2019                 & 1   & 180             & M     & 21,000\,$\pm$\,1000  & 3.50\,$\pm$\,0.10  & 48\,$\pm$\,2 &14 & Hybrid  \\
 HD\,75290          & B3/5\,V         & Dec. 2017                 & 1   & 150             & H     & 15,000\,$\pm$\,1000  & 4.50\,$\pm$\,0.10  & 297\,$\pm$\,8 &14 & SPB \\
 HD\,86248          & B2\,II        & Jan., June 2019           & 2   & 80,95           & M     & 22,000\,$\pm$\,500   & 3.70\,$\pm$\,0.10  & 41\,$\pm$\,1  &10 & BCEP \\
 HD\,87592          & B1\,IV        & Dec. 2017                 & 2   & 170,140         & H     & 25,000\,$\pm$\,1000  & 3.75\,$\pm$\,0.15  & 18\,$\pm$\,2 &3  & Hybrid \\
 HD\,90288          & B2\,III/IV        & Jan. 2018                 & 1   & 130             & H     & 25,000\,$\pm$\,1000  & 4.25\,$\pm$\,0.10  & 143\,$\pm$\,5 &3 & Hybrid \\
 HD\,92291          & B1\,Ib/II     & Feb. 2018                 & 1   & 150             & H     & 23,000\,$\pm$\,1000  & 3.60\,$\pm$\,0.10  & 31\,$\pm$\,3 &3 & BCEP  \\
 HD\,94065          & B0/1\,III/IV  & Feb. 2018                 & 1   & 180             & H     & 23,000\,$\pm$\,1000  & 3.50\,$\pm$\,0.10  & 37\,$\pm$\,3 &3  & Hybrid  \\
 HD\,94345          & B0/1\,IV      & Jan., Apr. 2018           & 2   & 100,170         & H     & 22,000\,$\pm$\,1000  & 3.25\,$\pm$\,0.10  & 42\,$\pm$\,3 &3  & Hybrid  \\
 HD\,96882          & B1\,II/III    & June 2019                 & 1   & 160             & H     & 24,000\,$\pm$\,1000  & 3.75\,$\pm$\,0.10  & 92\,$\pm$\,3 &3  & Hybrid \\
 HD\,96901          & B1\,III/IV    & Feb. 2018                 & 1   & 160             & H     & 24,000\,$\pm$\,500   & 3.75\,$\pm$\,0.10  & 75\,$\pm$\,3 &3  & Hybrid \\
 HD\,96927          & B3\,III       & Jan., June 2019           & 2   & 140,150         & H     & 23,000\,$\pm$\,500   & 3.70\,$\pm$\,0.10  & 63\,$\pm$\,2 &14 & BCEP  \\
 HD\,101838         & B0.5/1\,III   & Jan., May 2019            & 2   & 100,210         & M     & 23,000\,$\pm$\,1000  & 3.30\,$\pm$\,0.15  & 183\,$\pm$\,2 &3 & BCEP+EB  \\
 HD\,102505         & B1\,IV        & Apr. 2018                 & 2   & 160,160         & H     & 26,000\,$\pm$\,1000  & 3.75\,$\pm$\,0.10  & 91\,$\pm$\,4 &3  & Hybrid \\
 HD\,106345         & B2\,III       & Apr. 2019                 & 1   & 150             & H     & 22,000\,$\pm$\,500   & 3.75\,$\pm$\,0.10  & 21\,$\pm$\,1 &3 & Hybrid  \\
 HD\,112481         & B2\,II/III    & Feb., July 2019           & 2   & 80,175          & H     & 24,000\,$\pm$\,500   & 3.50\,$\pm$\,0.10   & 42\,$\pm$\,1 &14 & Hybrid  \\
 HD\,119910         & B2\,III/IV    & June, July 2019           & 2   & 170,200         & H     & 21,000\,$\pm$\,500   & 3.75\,$\pm$\,0.10  & 18\,$\pm$\,2 & 3  & Hybrid  \\
 HD\,126357         & B1\,III       & Feb. 2018                 & 1   & 160             & H     & 25,000\,$\pm$\,1000  & 3.75\,$\pm$\,0.10  & 71\,$\pm$\,3 & 3  & Hybrid \\
 HD\,129929         & B2            & Feb., June 2019           & 2   & 90,140          & H     & 25,000\,$\pm$\,1000  & 3.90\,$\pm$\,0.10  & 19\,$\pm$\,1 & 10  & Hybrid  \\
 HD\,131805         & B5\,V         & June, July 2019           & 2   & 60,120          & H     & 17,000\,$\pm$\,1000  & 4.00\,$\pm$\,0.20  & 270\,$\pm$\,7 &3 & SPB \\
 HD\,140543         & B1\,Iab/b     & Mar., June 2019           & 2   & 130,150         & H     & 28,000\,$\pm$\,1000  & 3.30\,$\pm$\,0.20  & 324\,$\pm$\,10 &12 & Hybrid \\
 HD\,145794         & B2\,II/III    & Apr. 2018                 & 1   & 250             & H     & 17,000\,$\pm$\,1000  & 3.50\,$\pm$\,0.10  & 148\,$\pm$\,5 &14 & BCEP \\
 HD\,146442         & B2\,III       & May, June 2019            & 2   & 120,130         & M     & 22,000\,$\pm$\,500   & 3.70\,$\pm$\,0.10  & 44\,$\pm$\,1 &14  & BCEP \\
 HD\,147985         & B1/2\,II/III  & Feb., Apr. 2019           & 2   & 110,190         & H     & 22,000\,$\pm$\,1000  & 4.00\,$\pm$\,0.10  & 126\,$\pm$\,4 &14 & Hybrid \\
 HD\,149100         & B2\,III       & May 2018                  & 1   & 210             & H     & 17,000\,$\pm$\,1000  & 3.90\,$\pm$\,0.10  & 127\,$\pm$\,5 &14  & Hybrid\\
 HD\,151158         & B2\,Ib/II     & Sep. 2019                 & 2   & 175             & H     & 21,500\,$\pm$\,2000    & 3.60\,$\pm$\,0.20  & 30\,$\pm$\,2 &14  & Non-BCEP \\
 HD\,151654         & F0\,V         & June 2019                 & 1   & 210             & H     & 13,000\,$\pm$\,1000  & 4.20\,$\pm$\,0.10  & 46\,$\pm$\,3 &13 & No TESS light curve \\
 HD\,152162         & B2\,III       & Mar., July 2019           & 2   & 125,150         & H     & 21,000\,$\pm$\,500   & 3.75\,$\pm$\,0.10  & 18\,$\pm$\,2 &14 & Hybrid  \\
 HD\,152372         & B2\,III/IV    & Mar., June 2019           & 2   & 130,150         & H     & 23,000\,$\pm$\,500   & 3.90\,$\pm$\,0.10  & 19\,$\pm$\,1 &14  & Hybrid  \\
 HD\,153772         & B2\,V         & June, July 2019           & 2   & 90,190          & H     & 21,000\,$\pm$\,1000  & 4.10\,$\pm$\,0.10  & 72\,$\pm$\,1 & 14  & BCEP \\
 HD\,156172         & O9\,V         & July, Sep. 2019           & 2   & 110,180         & H     & 30,000\,$\pm$\,3,000  & 4.00\,$\pm$\,0.10  & 280\,$\pm$\,11 & 4 & BCEP\\
 HD\,156321         & B2\,V         & Sep. 2019                 & 2   & 20,100          & H     & 25,000\,$\pm$\,500   & 4.00\,$\pm$\,0.10  & 123\,$\pm$\,3 & 6 & Hybrid \\
 HD\,156662         & B2\,III       & Feb. 2018                 & 1   & 165             & H     & 22,000\,$\pm$\,1000  & 3.50\,$\pm$\,0.10  & 109\,$\pm$\,4 & 14 & BCEP \\
 HD\,160233         & B1\,II        & Apr. 2018                 & 1   & 130             & H     & 24,000\,$\pm$\,1000  & 3.90\,$\pm$\,0.10  & 135\,$\pm$\,5 &13 & No TESS light curve \\
 HD\,166540         & B1\,Ib        & Apr. 2018                 & 2   & 90              & H     & 29,000\,$\pm$\,1000  & 3.90\,$\pm$\,0.15  & 35\,$\pm$\,3 & 12 & Hybrid  \\
 HD\,167451         & B1\,Ib        & Mar., May 2019            & 2   & 120,120         & H     & 28,000\,$\pm$\,1500  & 3.25\,$\pm$\,0.15  & 69\,$\pm$\,2 & 12  & BCEP \\
 HD\,167743         & B2\,Ib        & Apr., June 2019           & 2   & 100,140         & M     & 23,000\,$\pm$\,1000  & 3.75\,$\pm$\,0.10  & 58\,$\pm$\,1 &12  & BCEP \\
 HD\,168050         & B3/5\,Ib      & Mar., April 2019          & 2   & 40,80           & M     & 23,000\,$\pm$\,1500  & 4.00\,$\pm$\,0.20  & 62\,$\pm$\,3 &12  & BCEP+EB  \\
 HD\,171305         & B2\,II        & June, Aug. 2019           & 2   & 90,100          & H     & 21,000\,$\pm$\,500   & 3.50\,$\pm$\,0.10  & 16\,$\pm$\,1  &13  & BCEP \\
 HD\,178987         & B2\,II        & Mar., Apr. 2019           & 2   & 100,140         & M     & 21,000\,$\pm$\,1000  & 3.50\,$\pm$\,0.10  & 135\,$\pm$\,2  &14 & BCEP \\
 HD\,186610         & B2\,Ib/II     & Mar., May 2019            & 2   & 80,130          & M     & 20,000\,$\pm$\,500   & 3.25\,$\pm$\,0.10  & 37\,$\pm$\,1 &13  & BCEP \\
 HD\,187536         & B3\,II            & Aug. 2022                 & 3   & 110, 85, 155    & L     & 24,000\,$\pm$\,1000  & 3.50\,$\pm$\,0.10  & 173\,$\pm$\,4 &10 & Hybrid \\
 HD\,357657         &  B             & July, Aug. 2022           & 2   & 90, 105         & L     & 24,000\,$\pm$\,1000  & 4.00\,$\pm$\,0.10   & 38\,$\pm$\,2 &6  & Hybrid \\
\enddata
\tablenotetext{a}{H, M and L stand for the high, medium and low resolution setups, respectively, as described in Section \ref{sec:spectra}.}
\tablenotetext{b}{Spectral type references:
(1) \citet{1970ApJ...161..855S};
(2) \citet{2003AJ....125.2531R};
(3) \citet{1975mcts.book.....H};
(4) \citet{1977ApJS...35..111G};
(5) \citet{1983PASP...95..361F};
(6) \citet{SIMBAD};
(7) \citet{2013MNRAS.431..240O};
(8) \citet{1975AJ.....80..212H};
(9) \citet{1952BOTT....1b...1M};
(10) \citet{1982mcts.book.....H};
(11) \citet{2011ApJS..193...24S};
(12) \citet{1988mcts.book.....H};
(13) \citet{1999MSS...C05....0H};
(14) \citet{1978mcts.book.....H}.}
\tablenotetext{c}{The classifications assigned in this work.}
\end{deluxetable*}
\endgroup

\onecolumngrid
\section{Visual Classification} \label{sec:vis_class_app}
\begin{figure*}
    \centering
    \includegraphics[width=1\linewidth]{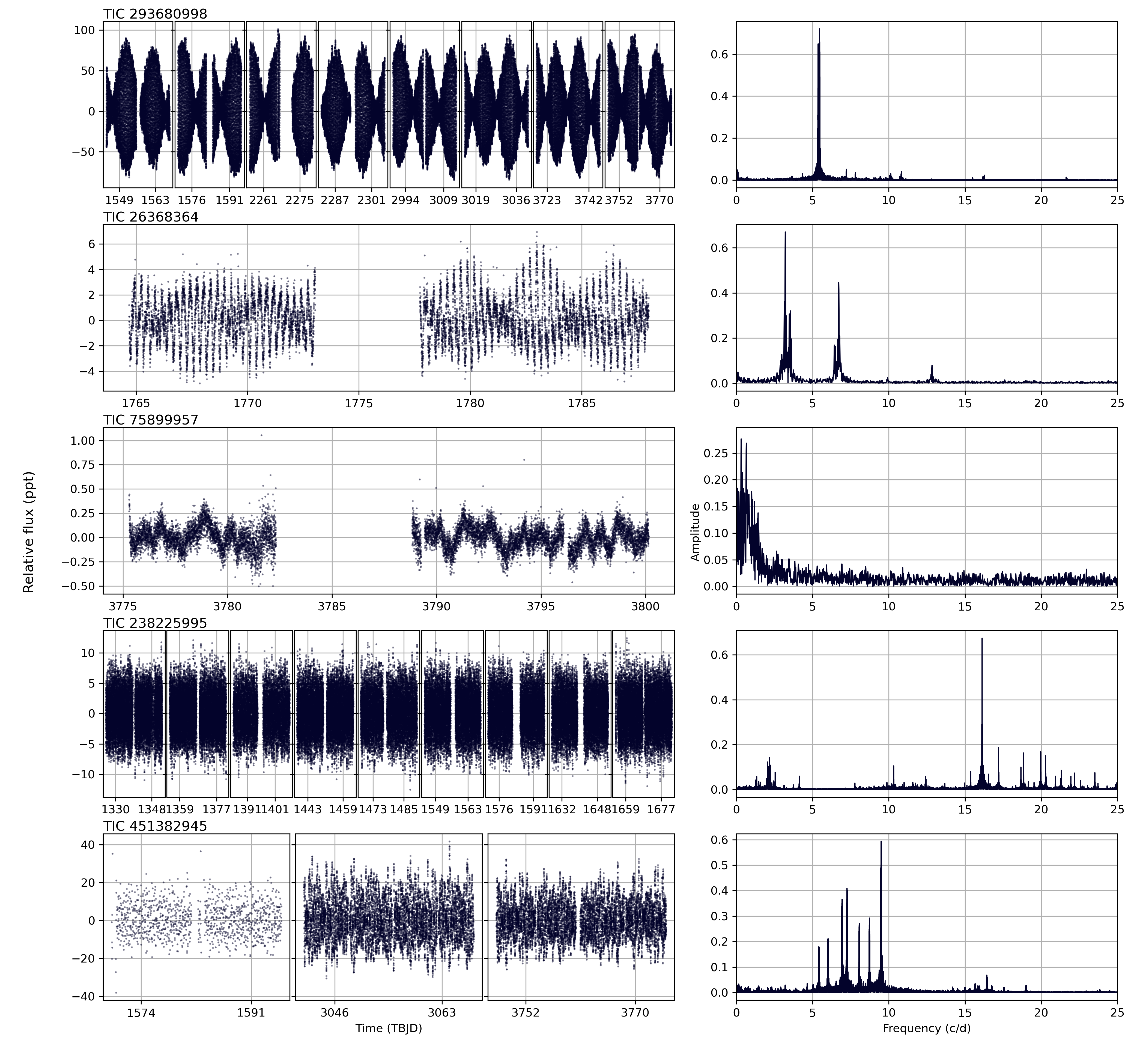}
    \caption{Examples of the four types of impostors described in Section~\ref{sec:visual_inspection}, with a \bcep\ pulsator shown in the top panel for comparison. The left panels show the TESS light curves and the right panels show their corresponding Lomb-Scargle periodograms. The first panel (from the top) is a potential \bcep\ star that was found to be an SPB star with rotation that shifts the ``real'' SPB pulsation frequency to higher frequencies through harmonics. The second panel is for a GDOR star, the third displays a light curve showing GDOR + DSCT pulsations, and the fourth is for a star displaying DSCT pulsations.}
    \label{fig:impostors}
\end{figure*}

\begin{figure*}
    \centering
    \includegraphics[width=1\linewidth]{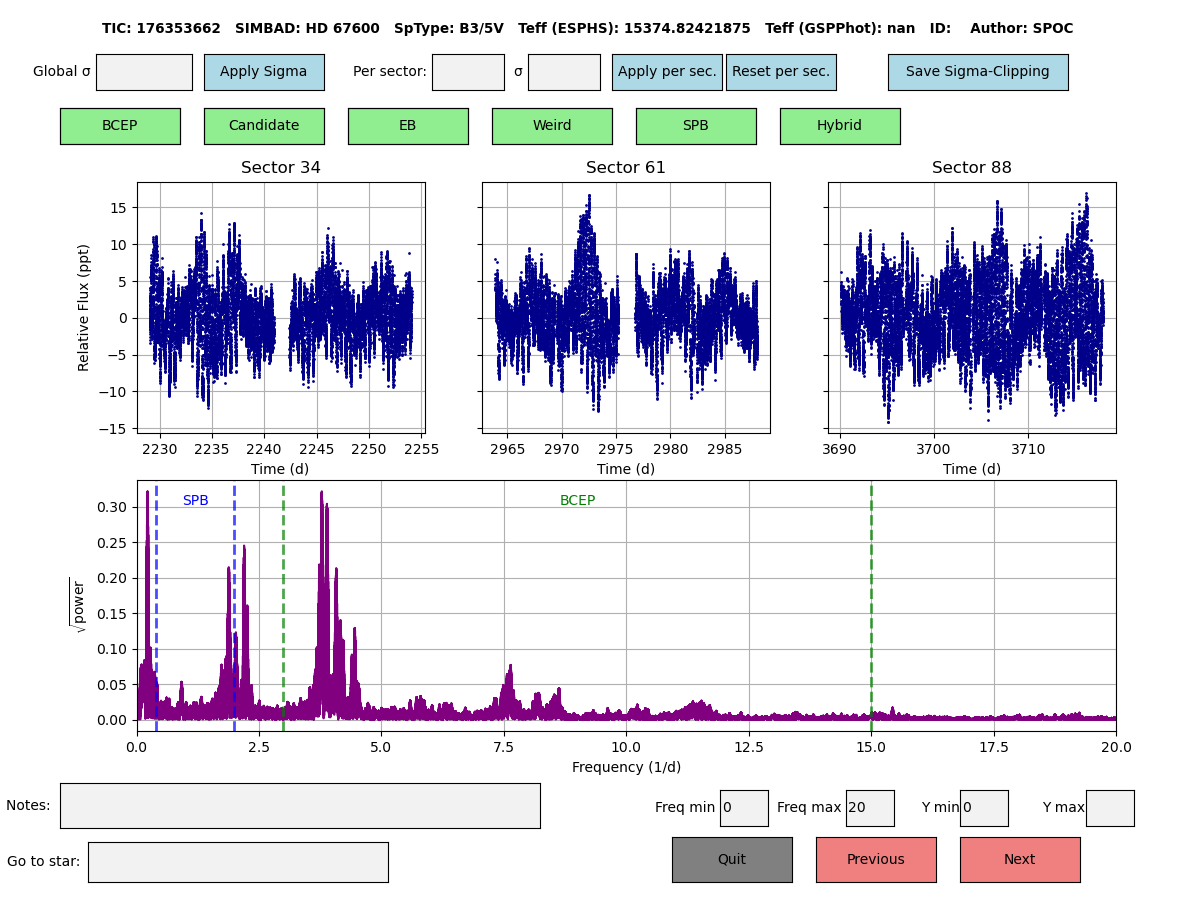}
    \caption{A screenshot of the visual-classification GUI developed for this work. The top annotations display the stellar metadata (TIC, SIMBAD ID, spectral type, $T_{\mathrm{eff}}$ from the ESP-HS and GSP-Phot pipelines and the TESS pipeline from which the light curve is obtained). Below these annotations are the sigma-clipping controls, which allow either global sigma clipping of the full light curve or sector-specific clipping, with the option to save the processed data. Below these controls are the classification buttons. The top panel of plots shows the TESS light curve, divided into sector-by-sector subplots. The bottom panel displays the Lomb–Scargle periodogram together with the \bcep\ and SPB pulsation frequency ranges. At the bottom of the interface, a text box is provided for notes, along with controls for adjusting the x- and y-axis limits, and for loading a specific star.}
    \label{fig:visual_inspection_HD67600}
\end{figure*}

Figure~\ref{fig:impostors} shows examples of possible \bcep\ impostor types encountered during the classification process described in Section~\ref{sec:visual_inspection}. In Figure~\ref{fig:visual_inspection_HD67600} we present a screenshot of the GUI used for the visual classification described in Section~\ref{sec:visual_inspection}.

\section{Description of the Catalog} \label{sec:catalog_description}
The machine-readable catalog is provided electronically. Table~\ref{tab:catalog_columns} describes the catalog columns, and Table~\ref{tab:catalog} presents the first nine columns and first 45 rows of the catalog. The catalog contains one row per star. When both SPOC and QLP light curves are available, we retain QLP if \texttt{QLP\_OVER\_SPOC}~$=1$ and otherwise we retain SPOC. Empty entries indicate unavailable data or no derived parameters.

\begin{table*}
\centering
\small
\caption{Description of the columns in the machine-readable catalog.
\label{tab:catalog_columns}}
\begin{tabular}{p{0.48\textwidth}p{0.42\textwidth}}
\hline
\hline
\textbf{Columns} & \textbf{Description} \\
\hline
{\raggedright\texttt{target\_id}, \texttt{TIC}, \texttt{TIC\_sep}, \texttt{source\_id}, \texttt{Gaia\_sep}, \texttt{designation}, \texttt{ra}, \texttt{dec}, \texttt{RA\_Simbad}, \texttt{Dec\_Simbad}, \texttt{TIC\_MULT\_MATCH}, \texttt{Gaia\_MULT\_MATCH}, \texttt{number\_of\_gaia\_candidates}, \texttt{solution\_id}\par}
&
Identifiers, coordinates, cross-match separations (arcsec), multiple-match flags, number of Gaia candidates, and Gaia DR3 solution identifier. \\
\hline
{\raggedright\texttt{Classification}, \texttt{SpType\_SIMBAD}, \texttt{Vmag\_SIMBAD}, \texttt{vsini\_SIMBAD}, \texttt{SigmaClipping\_classification}, \texttt{SigmaValue\_classification}, \texttt{Notes\_classification}, \texttt{cluster?}\par}
&
Final classification; SIMBAD spectral type, $V$ magnitude, and $v\sin i$; visual-classification clipping information and notes; and cluster-membership flag. \texttt{SigmaClipping\_classification} indicates whether sigma clipping was applied, and \texttt{SigmaValue\_classification} gives either the global clipping value or sector-specific values in the form \texttt{S<sector>=<value>}. \\
\hline
{\raggedright\texttt{Classification\_history\_literature}, \texttt{SIMBAD\_Classification}, \texttt{Notes}, \texttt{Ref.}, \texttt{Year}, \texttt{Link}, \texttt{Year\_earliest}, \texttt{Year\_history}, \texttt{Ref\_earliest}, \texttt{Ref\_history}\par}
&
Literature classifications, references, publication years, links, and notes. \\
\hline
{\raggedright\texttt{TESS\_lc}, \texttt{TESS\_pipeline}, \texttt{TESS\_sectors}, \texttt{TESS\_exp\_time}, \texttt{QLP\_OVER\_SPOC}\par}
&
\texttt{TESS\_lc} indicates whether a TESS light curve is available (1) or unavailable (0); adopted pipeline, analyzed sectors, exposure time (s), and flag indicating that QLP was adopted over SPOC. \\
\hline
{\raggedright\texttt{ROT}, \texttt{ELL}, \texttt{EB}, \texttt{HEARTBEAT}, \texttt{ONLY\_SLF}, \texttt{SLF}, \texttt{DSCT}, \texttt{GDOR}, \texttt{CONST}, \texttt{QLP\_BAD}, \texttt{SAT}, \texttt{SPEC\_WRONG}, \texttt{BLEND}, \texttt{NEARBY\_STAR}, \texttt{LOW\_FREQ\_HARMONICS}, \texttt{MANY\_FREQS}, \texttt{HIGH\_FREQS}, \texttt{COMB}\par}
&
Visual-classification and data-quality flags defined in Section~\ref{sec:visual_inspection}. A value of 1 denotes a secure identification and 0.5 a tentative identification. \\
\hline
{\raggedright\texttt{parallax}, \texttt{parallax\_error}, \texttt{parallax\_over\_error}, \texttt{pm}, \texttt{pmra}, \texttt{pmdec}, \texttt{ruwe}, \texttt{phot\_*\_mean\_mag}, \texttt{phot\_*\_mean\_flux\_over\_error}, \texttt{bp\_rp}, \texttt{bp\_g}, \texttt{g\_rp}, \texttt{radial\_velocity}, \texttt{radial\_velocity\_error}\par}
&
Gaia DR3 astrometric, photometric, color, and radial-velocity parameters used in this work. Here \texttt{phot\_*\_mean\_mag} and \texttt{phot\_*\_mean\_flux\_over\_error} refer to the $G$, $BP$, and $RP$ bands. \\
\hline
{\raggedright\texttt{teff\_gspphot}, \texttt{logg\_gspphot}, \texttt{mh\_gspphot}, and associated \texttt{\_lower} and \texttt{\_upper} columns; \texttt{teff\_esphs}, \texttt{logg\_esphs}, \texttt{ag\_esphs}, \texttt{ebpminrp\_esphs}, and associated \texttt{\_uncertainty} columns\par}
&
Gaia DR3 GSP-Phot and ESP-HS atmospheric parameters, reported bounds, and uncertainties. \\
\hline
{\raggedright\texttt{bayestar\_converged}, \texttt{bayestar\_reliable\_dist}, \texttt{A\_G\_bayestar}, \texttt{M\_G\_bayestar}, \texttt{M\_G\_bayestar\_err}, \texttt{M\_BP\_bayestar}, \texttt{M\_RP\_bayestar}, \texttt{M\_G\_esphs}, \texttt{M\_G\_esphs\_err}, \texttt{log\_L\_divide\_Lsun\_*}\par}
&
Bayestar19 flags and extinction, extinction-corrected absolute magnitudes, and luminosities used for the HR diagrams and luminosity-method comparisons. The \texttt{\_solar} luminosities use solar-metallicity bolometric corrections; the corresponding unsuffixed weighted-mean luminosities use the Gaia GSP-Phot metallicity. \\
\hline
{\raggedright\texttt{p\_orb\_*}, \texttt{depth\_*\_measured}, \texttt{depth\_*\_final\_model}, \texttt{t\_*\_measured}, \texttt{t\_*\_final\_model}, \texttt{e}, \texttt{w}, \texttt{i}, \texttt{r\_sum}, \texttt{r\_rat}, \texttt{sb\_rat}, \texttt{ratio1\_1--ratio4\_3}, \texttt{EB\_cand\_flag}, \texttt{EB\_flag}, \texttt{ROTMOD\_flag}, \texttt{EB\_score}\par}
&
STAR SHADOW orbital, eclipse, EB-model, and EB/rotational-modulation parameters.\\
\hline
{\raggedright\texttt{f1--f30}, \texttt{a1--a30}, \texttt{ph1--ph30}, \texttt{f1\_final--f30\_final}, \texttt{a1\_final--a30\_final}, \texttt{ph1\_final--ph30\_final}\par}
&
First and final STAR SHADOW frequency, amplitude, and phase sets. \\
\hline
\end{tabular}
\end{table*}

\begin{deluxetable*}{lcccccccc}
\tablecolumns{9}
\tabletypesize{\scriptsize}
\rotate
\tablecaption{Excerpt of the catalog showing the first 45 rows and first 9 columns. \label{tab:catalog}}
\tablehead{\colhead{{target\_id}} &
\colhead{{TIC}} &
\colhead{{TIC\_sep}} &
\colhead{{source\_id}} &
\colhead{{Gaia\_sep}} &
\colhead{{designation}} &
\colhead{{ra}} &
\colhead{{dec}} &
\colhead{{Classification}}
}
\startdata
HD~217035 & 13212836.0 & 0.0297427741602749 & \texttt{2207297149288415232} & 0.1928143579219829 & Gaia DR3 2207297149288415232 & 344.1289404042378 & 62.868366731474325 & Weird \\
TIC 329992262 & 329992262.0 & 0.0795664420220739 & \texttt{2198645573310089600} & 0.079566442025338 & Gaia DR3 2198645573310089600 & 333.14160464395064 & 57.25803726746035 & BCEP \\
* 23 Tau & 61139113.0 & 0.8336416088740166 & \texttt{65205373152172032} & 0.8023603560181609 & Gaia DR3 65205373152172032 & 56.58167144917715 & 23.94814381017594 & SPB \\
NGC 6231 239 & 339567650.0 & 0.1986717937769871 & \texttt{5966508992127646464} & 0.1976138585929844 & Gaia DR3 5966508992127646464 & 253.5650733309733 & -41.86926956956006 & Weird \\
V* V1205 Sco & 339567743.0 & 0.0361442906564723 & \texttt{5966508987825250048} & 0.036911038270429 & Gaia DR3 5966508987825250048 & 253.56556984806772 & -41.86108233261907 & Weird \\
* gam Peg & 51942308.0 & 0.1466761237982191 & \texttt{2769202742740232960} & 0.1715451775218926 & Gaia DR3 2769202742740232960 & 3.3089703843753013 & 15.183550741446307 & Hybrid \\
V* V909 Cas & 390122587.0 & 0.0256695636960721 & \texttt{510001703986507520} & 0.0259140491112362 & Gaia DR3 510001703986507520 & 24.160629451723917 & 61.43163626126622 & BCEP \\
V* V1155 Cas & 399436828.0 & 0.0167669529918569 & \texttt{511217592043001216} & 0.0181879021990801 & Gaia DR3 511217592043001216 & 26.662482725248275 & 61.23501371190087 & BCEP \\
BD+56 501 & 264613043.0 & 0.0266511533340822 & \texttt{458372795471717888} & 0.0232339562085865 & Gaia DR3 458372795471717888 & 34.62430147708934 & 57.15087308127002 & Hybrid \\
BD+56 508 & 264731122.0 & 0.0251906605466908 & \texttt{458379701779206400} & 0.0221381372097867 & Gaia DR3 458379701779206400 & 34.70007084831595 & 57.28551863787359 & BCEP \\
BD+56 520 & 264730502.0 & 0.0128627568456106 & \texttt{458377743274230784} & 0.0180362613550536 & Gaia DR3 458377743274230784 & 34.750863931250336 & 57.14563958733677 & \nodata \\
V* V619 Per & 348607672.0 & 0.0210737028890237 & \texttt{458454606008606336} & 0.0206041998312068 & Gaia DR3 458454606008606336 & 35.51159515649917 & 57.14027850945556 & BCEP \\
V* V595 Per & 348607604.0 & 0.0241940445549815 & \texttt{458407670603899648} & 0.0282639679739983 & Gaia DR3 458407670603899648 & 35.53570352714599 & 57.12454836175567 & Weird \\
* del Cet & 328161938.0 & 0.1957591157512272 & \texttt{2498783458525988992} & 0.2142599834166615 & Gaia DR3 2498783458525988992 & 39.87071163754866 & 0.3284966543857036 & Hybrid \\
V* KP Per & 65166720.0 & 0.0981306767959816 & \texttt{241678947918118912} & 0.0972575391803758 & Gaia DR3 241678947918118912 & 53.16241159653567 & 44.85573265544764 & BCEP \\
* eps Per & 187107884.0 & 0.4364421852004821 & \texttt{229898024422755840} & 0.463694421894351 & Gaia DR3 229898024422755840 & 59.463595170524904 & 40.0100950978535 & BCEP \\
* nu. Eri & 229087301.0 & 0.0888652771431315 & \texttt{3204884534485995264} & 0.0766973876220919 & Gaia DR3 3204884534485995264 & 69.07976184342182 & -3.3524812188642006 & Hybrid \\
TIC 4254645 & 4254645.0 & 0.0508950229249243 & \texttt{3214077516844988800} & 0.0508950229256123 & Gaia DR3 3214077516844988800 & 81.11923705133687 & -2.3971524975189644 & EB \\
* psi02 Ori & 264540478.0 & 0.0226811299659237 & \texttt{3235897703054775936} & 0.0233310322647902 & Gaia DR3 3235897703054775936 & 81.70929343540975 & 3.0956682931066135 & EB \\
* bet CMa & 34590771.0 & 0.0646075947169533 & \nodata & \nodata & \nodata & \nodata & \nodata & BCEP \\
* ksi01 CMa & 47763235.0 & 0.0976425604404019 & \texttt{2924859847973986048} & 0.1107977994277122 & Gaia DR3 2924859847973986048 & 97.9640099410608 & -23.418394841586725 & BCEP \\
* 15 CMa & 78897024.0 & 0.1519963835896338 & \texttt{2932263546963330176} & 0.1180243757102372 & Gaia DR3 2932263546963330176 & 103.38708452690732 & -20.224232345171853 & BCEP \\
* 19 Mon & 269087549.0 & 0.0722039668161381 & \texttt{3101768554465447936} & 0.0728462047908323 & Gaia DR3 3101768554465447936 & 105.72822063907358 & -4.239222040833818 & Hybrid \\
* 27 CMa & 65803653.0 & 0.1073101367779548 & \texttt{5610288149803056896} & 0.131030112807695 & Gaia DR3 5610288149803056896 & 108.5633486897137 & -26.35248705694927 & Hybrid \\
V* V350 Pup & 112923257.0 & 0.0969961120344361 & \texttt{5591456230038742016} & 0.0947352858055188 & Gaia DR3 5591456230038742016 & 112.64195532159609 & -34.09057544918967 & Hybrid \\
V* PT Pup & 349043273.0 & 0.1278088094289274 & \texttt{5716409335624704768} & 0.1190990386233505 & Gaia DR3 5716409335624704768 & 114.17094973342256 & -19.702323468770643 & BCEP \\
V* QU Pup & 143488434.0 & 0.1538874939286954 & \texttt{5532548142233447424} & 0.158404749691211 & Gaia DR3 5532548142233447424 & 117.91812003760556 & -42.88816971673171 & BCEP \\
V* V372 Car & 340985103.0 & 0.1564823229392656 & \texttt{5488339734612857984} & 0.1555052155381935 & Gaia DR3 5488339734612857984 & 118.12388808016225 & -54.367125332398615 & Hybrid \\
HD~71913 & 56630835.0 & 0.1058910789731094 & \texttt{5543515083529532928} & 0.1043260320653663 & Gaia DR3 5543515083529532928 & 127.17916162880256 & -34.73159605638968 & BCEP \\
V* KK Vel & 31921921.0 & 0.141989633579199 & \texttt{5330841772046764928} & 0.1382864181840643 & Gaia DR3 5330841772046764928 & 136.92711698409929 & -44.63242192642359 & Hybrid \\
V* IL Vel & 293680998.0 & 0.0850549785229819 & \texttt{5311655328520521600} & 0.0843976176997563 & Gaia DR3 5311655328520521600 & 139.37976450535623 & -52.83873184173844 & BCEP \\
HD~90288 & 464296873.0 & 0.1466273724611769 & \texttt{5351762420316030464} & 0.1503278132543991 & Gaia DR3 5351762420316030464 & 155.9870673697272 & -57.46454346979854 & Hybrid \\
V* V400 Car & 457975914.0 & 0.1331817719555949 & \texttt{5351451567771333888} & 0.1315068373452932 & Gaia DR3 5351451567771333888 & 158.70278662979692 & -58.150312241218096 & BCEP \\
V* V401 Car & 457975502.0 & 0.1346907660212208 & \texttt{5351448127473759872} & 0.1340768397027906 & Gaia DR3 5351448127473759872 & 158.87515622648337 & -58.20225711685458 & BCEP \\
V* V403 Car & 458076392.0 & 0.1318354701154205 & \texttt{5351448784632614656} & 0.1328591180272502 & Gaia DR3 5351448784632614656 & 158.91953343554542 & -58.21228556917196 & BCEP \\
CD-57 3347 & 458076501.0 & 0.1342071368788862 & \texttt{5351447238444368768} & 0.135016779321505 & Gaia DR3 5351447238444368768 & 158.93034567477468 & -58.22599652116463 & BCEP \\
MCW 1182 & 458076594.0 & 0.1331674702439058 & \texttt{5351447234127452672} & 0.133338364404774 & Gaia DR3 5351447234127452672 & 158.95098085669855 & -58.23792338618754 & BCEP \\
V* V405 Car & 458076374.0 & 0.1332548972972942 & \texttt{5351448750272913664} & 0.1339167444294793 & Gaia DR3 5351448750272913664 & 158.95082755114157 & -58.20920818265427 & BCEP \\
CPD-57 3517 & 458076652.0 & 0.1323882927379847 & \texttt{5351447032285937664} & 0.1339501522646328 & Gaia DR3 5351447032285937664 & 158.97343952899308 & -58.24660999240357 & BCEP \\
V* V440 Car & 458076425.0 & 0.1370882697304282 & \texttt{5351448544114492672} & 0.1345569093731577 & Gaia DR3 5351448544114492672 & 158.97875274657133 & -58.21641354253266 & \nodata \\
V* V406 Car & 458076353.0 & 0.1344753120534328 & \texttt{5351448647193741440} & 0.1354210456896976 & Gaia DR3 5351448647193741440 & 158.99081222124084 & -58.2059036528582 & Hybrid \\
HD~92007 & 458076693.0 & 0.131191910934169 & \texttt{5351448441035539328} & 0.1336269101352186 & Gaia DR3 5351448441035539328 & 159.0065737483128 & -58.25264812418845 & Hybrid \\
TIC 458076434 & 458076434.0 & 0.1280109370613358 & \texttt{5350697989986480896} & 0.1280109370670217 & Gaia DR3 5350697989986480896 & 159.03466004877737 & -58.21786200091148 & BCEP+EB \\
V* KZ Mus & 327656776.0 & 0.1099289858190243 & \texttt{5842913572934429696} & 0.1101050132517212 & Gaia DR3 5842913572934429696 & 189.82975577557767 & -71.62179009925686 & BCEP \\
* bet Cru & 405567821.0 & 0.170459322129786 & \nodata & \nodata & \nodata & \nodata & \nodata & BCEP \\
\enddata
\tablecomments{A subset of the columns and rows of the catalog table is shown here. The full table is available in machine-readable format. All columns are described in Table~\ref{tab:catalog_columns}.}
\end{deluxetable*}
\onecolumngrid

\section{Stellar Parameters Statistics} \label{sec:stellar_star_app}
Here we present the full statistical analysis and the resulting values of the statistical tests we perform on the stellar- and pulsation-parameter distributions presented in Section~\ref{sec:stellar_pulsation_stat}. 

We calculate the KDEs using the \texttt{gaussian\_kde} implementation in SciPy \citep{2020SciPy-NMeth} and construct the empirical cumulative distribution functions (ECDFs) directly from the ordered parameter values.
To compare the BCEP and hybrid distributions quantitatively, we apply two-sample Kolmogorov-Smirnov (KS) tests \citep{an1933sulla,smirnov1939estimation,1958ArM.....3..469H} using \texttt{ks\_2samp} from SciPy \citep{2020SciPy-NMeth} and report the raw $p$-value for each parameter. We additionally report $p$-values adjusted using the Benjamini-Hochberg procedure \citep{Benjamini_Hochberg1995}, which is described below.

\begin{figure*}
    \centering
    \includegraphics[width=0.6\textwidth]{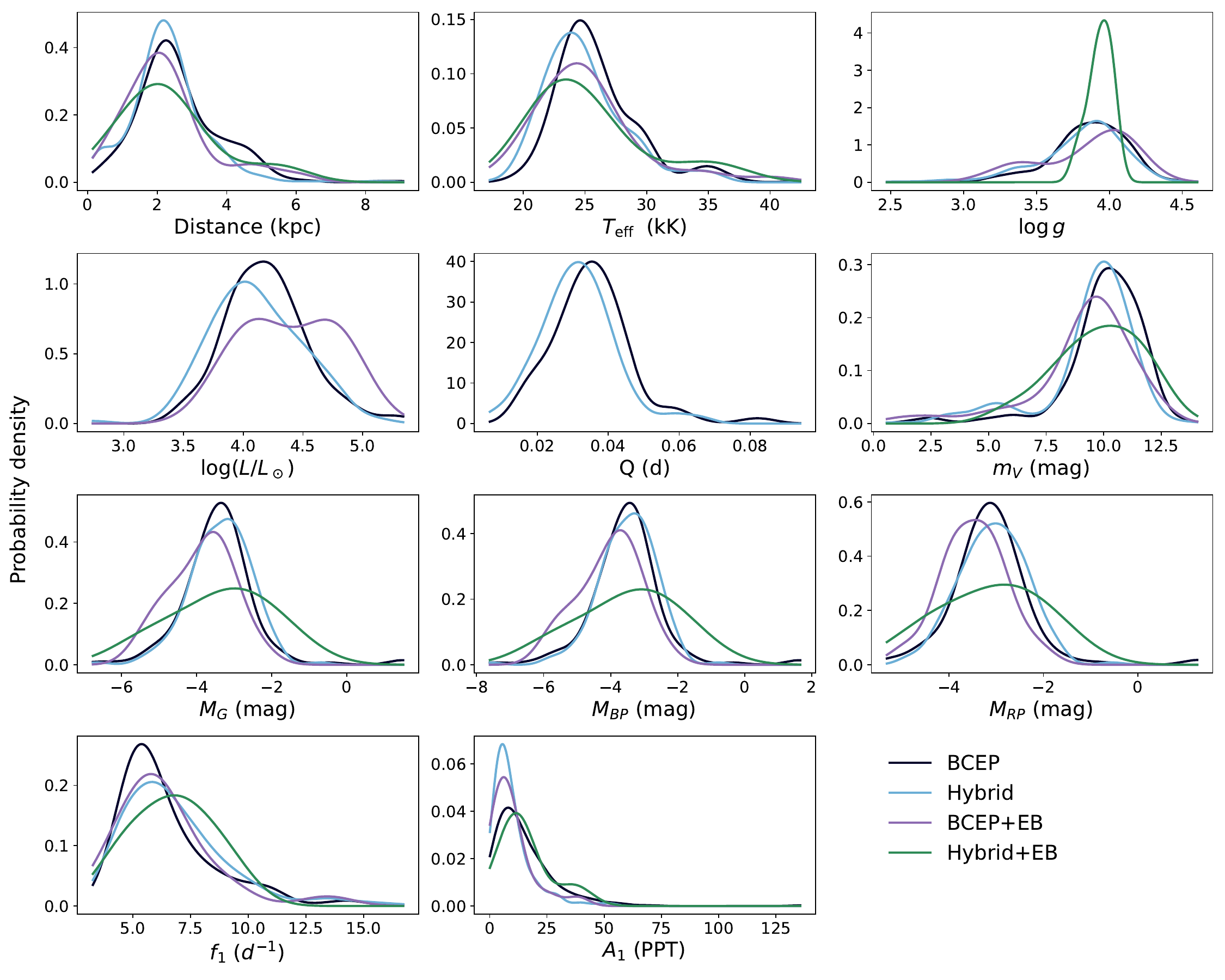}
    \caption{Kernel density estimates of the stellar and pulsation parameter distributions for the different variability classifications.}
    \label{fig:stellar_stat_kernel}
\end{figure*}

\begin{figure*}
    \centering
    \includegraphics[width=0.6\textwidth]{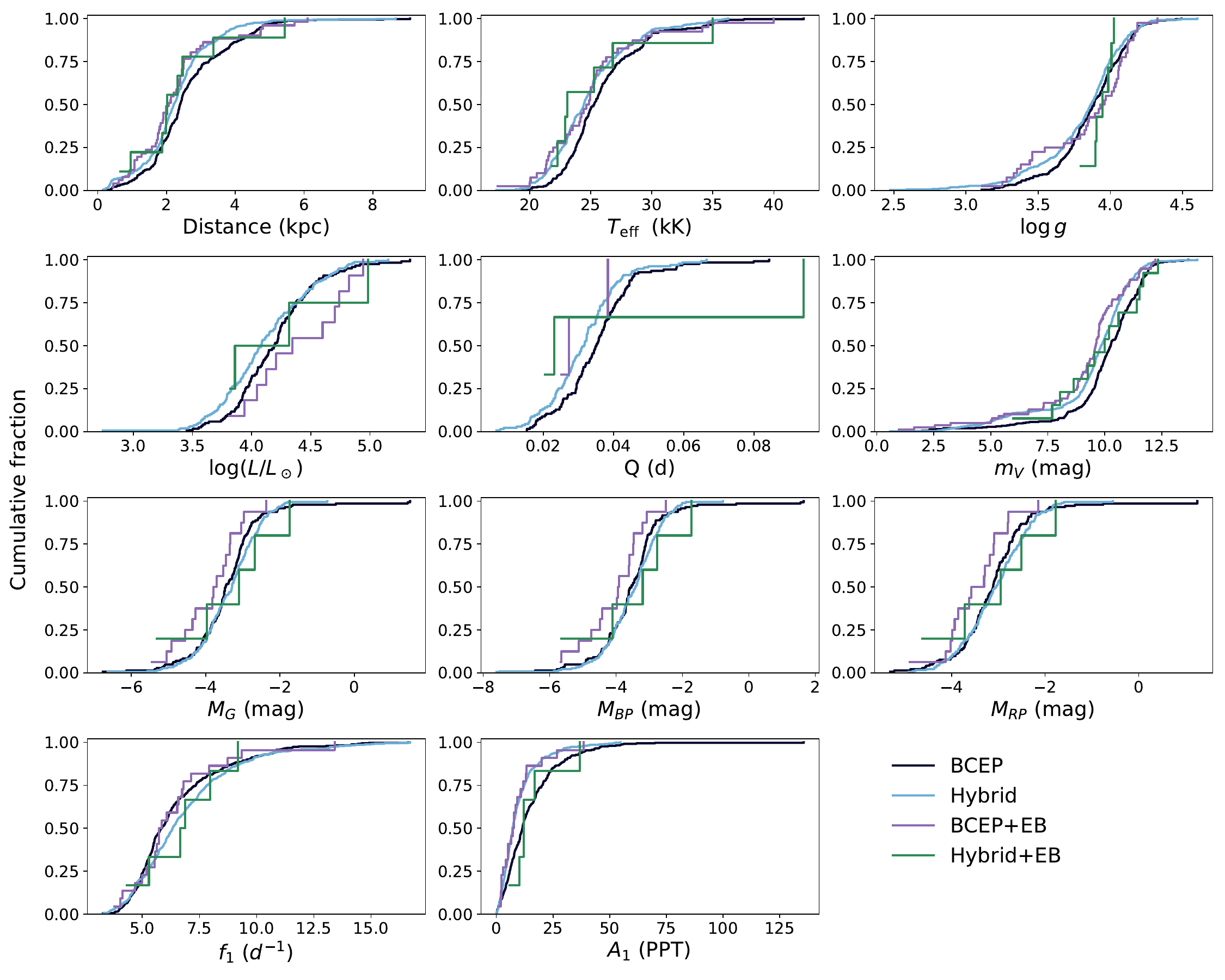}
    \caption{Empirical cumulative distribution functions of the stellar and pulsation parameters for the different variability classifications.}
    \label{fig:stellar_stat_cumulative}
\end{figure*}

We fit one- and two-component Gaussian-mixture models using the \texttt{GaussianMixture} implementation in scikit-learn \citep{scikit-learn} and compare the models using the BIC. A preference for two components may indicate internal structure, but does not by itself establish bimodality because it may also result from skewness or extended tails. We also examine the KDEs and the ECDFs, which we present in Figures \ref{fig:stellar_stat_kernel} and \ref{fig:stellar_stat_cumulative}. We present the complete set of values from the statistical tests in Table~\ref{tab:stellar_parameter_statistics} and a summary of the ranges, medians, and highest peak centers for each parameter and classification in Table~\ref{tab:stellar_parameter_vals_summ}. We restrict the quantitative comparison to the BCEP and hybrid classes because the corresponding samples of stars in EBs are substantially smaller. For the EB populations, we therefore discuss the distributions descriptively and report peak values only when the sample contains at least 40 objects. The unreported BCEP+EB peaks, correspond to samples of $N=3$--22, whereas the hybrid+EB samples contain $N=3$--13 objects.

Figure~\ref{fig:stellar_stat_kernel} shows the KDEs, which provide a smoothed representation of the distribution shapes and highlight their peaks, asymmetries, and internal structure. KDEs are not shown for $\log(L/L_\odot)$ in the hybrid+EB class or for $Q$ in the BCEP+EB and hybrid+EB classes, because these samples contain only four, three, and three objects, respectively. Figure~\ref{fig:stellar_stat_cumulative} presents the corresponding ECDFs, which allow the BCEP and hybrid distributions to be compared independently of the histogram binning.

Table~\ref{tab:stellar_parameter_statistics} lists the complete results of the statistical tests for the BCEP and hybrid classes. For each parameter and classification, we report the BIC values of the one- and two-component Gaussian-mixture models and their difference, $\Delta BIC$. Positive $\Delta BIC$ values favor the two-component model. We also report the KS statistic and the corresponding raw and Benjamini-Hochberg adjusted $p$-values for the comparison between the BCEP and hybrid distributions. The KS statistic and its corresponding raw and Benjamini-Hochberg-adjusted $p$-values are repeated for the two classification rows because each KS test compares the two classes directly. We present the interpretations of these results in Section~\ref{sec:stellar_pulsation_stat}.

\begin{deluxetable}{llccc}
\tablewidth{0.8\linewidth}
\tabletypesize{\scriptsize}
\tablecaption{Summary of the value ranges, medians and highest peaks for each stellar and pulsation parameter for the different classes and for all populations combined. \label{tab:stellar_parameter_vals_summ}}
\tablehead{
\colhead{Parameter} &
\colhead{Class} &
\colhead{Range} &
\colhead{Median} &
\colhead{Peak}
}
\startdata
Distance (kpc)
    & BCEP       & 0.28--9.08 & 2.39 & 2.38 \\
    & Hybrid     & 0.15--8.67 & 2.21 & 1.91 \\
    & BCEP+EB   & 0.44--6.11 & 2.06 & 2.38 \\
    & Hybrid+EB & 0.65--5.43 & 2.02 & -- \\
    & All        & 0.15--9.08 & 2.31 & 2.38 \\ \hline
$m_V$ (mag)
    & BCEP       & 1.25--13.66 & 10.26 & 10.61 \\
    & Hybrid     & 0.58--14.06 & 9.83 & 9.62 \\
    & BCEP+EB   & 0.97--12.21 & 9.59 & 9.62 \\
    & Hybrid+EB & 5.99--12.33 & 10.00 & -- \\
    & All        & 0.58--14.06 & 9.96 & 9.62 \\ \hline
$M_G$ (mag)
    & BCEP       & ($-6.76$)--1.50 & $-3.46$ & $-3.17$ \\
    & Hybrid     & ($-6.63$)--($-0.73$) & $-3.29$ & $-3.17$ \\
    & BCEP+EB   & ($-5.44$)--($-2.37$) & $-3.75$ & -- \\
    & Hybrid+EB & ($-5.31$)--($-1.74$) & $-3.10$ & -- \\
    & All        & ($-6.76$)--1.50 & $-3.37$ & $-3.17$ \\ \hline
$M_{BP}$ (mag)
    & BCEP       & ($-7.50$)--1.65 & $-3.54$ & $-3.26$ \\
    & Hybrid     & ($-7.60$)--($-0.78$) & $-3.42$ & $-3.26$ \\
    & BCEP+EB   & ($-5.66$)--($-2.51$) & $-3.93$ & -- \\
    & Hybrid+EB & ($-5.64$)--($-1.73$) & $-3.20$ & -- \\
    & All        & ($-7.60$)--1.65 & $-3.49$ & $-3.26$ \\ \hline
$M_{RP}$ (mag)
    & BCEP       & ($-5.31$)--1.25 & $-3.11$ & $-3.00$ \\
    & Hybrid     & ($-4.92$)--($-0.55$) & $-3.05$ & $-3.00$ \\
    & BCEP+EB   & ($-4.89$)--($-2.14$) & $-3.44$ & -- \\
    & Hybrid+EB & ($-4.62$)--($-1.78$) & $-2.95$ & -- \\
    & All        & ($-5.31$)--1.25 & $-3.11$ & $-3.00$ \\ \hline
$T_{\text{eff}}$ (kK)
    & BCEP       & 18.91--42.45 & 25.29 & 24.41 \\
    & Hybrid     & 17.29--36.26 & 24.44 & 23.58, 24.41 \\
    & BCEP+EB   & 17.37--40.00 & 24.74 & 25.25 \\
    & Hybrid+EB & 21.84--34.97 & 23.09 & -- \\
    & All        & 17.29--42.45 & 24.76 & 24.41 \\ \hline
$\log g$
    & BCEP       & 3.11--4.49 & 3.89 & 3.86 \\
    & Hybrid     & 2.47--4.60 & 3.86 & 3.93 \\
    & BCEP+EB   & 3.11--4.32 & 3.94 & 4.07, 4.14 \\
    & Hybrid+EB & 3.79--4.02 & 3.94 & -- \\
    & All        & 2.47--4.60 & 3.88 & 3.93 \\ \hline
$\log(L/L_\odot)$
    & BCEP       & 3.45--5.34 & 4.19 & 4.18 \\
    & Hybrid     & 2.74--5.15 & 4.08 & 4.04 \\
    & BCEP+EB   & 3.81--4.94 & 4.35 & -- \\
    & Hybrid+EB & 3.82--4.98 & 4.09 & -- \\
    & All        & 2.74--5.34 & 4.13 & 4.18 \\ \hline
$Q$ (d)
    & BCEP       & 0.015--0.084 & 0.035 & 0.036 \\
    & Hybrid     & 0.007--0.067 & 0.032 & 0.036 \\
    & BCEP+EB   & 0.025--0.038 & 0.027 & -- \\
    & Hybrid+EB & 0.021--0.094 & 0.023 & -- \\
    & All        & 0.007--0.094 & 0.033 & 0.036 \\ \hline
$f_1$ ($d^{-1}$)
    & BCEP       & 3.28--16.63 & 5.84 & 5.08 \\
    & Hybrid     & 3.31--16.71 & 6.32 & 6.12 \\
    & BCEP+EB   & 3.78--13.41 & 5.80 & -- \\
    & Hybrid+EB & 4.34--9.18 & 6.77 & -- \\
    & All        & 3.28--16.71 & 6.11 & 5.60 \\ \hline
$A_1$ (ppt)
    & BCEP       & 0.14--135.58 & 11.51 & 5.62 \\
    & Hybrid     & 0.07--54.81 & 7.11 & 5.62 \\
    & BCEP+EB   & 0.91--38.50 & 7.50 & -- \\
    & Hybrid+EB & 5.78--36.84 & 12.05 & -- \\
    & All        & 0.07--135.58 & 8.70 & 5.62 \\
\enddata
\tablecomments{Peak refers to the highest-bin center. When two bins share the highest count, all corresponding centers are listed. Peaks are not reported for samples with fewer than 40 data points.}
\end{deluxetable}

\begin{deluxetable*}{llccccccc}
\tabletypesize{\scriptsize}
\tablecaption{Gaussian-mixture BIC comparisons and two-sample KS-test results for the BCEP and hybrid distributions.\label{tab:stellar_parameter_statistics}}
\tablehead{
\colhead{Parameter} &
\colhead{Class} &
\colhead{$N$} &
\colhead{BIC(1)} &
\colhead{BIC(2)} &
\colhead{$\Delta BIC$} &
\colhead{$D_{\rm KS}$} &
\colhead{$p_{\rm raw}$} &
\colhead{$p_{\rm BH}$}
}
\startdata
Distance
    & BCEP   & 329 & 1072.68 & 1041.42 & 31.26 & 0.12 & 0.005889 & 0.011266 \\
    & Hybrid & 454 & 1374.78 & 1336.47 & 38.31 & 0.12 & 0.005889 & 0.011266 \\
$T_{\text{eff}}$
    & BCEP   & 278 & 1486.17 & 1431.51 & 54.66 & 0.19 & $9.612\times10^{-6}$ & $3.574\times10^{-5}$ \\
    & Hybrid & 380 & 2006.24 & 1961.24 & 45.00 & 0.19 & $9.612\times10^{-6}$ & $3.574\times10^{-5}$ \\
$\log g$
    & BCEP   & 278 & 15.74 & 17.82 & $-2.08$ & 0.08 & 0.188219 & 0.226449 \\
    & Hybrid & 380 & 180.73 & 130.09 & 50.64 & 0.08 & 0.188219 & 0.226449 \\
$\log(L/L_\odot)$
    & BCEP   & 120 & 94.34 & 102.19 & $-7.84$ & 0.15 & 0.069025 & 0.108467 \\
    & Hybrid & 159 & 153.28 & 167.98 & $-14.70$ & 0.15 & 0.069025 & 0.108467 \\
$Q$
    & BCEP   & 125 & $-756.46$ & $-765.19$ & 8.73 & 0.21 & 0.006145 & 0.011266 \\
    & Hybrid & 135 & $-838.82$ & $-831.82$ & $-7.00$ & 0.21 & 0.006145 & 0.011266 \\
$m_V$
    & BCEP   & 354 & 1423.13 & 1269.16 & 153.97 & 0.17 & $9.748\times10^{-6}$ & $3.574\times10^{-5}$ \\
    & Hybrid & 527 & 2247.85 & 1945.31 & 302.54 & 0.17 & $9.748\times10^{-6}$ & $3.574\times10^{-5}$ \\
$M_G$
    & BCEP   & 143 & 416.51 & 381.98 & 34.52 & 0.12 & 0.200849 & 0.226449 \\
    & Hybrid & 185 & 456.76 & 467.28 & $-10.53$ & 0.12 & 0.200849 & 0.226449 \\
$M_{BP}$
    & BCEP   & 143 & 438.70 & 400.24 & 38.46 & 0.12 & 0.205863 & 0.226449 \\
    & Hybrid & 185 & 478.43 & 483.74 & $-5.31$ & 0.12 & 0.205863 & 0.226449 \\
$M_{RP}$
    & BCEP   & 143 & 371.62 & 340.28 & 31.35 & 0.11 & 0.229426 & 0.229426 \\
    & Hybrid & 185 & 406.06 & 422.80 & $-16.73$ & 0.11 & 0.229426 & 0.229426 \\
$f_1$
    & BCEP   & 342 & 1509.10 & 1366.52 & 142.58 & 0.13 & 0.002541 & 0.006988 \\
    & Hybrid & 421 & 1920.03 & 1824.35 & 95.68 & 0.13 & 0.002541 & 0.006988 \\
$A_1$
    & BCEP   & 342 & 2766.13 & 2625.03 & 141.10 & 0.25 & $6.356\times10^{-11}$ & $6.991\times10^{-10}$ \\
    & Hybrid & 421 & 3046.87 & 2814.42 & 232.45 & 0.25 & $6.356\times10^{-11}$ & $6.991\times10^{-10}$ \\
\enddata
\tablecomments{BIC(1) and BIC(2) are the Bayesian information criteria for the one- and two-component Gaussian-mixture models, respectively, and $\Delta BIC={\rm BIC}(1)-{\rm BIC}(2)$; $D_{\rm KS}$ is the two-sample Kolmogorov--Smirnov statistic comparing the BCEP and hybrid samples.}
\end{deluxetable*}
\onecolumngrid

\section{Velocity Statistics} \label{sec:app_vel}
We define stars as kinematic outliers as those having tangential velocities exceeding three standard deviations above the sample median ($183.51~\mathrm{km\,s^{-1}}$), as well as stars with directional deviations exceeding three standard deviations from the median ($46.30^{\circ}$) and tangential velocities greater than $20~\mathrm{km\,s^{-1}}$ (see Section~\ref{sec:coords_and_vel}). These criteria identify 16 stars with exceptional tangential velocities and 39 with exceptional directions; two stars satisfy both criteria. These stars are listed in Table~\ref{tab:tangential_velocity_candidates}.

\begin{deluxetable}{lcc}
\tabletypesize{\scriptsize}
\tablecaption{Stars with exceptional tangential velocities or directions relative to the Galactic plane.
\label{tab:tangential_velocity_candidates}}
\tablehead{
\colhead{TIC} &
\colhead{$v_{\rm tan}$} &
\colhead{$\theta_{\rm GP}$} \\
\colhead{} &
\colhead{($\mathrm{km\,s^{-1}}$)} &
\colhead{($\mathrm{deg}$)}
}
\startdata
\multicolumn{3}{c}{Exceptional tangential velocity only} \\[2pt]
352229000 & $316.878$ & $-42.632$ \\
130720430 & $272.716$ & $+6.249$ \\
332791486 & $266.086$ & $+4.430$ \\
4030832   & $252.942$ & $-5.609$ \\
446463792 & $250.211$ & $+2.802$ \\
439747589 & $238.266$ & $-9.687$ \\
41548989  & $228.003$ & $+5.226$ \\
384540878 & $225.792$ & $-14.069$ \\
80333031  & $219.183$ & $-12.980$ \\
466208395 & $212.342$ & $-0.343$ \\
401975322 & $208.417$ & $-36.620$ \\
49217799  & $202.705$ & $-35.270$ \\
213272222 & $193.390$ & $-3.627$ \\
303416677 & $192.022$ & $-0.524$ \\
\hline
\multicolumn{3}{c}{
Exceptional direction with $v_{\rm tan}>20\ \mathrm{km\,s^{-1}}$
} \\[2pt]
468498193 & $85.444$ & $-81.922$ \\
301263808 & $40.395$ & $-78.744$ \\
173767386 & $38.103$ & $-76.301$ \\
71737416  & $26.432$ & $-75.601$ \\
60433558  & $32.519$ & $-74.453$ \\
433182985 & $28.830$ & $-73.656$ \\
65564230  & $59.189$ & $-71.955$ \\
34003157  & $81.070$ & $-71.056$ \\
407463080 & $27.165$ & $+70.214$ \\
374982274 & $20.477$ & $-69.734$ \\
198165966 & $56.492$ & $+68.172$ \\
15644631  & $26.194$ & $-67.711$ \\
374478187 & $28.005$ & $-64.951$ \\
121679880 & $22.886$ & $-64.901$ \\
247649462 & $21.041$ & $-63.292$ \\
422666117 & $127.441$ & $+61.320$ \\
113404375 & $20.242$ & $-60.407$ \\
380870688 & $22.871$ & $-58.318$ \\
196639455 & $22.352$ & $-57.203$ \\
308954763 & $31.723$ & $-56.910$ \\
349351160 & $88.306$ & $-55.204$ \\
28565824  & $28.571$ & $+55.112$ \\
347378183 & $22.773$ & $-54.252$ \\
314393072 & $24.025$ & $-53.587$ \\
246097860 & $118.657$ & $-53.456$ \\
390432717 & $75.082$ & $-53.035$ \\
195407015 & $26.632$ & $-52.789$ \\
452188490 & $59.663$ & $-52.303$ \\
435123631 & $46.919$ & $+51.610$ \\
211408145 & $166.723$ & $-51.074$ \\
34803944  & $86.095$ & $+50.109$ \\
65516748  & $85.088$ & $-49.977$ \\
359424879 & $61.295$ & $-49.152$ \\
76426958  & $82.172$ & $-48.762$ \\
435742094 & $54.003$ & $+48.258$ \\
217755761 & $33.306$ & $-47.553$ \\
458246279 & $25.792$ & $-47.099$ \\
\hline
\multicolumn{3}{c}{Both criteria} \\[2pt]
98084512  & $244.952$ & $+88.314$ \\
57384424  & $229.423$ & $-67.131$ \\
\enddata
\tablecomments{
The angle $\theta_{\rm GP}$ gives the signed direction of the tangential
velocity relative to the Galactic plane. Positive and negative values
indicate motion toward positive and negative Galactic latitude,
respectively.
}
\end{deluxetable}
\onecolumngrid

\begin{figure*}
    \centering
    \includegraphics[width=1\textwidth]{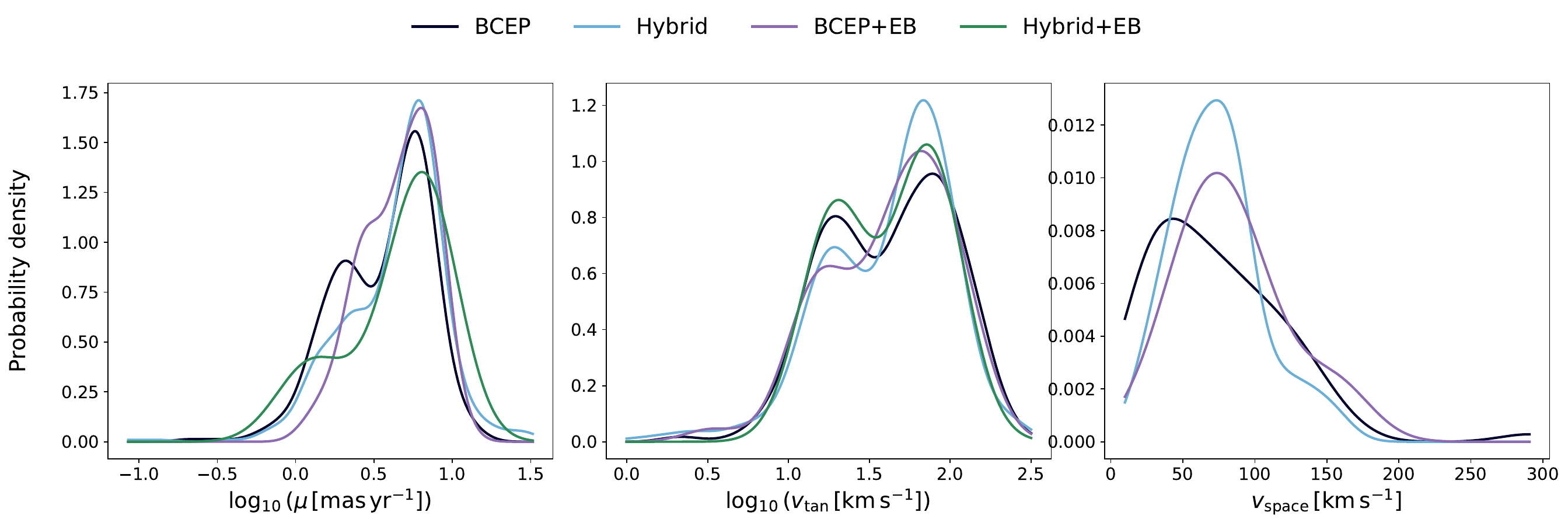}
    \caption{Kernel density estimates of the proper-motion (left),
    tangential-velocity (middle), and space-velocity (right) distributions
    for the different classifications.}
    \label{fig:velocity_kdes}
\end{figure*}

Figure~\ref{fig:velocity_kdes} presents kernel density estimates of the proper-motion, tangential-velocity, and space-velocity distributions. These estimates provide a complementary visualization of the distribution shapes and possible bimodality discussed in Section~\ref{sec:coords_and_vel}.

\end{document}